\documentclass[a4paper,11pt]{article}

\pdfoutput=1

\usepackage{jheppub}
\usepackage{amsmath}
\usepackage{xspace}
\usepackage{hyperref}
\usepackage{mdwlist}

\newcommand{\eq}[1]{eq.~\eqref{eq:#1}}
\newcommand{\eqs}[2]{eqs.~\eqref{eq:#1} and \eqref{eq:#2}}
\renewcommand{\sec}[1]{sec.~\ref{sec:#1}}

\newcommand{\app}[1]{app.~\ref{app:#1}}
\newcommand{\fig}[1]{fig.~\ref{fig:#1}}

\newcommand{\nnu}{\nonumber\\}
\newcommand{\nn}{\nonumber}
\newcommand{\bef}{\begin{figure}[t]\centering}
\newcommand{\eef}{\end{figure}}
\def\bea#1\eea{\begin{align}#1\end{align}}
\def \be  {\begin{equation}}
\def \ee  {\end{equation}}
\def \ba  {\begin{eqnarray}}
\def \ea  {\end{eqnarray}}

\newcommand{\f}{\frac}
\newcommand{\ord}[1]{\mathcal{O}(#1)}

\newcommand{\df}{\mathrm{d}}

\newcommand{\sdt}{\!\cdot\!}

\newcommand{\al}{\alpha}
\newcommand{\bt}{\beta}
\newcommand{\ga}{\gamma}

\newcommand{\de}{\delta}
\newcommand{\eps}{\epsilon}

\newcommand{\si}{\sigma}

\newcommand{\cG}{{\mathcal G}}

\newcommand{\bn}{\bar{n}}

\newcommand{\bnslash}{\bar{n}\!\!\!\slash}

\newcommand{\wtb}{\widetilde{\beta}}

\newcommand\as{\alpha_s}
\newcommand{\lqcd}{\Lambda_\mathrm{QCD}}

\newcommand{\Pythia}{\textsc{Pythia}\xspace}

\allowdisplaybreaks[2]

\title{The Jet Shape at NLL$'$}

\author[a,b]{Pedro Cal,}
\author[c]{Felix Ringer,}
\author[a,b]{Wouter J. Waalewijn}

\affiliation[a]{Institute for Theoretical Physics Amsterdam and Delta Institute for Theoretical Physics, University of Amsterdam, Science Park 904, 1098 XH Amsterdam, The Netherlands}
\affiliation[b]{Nikhef, Theory Group, Science Park 105, 1098 XG, Amsterdam, The Netherlands}
\affiliation[c]{Nuclear Science Division, Lawrence Berkeley National Laboratory, Berkeley, CA 94720, USA}
                                         
\emailAdd{p.cal@nikhef.nl}
\emailAdd{fmringer@lbl.gov}
\emailAdd{w.j.waalewijn@uva.nl}

\abstract{The jet shape is the fraction of the jet energy within a cone $r$ centered on the jet axis. We calculate the jet shape distribution at next-to-leading logarithmic accuracy plus next-to-leading order (NLL$'$), accounting for logarithms of both the jet radius $R$ and the ratio $r/R$. This is the first phenomenological study that takes the recoil of the jet axis due to soft radiation into account, which is needed to reach this accuracy, but complicates the calculation of collinear radiation and requires the treatment of rapidity logarithms and non-global logarithms. We present numerical results, finding good agreement with ATLAS and CMS measurements of the jet shape in an inclusive jet sample, $pp \to {\rm jet}+X$, for different kinematic bins. The effect of the underlying event and hadronization are included using a simple one-parameter model, since they are not part of our perturbative calculation.
}

\preprint{NIKHEF 2018-035}

\begin{document}
\maketitle

\section{Introduction}
\label{sec:intro}

The jet shape is a classic jet substructure observable that maps out the transverse energy profile of jets~\cite{Ellis:1992qq}. It is one of the most frequently studied jet substructure observables, which has been measured at a variety of collider experiments over the past decades. The existing data sets include jet shape measurements in $pp$, $p\bar p$, $ep$, $e^+e^-$ and heavy-ion collisions, and the range of different center-of-mass energies ($E_{\rm cm}$) makes the jet shape a unique testing ground for precision QCD studies. In order to achieve a meaningful comparison between the experimental data and theoretical calculations, we need to be able to make precise predictions within perturbative QCD, which is the goal of this work. 

Jet shapes have been used to constrain parton shower event generators, including their models of hadronization and the underlying event contribution, see e.g.~ref.~\cite{ATLAS:2011gmi}. Furthermore, QCD predicts that the distribution of particles in gluon jets is broader than in quark jets, making the jet shape a useful observable to discriminate between quark and gluon jets (see ref.~\cite{Gras:2017jty} for a recent review). More generally, jet substructure techniques have started to play an important role in the search for physics beyond the standard model~\cite{Adams:2015hiv}, where the separation of boosted objects from QCD jets requires a sophisticated understanding of jet substructure. For recent reviews of jet substructure techniques and their applications, see refs.~\cite{Larkoski:2017jix,Asquith:2018igt}.

\begin{figure}[t]
    \centering
 \includegraphics[width=0.5\textwidth]{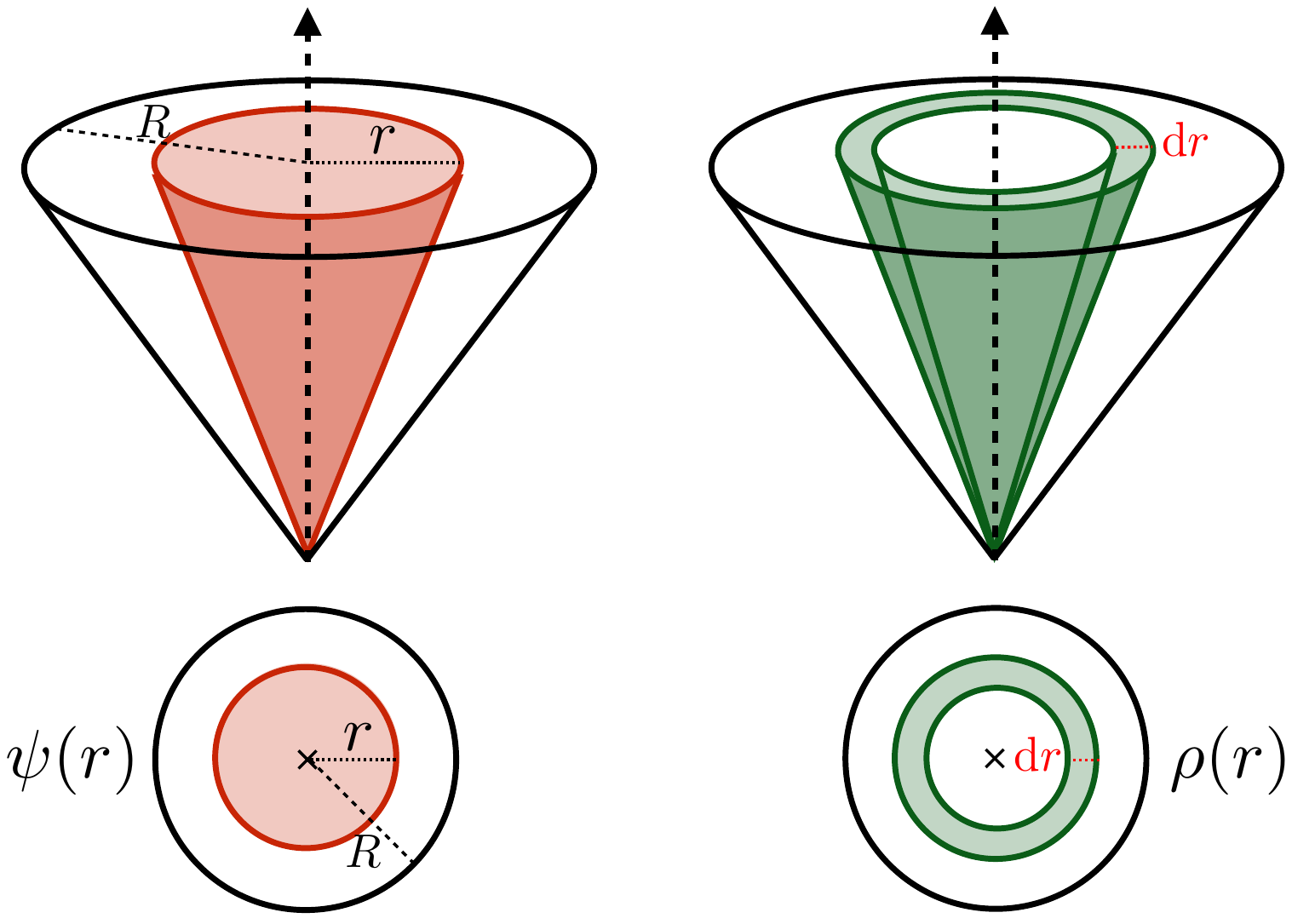} \\
    \caption{The integrated (differential) jet shape, shown in the left (right) panel, is the fraction of the jet transverse momentum contained in a circle (annulus) in $(\eta,\phi)$ coordinates, centered on the jet axis.}
    \label{fig:jetshape}
\end{figure}

We consider inclusive jet production, $pp\to{\rm jet}+X$, where any observed jet in a given transverse momentum $p_T$ and rapidity $\eta$ interval is taken into account, and we sum over everything else ($X$) in the final state. Given such an identified jet with radius $R$, the integrated jet shape $\psi(r)$ and the differential jet shape $\rho(r)$ are defined as follows
\be\label{eq:def}
\psi(r)=\frac{\sum_{r_i<r} \, p_{Ti}}{\sum_{r_i<R} \, p_{Ti}}\,,\qquad \rho(r)=\frac{\df \psi(r)}{\df r}
\,,\ee
see \fig{jetshape}.
Here $r_i$ denotes the distance in the $(\eta,\phi)$ plane of particle $i$ in the jet to the jet axis, and $p_{Ti}$ is its transverse momentum with respect to the beam axis. The dependence on the $p_T$, $\eta$ and radius $R$ of the observed jet is left implicit. The integrated jet shape is normalized by construction, i.e.~$\psi(R)=1$. Note that it is important for the jet shape which jet axis is chosen. We will focus on the standard jet axis, which is consistent with the currently existing data sets. For a discussion of the jet shape using the winner-take-all axis~\cite{Bertolini:2013iqa}, see refs.~\cite{Kang:2017mda,Neill:2018wtk}.

The first jet shape measurements were already performed by the OPAL collaboration at LEP~\cite{Akers:1994wj}. It has also been measured at the Tevatron~\cite{Abe:1992wv,Abachi:1995zw,Acosta:2005ix}, and at HERA in both deep inelastic scattering~\cite{Aid:1995ma,Adloff:1997gq,Adloff:1998ni,Breitweg:1998gf} and photoproduction~\cite{Breitweg:1997gg}. At the LHC, jet shape measurements on an inclusive jet sample were performed by ATLAS~\cite{Aad:2011kq} and CMS~\cite{Chatrchyan:2012mec} at $E_{\rm cm}=$ 7~TeV. The jet shape has also been measured for heavy-flavor jets~\cite{Aaltonen:2008de,Aad:2013fba}. We leave an extensive comparison to all these data sets for future work, focussing only on the LHC data in this paper. In recent years, jet shapes have also received a growing attention in heavy-ion collisions as a probe of the properties of the quark-gluon plasma. The transverse energy profile of jets that traverse the hot and dense QCD medium gets modified in comparison to jets in $pp$ collisions. See refs.~\cite{Chatrchyan:2013kwa,CMS:2018ogb} for recent experimental results from the LHC.

In this work we build upon the framework for jet shapes developed by some of us in the context of subjet distributions~\cite{Kang:2017mda}, and extend it to full next-to-leading logarithmic (NLL$'$) accuracy. We employ Soft Collinear Effective Theory (SCET)~\cite{Bauer:2000ew, Bauer:2000yr, Bauer:2001ct, Bauer:2001yt,Beneke:2002ph} in order to achieve the resummation of $\ln R$ and $\ln(r/R)$. The starting point is our factorization of the cross section differential in jet transverse momentum $p_T$, rapidity $\eta$, and the energy fraction $z_r$ contained in a subjet of radius $r$ centered along the jet axis, 
\be\label{eq:factorization1}
\frac{\df\sigma}{\df p_T \,\df \eta\, \df z_r}= \sum_{a,b,c} f_a(x_a,\mu)\otimes f_b(x_b,\mu)\otimes {\cal H}_{ab}^c\left(x_a, x_b, \eta, p_T/z, \mu\right)\otimes \cG_c^{\rm jet}(z, z_r, p_T R , r/R, \mu) \,.
\ee
Here $f_{a,b}$ denote the PDFs for the incoming protons, the hard functions ${\cal H}_{ab}^c$ describe the hard scattering $ab \to c+X$ and $\otimes$ denote appropriate integrals over the momentum fractions $x_{a,b}$ and $z$ (see \eq{fact} for more details). The production of the jet, including the measurement of the energy fraction $z_r$, is captured by the jet function $\cG_c^{\rm jet}$. The factorization in \eq{factorization1} holds for narrow jets and is analogous to inclusive hadron production, with fragmentation functions replaced by jet functions~\cite{Kaufmann:2015hma,Kang:2016mcy,Dai:2016hzf}. The integrated jet shape $\psi(r)$ in \eq{def} is then given by the energy average of the $z_r$ differential cross section normalized by the inclusive jet cross section
\be \label{eq:jetshape}
\psi(r)=\int_0^1 dz_r\, z_r \f{\df\sigma}{\df p_T \,\df \eta\, \df z_r} \, \Big/ \, \f{\df\sigma}{\df p_T \,\df \eta} \,.
\ee
The expression for $\psi(r)$ involves single logarithms in the jet radius parameter $\alpha_s^n\ln^nR$ and double logarithms in the ratio of the two jet radii $\alpha_s^n\ln^{2n}(r/R)$, which can be large and will be resummed. The resummation of logarithms in the jet radius parameter $R$ follows from the usual DGLAP evolution equations satisfied by the jet function $\cG_c^{\rm jet}$. This was found to be a characteristic feature of single-inclusive jet substructure observables, see e.g.~refs.~\cite{Dasgupta:2014yra,Kang:2016ehg,Dai:2016hzf,Neill:2016vbi}.

\begin{figure}[t]
    \centering
    \includegraphics[width=0.5\textwidth]{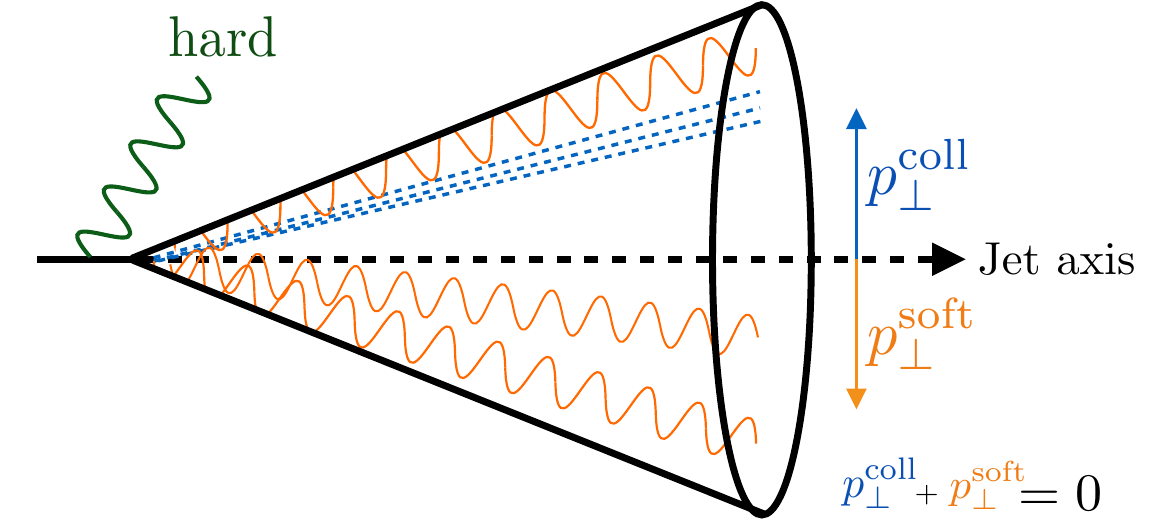}
    \caption{Illustration of the refactorized expression of the jet function ${\cal G}_c$ in \eq{refact1} in the limit $r\ll R$. Hard radiation at the jet scale $p_T R$ is allowed outside the observed jet. The collinear radiation (blue) is offset from the jet axis, due to the recoil from the soft radiation (orange) inside the jet, since the jet axis is along the jet momentum.\label{fig:refactorizationrecoil}}
\end{figure}

The resummation of logarithms in $r/R$ requires a treatment within SCET$_{\rm II}$ due to the recoil-effect of soft radiation, as pointed out in ref.~\cite{Kang:2017mda}. This resummation is thus similar to that encountered for transverse momentum dependent observables~\cite{Collins:1984kg,Becher:2010tm,Collins:2011zzd,GarciaEchevarria:2011rb,Chiu:2012ir,Kang:2017glf}.
It is accomplished by refactorizing the jet function $\cG_c^{\rm jet}$ in \eq{factorization1} in the limit $r\ll R$ as
\begin{align} \label{eq:refact1}
\cG_c^{\mathrm{jet}}(z, z_r, p_T R , r/R, \mu) &\stackrel{{\rm NLL}'}{=} \sum_d H_{cd}(z,p_{T} R,\mu)\,
\int\! \df^2 k_\perp\,
 C_d(z_r, p_T r, k_\perp,\mu,\nu) 
\\ & \qquad \times
 S_d^{\rm G}(k_\perp,\mu, \nu R)\, S_d^{\rm NG}\Big(\frac{r}{R}\Big)
\Big[1+ \mathcal{O}\Big(\frac{r}{R} \Big)\Big]\, .
\nn \end{align}
Here, the hard functions $H_{cd}$ describe how the parton $c$ coming from the hard-scattering produces a jet of size $R$ and of parton flavor $d$, carrying a fraction $z$ of the initial parton. The collinear function $C_d$ and the soft function $S_d^{\rm G}$ take into account collinear and (global) soft radiation inside the jet. We integrate over the transverse momentum $k_\perp$ generated by the soft radiation, accounting for its recoil on the collinear radiation. By solving the associated renormalization group (RG) equations of the different functions, we achieve the resummation of logarithms in $r/R$. This involves both the standard renormalization group evolution in the invariant mass scale $\mu$, as well as an evolution in the rapidity scale $\nu$, as discussed in \sec{framework}. We calculated the collinear functions $C_d$ for the first time at one-loop order, accounting for the dependence on the recoil $k_\perp$. We also include the contribution of non-global logarithms (NGLs)~\cite{Dasgupta:2001sh}, which are captured by the function $S_d^{\rm NG}$ in \eq{refact1}, but only affect the region where $r/R$ is very small. The simple (multiplicative) treatment of the NGLs in \eq{refact1} is the reason this equation is only valid to NLL$'$ accuracy. The refactorized cross section in eq.~(\ref{eq:refact1}) which is given in terms of hard, collinear and soft functions in the limit $r\ll R$ is illustrated in fig.~\ref{fig:refactorizationrecoil}. In particular, we show the effect of the recoil due to the soft radiation inside the jet.

The perturbative order of the various ingredients needed for the $\ln R$ and $\ln(r/R)$ resummation is summarized in table~\ref{tab:orders}. We work at NLL$'$ accuracy in the $\ln (r/R)$ resummation, as this is commensurate with an NLL resummation of $\ln R$. Going beyond this accuracy will be daunting, especially in the treatment of non-global logarithms.
We would like to stress that earlier jet shape calculations~\cite{Seymour:1997kj,Li:2011hy,Chien:2014nsa} are formally only accurate to leading-logarithmic (LL) order in their treatment of the logarithms of $r/R$. In particular, the contribution from the rapidity evolution, as well as the non-global logarithms, that both first enter at NLL accuracy, are included here for the first time. In addition, earlier calculations did not account for the inclusive jet sample, where the resummation of logarithms of the jet radius $R$ changes the ratio of quark and gluon jets, thereby affecting the spectrum already at LL accuracy.

Finally, comparisons between our predictions and LHC data suggested a significant effect of nonperturbative physics (underlying event and hadronization) on the jet shape, particularly for small jet $p_T$. This is consistent with the picture that arises from studying these effects in \Pythia~\cite{Sjostrand:2014zea}. We will include them in our analysis by using a simple one-parameter model, finding good agreement with the data.

\begin{table}[t]
  \centering
  \begin{tabular}{l l | c c c c c c}
  \hline \hline
  & & Fixed-order & $\beta$ & $\ga_\mu$ & $\ga_\nu$ & NGLs \\ \hline
  $\ln R$ & LL & tree & $1$-loop & $1$-loop & - & - \\
  & NLL & $1$-loop & $2$-loop & $2$-loop & - & - \\
  & NNLL & $2$-loop & $3$-loop & $3$-loop & - & - \\ \hline
  $\ln (r/R)$ & LL & tree & $1$-loop & $1$-loop & - & - \\
  & NLL & tree & $2$-loop & $2$-loop & $1$-loop & LL \\ 
  & NLL$'$ & $1$-loop & $2$-loop & $2$-loop & $1$-loop & LL \\   
  & NNLL & $1$-loop & $3$-loop & $3$-loop & $2$-loop & NLL \\
  \hline\hline
  \end{tabular}
  \caption{The perturbative ingredients needed at various orders in the $\ln R$ and $\ln (r/R)$ resummation. The columns correspond to the loop order of the fixed-order ingredients, the QCD beta function, the $\mu$ and $\nu$ anomalous dimensions, and the non-global logarithms. The non-cusp part of the $\mu$ anomalous dimension is only needed at one-loop order lower than indicated above.}
\label{tab:orders}
\end{table}

The remainder of this paper is organized as follows: The theoretical framework used to calculate the jet shape is discussed in \sec{framework}, with certain ingredients relegated to the appendices. We present a detailed derivation of the collinear function, relevant for the resummation of logarithms in $r/R$, in \sec{coll}. The details that enter our numerical evaluation of the cross section  are described in~\sec{implement}, and first (perturbative) results for quark and gluon jets are presented there. Nonperturbative effects are investigated in \sec{nonp}, for which two simple models are explored. In \sec{results} we show our final results, which we compare to available $pp$ data from the LHC, and we conclude in \sec{conclusions}.

\section{Framework}
\label{sec:framework}

In this section we present the theoretical framework that we use to obtain our results. In \sec{fac} we describe the factorization formulae and how they enable resummation. We then rearrange these formulae in \sec{rearrange}, to separate them into the inclusive production of jets and the jet shape itself. In \sec{hard} the one-loop hard function is given, and in \sec{soft} we discuss the soft function and non-global logarithms. The calculation of the collinear function at one-loop, accounting for the effect of recoil, is one of the main new results and presented separately in \sec{coll}. The one-loop expressions for the jet function for $r \lesssim R$ are given in \app{large_r}, and the anomalous dimensions are listed in \app{anom}.

\subsection{Factorization and resummation}
\label{sec:fac}

The cross section describing the measurement of the fraction $z_r$ of jet energy inside the cone of radius $r$ around the jet axis, in an inclusive sample of jets produced in $pp$ collisions, factorizes as follows 
\begin{align} \label{eq:fact}
  \frac{\df \si}{\df \eta\, \df p_T\, \df z_r}
  &= \sum_{a,b,c} 
  \int\! \frac{\df x_a}{x_a}\, f_a(x_a,\mu) \int\! \frac{\df x_b}{x_b}\, f_b(x_b,\mu)
  \int\! \frac{\df z}{z}\, {\cal H}_{ab}^c(x_a, x_b, \eta, p_T/z, \mu)
  \nn \\ & \quad\times
   \cG_c^{\rm jet}(z, z_r, p_T R , r/R, \mu) \big[1+\mathcal{O}(R^2)\big]
\,.\end{align}
The parton distribution functions $f_a$ and $f_b$ describe extracting a parton of flavor $a$ and $b$ out of the proton, and the function ${\cal H}_{ab}^c$~\cite{Jager:2002xm} encodes their hard scattering in which the parton with flavor $c$, rapidity $\eta$ and transverse momentum $p_T/z$ is produced. The subsequent formation of the jet with transverse momentum $z \times p_T/z = p_T$ moving in the same direction, as well as the jet shape measurement, is encoded in the jet function $\cG_c^{\rm jet}.$\footnote{In  ref.~\cite{Kang:2017mda} this was called the central subjet function and a hat was included on top of $\cG$, to distinguish it from other jet functions in that paper.} This collinear factorization requires that $R \ll 1$ to keep the $\mathcal{O}(R^2)$ power corrections small. In several examples it has been observed that these power corrections are still small for values of $R$ up to 0.7~\cite{Mukherjee:2012uz}. Note that this factorization formula is identical to that for the inclusive fragmentation of hadrons, with the fragmentation function replaced by our jet function.

The jet function $\cG_c^{\rm jet}$ describes the formation of the jet as well as the jet shape measurement through $z_r$, and has the following matrix-element definition in SCET,
\bea \label{eq:G_def}
\cG_q^{\mathrm{jet}}(z, z_r, p_T R , r/R, \mu) 
& = 16\pi^3\,\sum_X \frac{1}{2N_c}\, {\rm Tr} \Big[\frac{\bnslash}{2}
\langle 0|\delta\Big(2 - \frac{z\, \bar n\cdot {\mathcal P}}{p_T}\Big) \delta^2({\mathcal P}_\perp) \chi_n(0)  |X\rangle 
\langle X|\bar \chi_n(0) |0\rangle \Big]
\nnu & \quad \times
\sum_{J_R\in X} \de\big(p_T - p_T(J_R)\big) \delta\Big(z_r - \frac{p_T(j_r)}{p_T}\Big)
\,,\eea
for quark jets, and analogously for gluon jets. Here we exploit that the measurement is invariant under boosts along the beam axis, to set the jet rapidity $\eta$ equal to zero.\footnote{This avoids spurious factors of $\cosh \eta$ in intermediate expressions. They arise due to the difference between the energy and transverse momentum of the jet, and are compensated for by the angular size of the jet, which also depends on $\eta$ because the jet is defined in $(\eta,\phi)$ coordinates.} We will repeatedly make use of the following decomposition of a vector $p^\mu$ in light-cone coordinates 
\begin{align}
  p^\mu = \bn \sdt p\, \frac{n^\mu}{2} + n \sdt p\, \frac{\bn^\mu}{2} + p_\perp^\mu
\,,\end{align}
where $n^\mu = (1,0,0,1)$ and $\bn^\mu = (1,0,0,-1)$ are light-like vectors, and $p_\perp^\mu$ denotes the transverse components.
The collinear field $\bar \chi_n$ in \eq{G_def} describes the quark that initiates the jet, averaged over its spin and color configurations, leading to the factor $1/(2N_c)$. It contains a  Wilson line to ensure (collinear) gauge invariance. The $\delta(2 - z\, \bar n\cdot {\mathcal P}/p_T)$ fixes the quark field to have transverse momentum $p_T/z$ with respect to the \emph{beam} axis, and the $\delta^2({\mathcal P}_\perp)$ fixes our light-cone coordinates to be along the momentum of the initial parton. The last line describes the sum over all jets $J_R$ in the final state $|X \rangle$, with transverse momentum $p_T$, of which a fraction $z_r$ is inside the central subjet $j_r$. The functions $\cG_q^{\mathrm{jet}}$ and $\cG_g^{\mathrm{jet}}$ were calculated at one loop in ref.~\cite{Kang:2017mda}, and their expressions are collected in \app{large_r}.

The factorization in \eq{fact} separates the physics at scales
\begin{align}
    \mu_f &\sim \lqcd
    \,, &
    \mu_{\cal H} &\sim p_T
    \,, &
    \mu_{\cal G} &\sim p_T R
\,.\end{align}
By evolving $\cG$ from its natural scale $ \mu_{\cal G}$ to $\mu_{\cal H}$, the logarithms of $\mu_{\cal G}/\mu_{\cal H} \sim R$ are resummed. This involves the DGLAP evolution, 
\ba \label{eq:G_evo}
\mu\f{\df}{\df\mu}\,\cG_i^{\mathrm{jet}}(z, z_r, p_T R , r/R, \mu)=\sum_j\int_z^1\f{\df z'}{z'}\,\f{\as}{\pi} P_{ji}(z/z') \,\cG_j^{\mathrm{jet}}(z', z_r, p_T R , r/R, \mu)
\,,\ea
where the one-loop splitting functions are given by 
\begin{align} \label{eq:split}
P_{qq}(z) &= C_F\,\Big(\f{1+z^2}{1-z}\Big)_+ \,,
&
P_{gq}(z) &= C_F\,\f{1+(1-z)^2}{z}
\,, \nn \\
   P_{gg}(z) &= 2C_A \bigg[ \frac{z}{(1-z)}_+ + \frac{1-z}{z}+z(1-z)\bigg] + \frac{\beta_0}{2}\,\de(1-z)\,,
   &
   P_{qg}(z) &= T_F \big[z^2+(1-z)^2\big]
\,,\end{align}
and 
\begin{align} \label{eq:beta_0}
  \beta_0 = \frac{11}{3} C_A - \frac{4}{3} T_F n_f
\,.\end{align}

  \begin{table}
   \centering
   \begin{tabular}{l|c}
     \hline \hline
     Mode: & Scaling $(\bn \sdt p ,n \sdt p ,p_\perp)$  \\ \hline
     hard(-collinear) & $ p_T(1,R^2,R)$ \\
     collinear & $ p_T(1,r^2,r)$ \\
     (collinear-)soft & $ p_T(r/R,r R,r)$ \\ 
     \hline \hline
   \end{tabular}
   \caption{The parametric scalings of the momenta of the modes that enter in the factorization of the jet function for $r \ll R$.}
   \label{tab:modes}
   \end{table}

For $r \ll  R$, the jet function contains large logarithms of $r/R$ that also require resummation. This is achieved through a second factorization~\cite{Kang:2017mda}\footnote{This is identical to \eq{refact1} and repeated for convenience. Note that we adopted a slightly different convention for the arguments of the involved functions in comparison to ref.~\cite{Kang:2017mda}.}
\begin{align} \label{eq:refact}
\cG_c^{\mathrm{jet}}(z, z_r, p_T R , r/R, \mu) &\stackrel{{\rm NLL}'}{=} \sum_d H_{cd}(z,p_{T} R,\mu)\,
\int\! \df^2 k_\perp\,
 C_d(z_r, p_T r, k_\perp,\mu,\nu) 
\\ & \qquad \times
 S_d^{\rm G}(k_\perp,\mu, \nu R)\, S_d^{\rm NG}\Big(\frac{r}{R}\Big)
\Big[1+ \mathcal{O}\Big(\frac{r}{R} \Big)\Big]\, .
\nn \end{align}
The momentum scalings of the modes in SCET, corresponding to the various ingredients in \eq{refact}, are listed in table~\ref{tab:modes}. If we boost to a frame where the jet and out-of-jet region are complementary hemispheres, this is the usual power counting for hard, collinear and soft radiation, which is why we use this nomenclature instead of hard-collinear and collinear-soft.
The hard function $H_{cd}$ describes how the initial parton $c$ produces a jet of radius $R$ with parton flavor $d$ and a fraction $z$ of the initial transverse momentum with respect to the \emph{beam} axis. Within the jet, the parton $d$ can only undergo energetic splittings of angles of order $r$, otherwise the collinear radiation would lie outside the cone of radius $r$.\footnote{Of course there could be a splitting inside the jet of angular size R that is balanced in such a way that there is also collinear radiation inside the cone of size $r$ at the center of the jet, but such configurations give a power suppressed contribution. This is very similar to the power suppression of the contribution from two nearly back-to-back jets in Higgs production at small transverse momentum, discussed in e.g.~ref.~\cite{Chiu:2012ir}.} The fraction $z_r$ of this collinear radiation within the cone of size $r$ is described by the collinear function $C_d$. The collinear function also accounts for the transverse momentum offset $k_\perp$ of the initial collinear parton with respect to the \emph{jet} axis, due to recoil against soft radiation. In \sec{coll} we will present the first one-loop calculation of the collinear function for $k_\perp\neq 0$. The distribution of this recoil is encoded in the soft function, which we separate into a global contribution $S_d^{\rm G}$ and non-global logarithms $S_d^{\rm NG}$. Non-global logarithms~\cite{Dasgupta:2001sh} arise because only soft radiation \emph{inside} the jet affects the position of the axis, and our simple treatment of them is the reason why \eq{refact} only holds to NLL$'$ accuracy, see \sec{soft}. The $\mathcal{O}(r/R)$ power corrections in \eq{refact} can be extracted from $\cG_c^{\mathrm{jet}}$, and will be included.

Transverse momentum dependent observables generically suffer from rapidity divergences. We will employ the $\eta$-regulator~\cite{Chiu:2011qc,Chiu:2012ir}, for which $\nu$ denotes the corresponding rapidity renormalization scale. The factorization in \eq{refact} separates the jet function into ingredients at the scales
\begin{align} \label{eq:can}
  \mu_H &\sim p_T R
  \,, &
  \mu_C &\sim p_T\, r
  \,, &
  \mu_{S^{\rm G}} &\sim p_T\, r\,,  
  \nn \\
  & &
  \nu_C &\sim p_T
  \,,&
  \nu_{S^{\rm G}} &\sim p_T\, \frac{r}{R}\,.
\end{align}
By evaluating the ingredients at their natural scales and using the RG evolution to evolve them to a common scale, the global logarithms of $\mu_C/\mu_H \sim \mu_{S^{\rm G}}/\mu_H \sim \nu_{S^{\rm G}} / \nu_C \sim r/R$ are resummed. The RG equations are\footnote{The expressions in ref.~\cite{Kang:2017mda} contain a typo, as the convolution in $k_\perp$ for the $\nu$-RG equations was omitted.}
\begin{align} \label{eq:RGE}
\mu\f{\df}{\df\mu}\,H_{cd}(z,p_T R,\mu) &= \sum_e \int_z^1\f{\df z'}{z'}\,\gamma^H_{ce}\Big(\f{z}{z'},p_T R,\mu\Big)\,H_{ed}(z',p_T R,\mu)
\,. \nn \\
\mu\f{\df}{\df\mu}\,C_d(z_r,p_T r,k_\perp,\mu,\nu) &= \gamma^C_d(\mu, \nu / p_T)\,C_d(z_r,p_T r,k_\perp,\mu,\nu)
\,. \nn \\
\mu\f{\df}{\df\mu}\,S_d^{\rm G}(k_\perp, \mu,\nu R) &= \gamma^S_d(\mu, \nu R)\,S_d^{\rm G}(k_\perp,\mu,\nu R)
\,, \nn \\
\nu\f{\df}{\df\nu}\,C_d(z_r,p_T r,k_\perp,\mu,\nu) &= -\int\! \frac{\df^2 k_\perp'}{(2\pi)^2}\,
\gamma^\nu_d(k_\perp-k_\perp',\mu)\,C_d(z_r,p_T r,k_\perp',\mu,\nu)
\,. \nn \\
\nu\f{\df}{\df\nu}\,S_d^{\rm G}(k_\perp,\mu,\nu R) &= 
\int\! \frac{\df^2 k_\perp'}{(2\pi)^2}\,
\gamma^\nu_d(k_\perp-k_\perp',\mu)\,S_d^{\rm G}(k_\perp',\mu,\nu R)
\,. \end{align}
The anomalous dimensions are collected in \app{anom}. As is clear from \eq{refact}, the anomalous dimensions of the hard, collinear and soft function should combine to give the anomalous dimension of the jet function, which we checked. The $\mu$-evolution sums double logarithms, and the $\nu$-evolution, which was missed prior to our work in ref.~\cite{Kang:2017mda}, sums single logarithms. We also include the leading non-global logarithms to obtain the desired NLL$'$ accuracy, which will be discussed in \sec{soft}.

\subsection{Separating the jet production and jet shape}
\label{sec:rearrange}

We now rearrange our calculation in a way that simplifies the numerical implementation, effectively separating the jet production and the jet shape.\footnote{A similar rearrangement was carried out for hadron fragmentation inside a jet in ref.~\cite{Kaufmann:2015hma}.} We start by writing $\cG_c^{\rm jet}$ as  
\begin{align} \label{eq:rearrange}
  \cG_c^{\rm jet}(z, z_r, p_T R , r/R, \mu) 
  &= \sum_d J_{cd}(z, p_T R , \mu) \int\! \df z'\, \Bigl[\cG_d^{\rm jet}(z', z_r, p_T R , r/R, \mu) 
  \nn \\ & \quad - J_d^{(1)}(z', p_T R , \mu)\, \de(1-z_r)\Bigr] 
  + \ord{\al_s^2}
\,.\end{align}
Here $J_d$ is the semi-inclusive jet function~\cite{Kang:2016mcy,Dai:2016hzf} that enters in inclusive jet production.  $J_{cd}$ is directly related, except that it not only keeps track of the flavor $c$ of the initiating parton but also the flavor $d$ of the jet, so 
\begin{align} \label{eq:flavor_sum}
\sum_d J_{cd}(z, p_T R , \mu) = J_c(z, p_T R , \mu)
\,.\end{align}
Its one-loop expressions are given in \eq{J_cd}.
Note that \eq{rearrange} is not a factorization of physics at different scales, as the natural scale of $\mathcal{G}^{\rm jet}$ and $J$ is both $p_T R$. However, this is why it is justified to work to finite order in $\alpha_s$. We exploited that at one-loop order the nontrivial $z$-dependence cancels between $\cG^{\rm jet}_d$ and $J_d$, i.e.~their difference is proportional to $\de(1-z)$, since the splitting where one parton is outside the jet is treated the same in both calculations. This delta function is removed by the integral over $z$. Combining $J_{cd}$ with the rest of \eq{fact}, we identify this as the cross section for the inclusive production of jets of flavor $d$,
\begin{align} \label{eq:incl_jets}
  \frac{\df \si_d}{\df \eta\, \df p_T}
  &= \sum_{a,b,c} 
  \int\! \frac{\df x_a}{x_a}\, f_a(x_a,\mu) \int\! \frac{\df x_b}{x_b}\, f_b(x_b,\mu)
  \int\! \frac{\df z}{z}\, {\cal H}_{ab}^c(x_a, x_b, \eta, p_T/z, \mu)
  \nn \\ & \quad\times
   J_{cd}(z, p_T R, \mu) \big[1+\mathcal{O}(R^2)\big]
\,.\end{align}
By summing over $d$ and using \eq{flavor_sum}, this reproduces the inclusive jet cross section in refs.~\cite{Kang:2016mcy,Dai:2016hzf}.  Eq.~\eqref{eq:jetshape} then implies that this remainder corresponds to the jet shape $\psi_c(r)$ for a jet of flavor $c$, after taking the second Mellin moment of $z_r$,
\begin{align}
  \psi_d(r) = \int\! \df z\, \biggl[\int\! \df z_r\, z_r\, \cG_d^{\rm jet}(z, z_r, p_T R , r/R, \mu) - J_d^{(1)}(z, p_T R , \mu)\biggr] + \ord{\al_s^2}
\,.\end{align}
Using the expressions in \app{large_r}, the jet shape for $r\lesssim R$ is given by
\begin{align} \label{eq:psi_fo}
\psi_{q, r \lesssim R}(r) & = 1 + \f{\as C_F}{2\pi} \bigg[-\f{1}{2}L_{r/R}^2+\f{3}{2}L_{r/R}-\f{9}{2}
+\frac{6r}{R}-\f{3r^2}{2R^2} \bigg]
\,, \nnu
\psi_{g, r \lesssim R}(r) & = 1 + \f{\as}{2\pi} \bigg[ - \frac{C_A}{2}L_{r/R}^2 + \f{\beta_0}{2} L_{r/R} 
+C_A\Big( -\f{203}{36} +\f{8r}{R} - \f{3r^2}{R^2} + \f{8r^3}{9R^3} - \f{r^4}{4R^4}\Big)
\nn \\ & \quad
+ T_F n_f\Big(\f{41}{18} - \f{4r}{R} + \f{3r^2}{R^2} - \f{16r^3}{9R^3} + \f{r^4}{2R^4}\Big)\bigg]
\,, \end{align}
which is properly normalized, $\psi_{d,r \lesssim R} (R) = 1$.

For  $r \ll R$, the factorization of $\cG$ in \eq{refact} leads to the following expression for the jet shape
\begin{align} \label{eq:psi_fact}
\psi_{d,r \ll R}(r) &\stackrel{{\rm NLL}'}{=} \tilde H_d(p_T R,\mu)
\,
\int\! \df^2 k_\perp\, \int\! \df z_r\, z_r\,
 C_d(z_r, p_T r, k_\perp,\mu,\nu) 
\\ & \qquad \times
 S_d^{\rm G}(k_\perp,\mu, \nu R)\, S_d^{\rm NG}\Big(\frac{r}{R}\Big)
\Big[1+ \mathcal{O}\Big(\frac{r}{R} \Big)\Big]\, ,
\nn \end{align}
where
\begin{align} \label{eq:tilde_H}
 \tilde H_d(p_T R,\mu) = \int\! \df z \sum_e \Bigl[H_{de}(z,p_{T} R,\mu) - J_{de}^{(1)}(z, p_T R, \mu)\Bigr]
\,.\end{align}
The contribution of $H_{de}^{(1)}$ is removed by subtracting $J_{de}^{(1)}$, but there is a constant remainder $\propto \de_{de} \de(1-z)$ since $J_{de}^{(1)}$ also receives a contribution when both partons are inside the jet. This constant is contained in $\tilde H_d$, where it is multiplied by the Sudakov factor from the evolution kernels, ensuring that it's contribution vanishes for $r \to 0$ (as required). In our implementation of \eq{psi_fact}, we 1) include the evolution kernels from evolving the ingredients between their natural scales, 2) expand the fixed-order ingredients, i.e.~dropping cross terms such as $C^{(1)} S^{(1)}$, and 3) include the corrections contained in $\ord{r/R}$, which can be read of from \eq{psi_fo}.

\subsection{Hard function}
\label{sec:hard}

The hard function $H_{cd}$ in \eq{refact} is up to one-loop order given by~\cite{Kang:2017mda,Kang:2017glf}
\begin{align}
 H_{qq}(z,p_{T} R,\mu) 
  &= \de(1-z) + \f{\as}{2\pi} \bigg[C_F \delta(1-z)\Big(-\f{L_{R}^2}{2} - \frac32 L_R +\f{\pi^2}{12} \Big) 
\nnu
 & \quad
+L_{R} P_{qq}(z) -2C_F(1+z^2)\Big(\f{\ln(1-z)}{1-z}\Big)_+ -C_F(1-z)  \bigg] 
, \nnu
 H_{qg}(z,p_{T} R,\mu) 
 &=\f{\as}{2\pi}\bigg[\Big(L_{R} - 2 \ln(1-z) \Big) P_{gq}(z) - C_Fz \bigg]
, \nnu
H_{gq}(z,p_{T} R, \mu) 
 & =  \f{\as}{2\pi}\bigg[\Big(L_{R} - 2\ln(1-z) \Big)  P_{qg}(z) - T_F 2z(1-z) \bigg]
, \nnu
H_{gg}(z, p_{T} R, \mu) 
& = \de(1-z) + \f{\as}{2\pi}\bigg[ \delta(1-z)\Big(-C_A\f{L_R^2}{2} - \f{\beta_0}{2} L_R + C_A \frac{\pi^2}{12}\Big)
\nnu 
& \quad
+ L_R P_{gg}(z) - \frac{4C_A (1-z+z^2)^2}{z} \left(\frac{\ln(1-z)}{1-z}\right)_{+} \bigg]
.\end{align}
where the splitting functions are given in \eq{split}, $\beta_0$ is given in \eq{beta_0} and $L_R$ is 
\begin{align}
 L_R = \ln\Big(\f{\mu^2}{p_T^2 R^2} \Big) 
\,.\end{align}
The hard function is formally a matching coefficient. At one-loop order it is simply the contribution to $\cG_c$ from the region of phase space where the two partons produced by a splitting of $c$ are not clustered together. The second index $d$ denotes the parton inside the jet.

The hard function $\tilde H_d$ in \eq{tilde_H} is given by 
\begin{align}
\tilde H_q(p_{T} R,\mu) &= 1 + \frac{\al_s C_F}{2\pi} \biggl(- \frac12 L_R^2 - \frac32 L_R - \frac{13}{2} + \frac{3\pi^2}{4} \biggr) + \ord{\al_s^2}
 \,, \nn \\
\tilde H_g(p_{T} R,\mu) &= 1 + \frac{\al_s}{2\pi} \biggl[ C_A\bigg( - \frac12 L_R^2 - \frac{5}{12} + \frac{3\pi^2}{4}\bigg) + \beta_0 \bigg(-\frac12 L_R - \frac{23}{12}\bigg) \biggr] + \ord{\al_s^2}
\,.\end{align}
For completeness we mention that at one-loop order the semi-inclusive jet function with identified jet flavor $J_{cd}$ is 
\begin{align} \label{eq:J_cd}
 J_{cd}^{(1)}(z,p_T R,\mu) = H_{cd}^{(1)}(z,p_T R,\mu) - \de_{cd} \de(1-z) \tilde H_c^{(1)}(p_T R,\mu)
\,,\end{align}
in terms of the above equations.

\subsection{Soft function and non-global logarithms}
\label{sec:soft}

Up to NLL$'$ order, the global soft function for quark jets can be calculated from
\begin{align} \label{eq:S_def}
S_q^{\rm G}(k_\perp, \mu, \nu R)
\stackrel{{\rm NLL}'}{=}  \frac{1}{N_c} \sum_{X_s} \langle 0  |  {\rm\bar T}[Y_{\bn}^\dagger Y_n]\, |X\rangle \langle X | {\rm T}[Y_n^\dagger Y_{\bn} ]  |  0  \rangle \de^2\Big(k_\perp - \sum_{i\in {\rm jet}} k_{i,\perp}\Big)
\,,\end{align}
where the delta function sums the transverse momentum $k_{i,\perp}$ with respect to the \emph{jet axis} of soft radiation in $X$ that is inside the jet. Soft radiation does not resolve individual collinear splittings, and so soft radiation emitted by the collinear particles in the jet can be encoded by the eikonal Wilson line $Y_n^\dagger$. At NLL$'$, the one-loop calculation of the soft function comes with the tree-level hard function, so there is a single parton moving in the $\bn$ direction, resulting in $Y_{\bn}$. At higher orders, real emissions in the hard function (outside the jet) will result in additional Wilson lines~\cite{Larkoski:2015zka,Becher:2015hka}, see also ref.~\cite{Caron-Huot:2015bja}. This complicates the non-global logarithms, but is fortunately beyond the accuracy at which we are working. 

A one-loop calculation yields~\cite{Kang:2017mda}
\begin{align} \label{eq:S_nlo}
  S_q^{\rm G}(k_\perp,\mu,\nu R) = \de^2(k_\perp) \!+\! \frac{\al_s C_F}{2\pi^2} \bigg[-  \frac{1}{\mu^2}\,
\Big( \frac{\ln(k_\perp^2/\mu^2)}{k_\perp^2/\mu^2}\Big)_+\!
\!+\! \frac{1}{\mu^2}\,
 \frac{1}{(k_\perp^2/\mu^2)}_+\!\!\ln \frac{\nu^2 R^2}{4\mu^2}
 \!-\! \frac{\pi^2}{12} \de(\vec k_\perp^{\,2})\bigg]
\,.\end{align}
For $S_g^{\rm G}$ the Wilson lines are in the adjoint representation, instead of the fundamental representation, and the overall normalization in the definition is modified from $1/N_c$ to $1/(N_c^2-1)$. The one-loop result for $S_g^{\rm G}$ is simply given by replacing $C_F \to C_A$ in \eq{S_nlo}. 

The non-global logarithms arise from soft emission patterns that simultaneously probe the jet and out-of-jet region~\cite{Dasgupta:2001sh}. Since we consider $R\ll 1$, the NGLs are the same as in the hemisphere case~\cite{Banfi:2010pa}. Indeed, a direct calculation of the leading contribution at order $\al_s^2$ gives rise to
\begin{align}
   -\frac{\al_s^2 C_A C_i}{24\pi}\, \frac{1}{(p_TR)^2}\,
\Big( \frac{\ln(k_\perp^2/(p_TR)^2)}{k_\perp^2/(p_T R)^2}\Big)_+
,\end{align}
where the hard scale $p_T R$ arises from the emission outside the jet, and the color factor $C_i = C_F$ for quarks and $C_A$ for gluons. The integral in \eq{refact} with the tree-level collinear function in \eq{col_tree}, leads to
\begin{align}
   \int \df^2 k_\perp \Theta(k_\perp < p_T r)\times
   -\frac{\al_s^2 C_A C_i}{24\pi}\, \frac{1}{(p_TR)^2}\,
\Big( \frac{\ln(k_\perp^2/(p_TR)^2)}{k_\perp^2/(p_T R)^2}\Big)_+
 =    -\frac{\al_s^2 C_A C_i}{12}\, 
\ln^2\Big(\frac{R}{r}\Big)
.\end{align}
We obtain the same result if we had directly taken the NGL at order $\al_s^2$ of the hemisphere case~\cite{Dasgupta:2001sh}, with $R/r$ as the argument of the logarithm. 

Beyond order $\al_s^2$, we should write the NGLs in terms of plus distributions of $k_\perp/(p_T R)$, and convolve these with the global soft function and the collinear function. However, at NLL$'$ only the leading NGLs are required, and we may directly take the NGLs of the hemisphere case with the ratio $R/r$ as the argument of the logarithm, which is significantly simpler. This is justified because both the NGLs and the rapidity resummation are single logarithmic series of plus distributions in transverse momentum. All subtleties from convolutions of plus distributions in transverse momentum are subleading, i.e.~whether we first convolve these single logarithmic series with each other, and then integrate them against the tree-level collinear function that sets the upperbound $k_\perp = p_T r$, or directly integrate each of them up to $k_\perp = p_T r$, is the same to the accuracy that we are working.

\begin{figure}[t]
    \centering
    \includegraphics[width=0.5\textwidth]{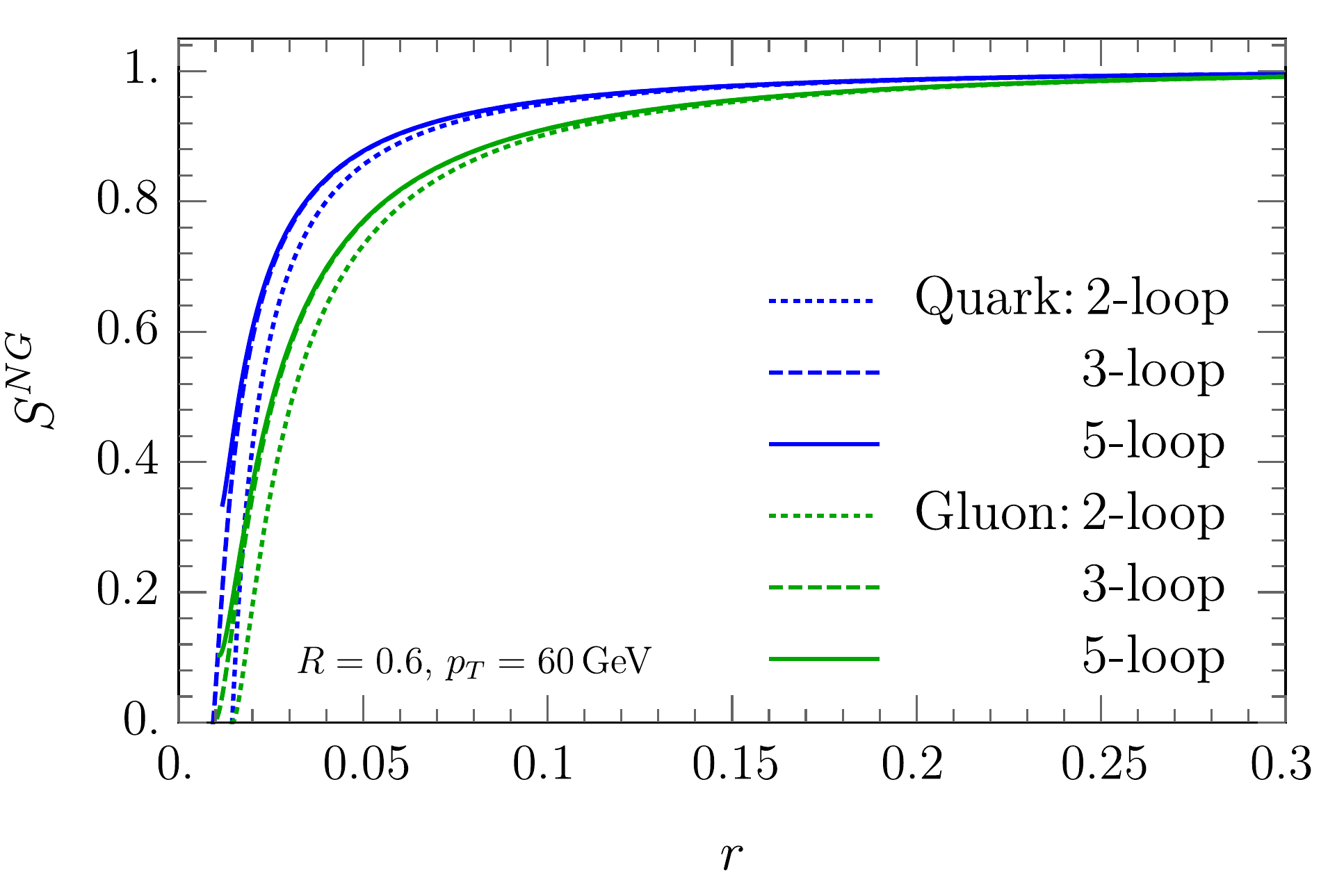}
    \vspace{-2ex}
    \caption{The non-global contribution to the soft function in \eq{S_NG} for quark jets (blue) and gluon jets (green) with R=0.6 at $p_T = 60$ GeV. Shown are the results up to two-loop (dotted), three-loop (dashed) and five-loop order (solid). \vspace{-1ex}}
    \label{fig:NGLsoft}
\end{figure}

The leading NGLs in the hemisphere case are described by a universal function, where in our case the argument of the logarithm is $R/r$. We will work in the large $N_c$ approximation (the leading NGLs without this approximation have been studied in ref.~\cite{Hatta:2013iba}). Rather than using the fit of ref.~\cite{Dasgupta:2001sh}, we employ the solution to the BMS equation~\cite{Banfi:2002hw} up to five-loop order~\cite{Schwartz:2014wha}, 
\begin{align} \label{eq:S_NG}
 S_{q}^{\rm NG}(\widehat L) = 1 - \frac{\pi^2}{24} \widehat L^2 + \frac{\zeta_3}{12} \widehat L^3 + \frac{\pi^4}{34560} \widehat L^4+ \Big(-\frac{\pi^2 \zeta_3}{360} + \frac{17\zeta_5}{480}\Big) \widehat L^5 + \ord{L^6}
\,,\end{align}
where
\begin{align}
\widehat L= \frac{\al_s N_c}{\pi} \ln \frac{R}{r}
\,.\end{align}
The advantage of using this result is that it allows us to test the perturbative convergence, which is excellent for the range of $r/R$ we are interested in. Specifically, there is a small difference from including the cubic term in \eq{S_NG}, but the effect of subsequent terms is not visible, as is clear from \fig{NGLsoft}. This figure also shows that the effect of the non-global contribution to the soft function is limited to rather small values of $r$. For the gluon case we have $S_{g}^{\rm NG} =  (S_{q}^{\rm NG})^2$, which follows from rewriting the adjoint Wilson line in terms of a fundamental and anti-fundamental Wilson line.\footnote{Alternatively, one can use non-abelian exponentation, and note that the webs only differ by a factor $C_F$ vs.~$C_A$, implying $\ln  S_{g}^{\rm NG} =  (C_A/C_F) \ln S_{q}^{\rm NG} =  \ln (S_{q}^{\rm NG})^2$, in the large $N_C$ approximation. We thank D.~Neill for discussions on this.} 

\section{Collinear function including recoil}
\label{sec:coll}

The definition of the collinear function is similar to \eq{G_def}, and is given by the following expression for the quark case,
\bea
 &C_q(z_r, p_T r, k_\perp,\mu,\nu) \\
 & \quad = 16\pi^3\,\sum_X \frac{1}{2N_c}\, {\rm Tr} \Big[\frac{\bnslash}{2}
\langle 0| \delta(2p_T \!-\! \bar n\cdot {\mathcal P}) \delta^2({\mathcal P}_\perp \!-\! k_\perp) \chi_n(0)  |X\rangle 
\langle X|\bar \chi_n(0) |0\rangle \Big]
\delta\Big(z_r - \frac{p_{T}(j_r)}{p_{T}}\Big)
\,.\nn \eea
As before, we take the jet rapidity $\eta$ equal to zero, since the measurement is invariant under boosts along the beam axis.  
In contrast to \eq{G_def}, all this collinear radiation is inside the jet, so there is no sum over jets in $X$ or measurement of $z$. The recoil due to soft radiation is accounted for through the $\delta^2({\mathcal P}_\perp - k_\perp)$. 
At tree level, the collinear function is given by 
\begin{align} \label{eq:col_tree}
  C_d(z_r, p_T r, k_\perp,\mu,\nu) = \de(1-z_r)\, \Theta(k_\perp < p_T r)
\,,\end{align}
for both quarks ($d=q$) and gluons ($d=g$). This simply states that as long as the recoil is not too large $(k_\perp < p_T r)$, the parton is inside the cone of radius $r$ around the jet axis, and $z_r = 1$.

In this section we calculate the collinear function at order $\alpha_s$, which involves the collinear splitting of an initial quark or gluon into two partons. We give the collinear phase-space and matrix elements in \sec{psme}, and the geometry of the setup is described in \sec{geo}. The resulting integrals are performed in detail for the quark case in \sec{coll_q}, and results for the gluon case are presented in \sec{coll_g}. We verify in \app{check} that our result satisfies the rapidity renormalization group equation.

\subsection{Collinear phase-space and matrix elements}
\label{sec:psme}

Due to the recoil of the soft radiation, the jet axis is not aligned with the initial parton, requiring us to take the azimuthal dependence into account in the phase-space integration of the collinear-matrix elements, 
\begin{align}
\int \mathrm{d}\Phi_2 \sigma^{c}_{2,q}&=\frac{\alpha_s}{2\pi^{2}} \frac{(e^{\gamma_E}\mu^{2})^{\epsilon}}{\Gamma(1-\epsilon)}\int_0^{2\pi} \mathrm{d\phi} \int_{0}^{1} \mathrm{d}x \; \hat{P}_{qq}(x)\int \frac{\mathrm{d}q_\perp}{q_{\perp}^{1+2\epsilon}}, \nn \\
\int \mathrm{d}\Phi_2 \sigma^{c}_{2,g}&=\frac{\alpha_s}{2\pi^{2}} \frac{(e^{\gamma_E}\mu^{2})^{\epsilon}}{\Gamma(1-\epsilon)}\int_0^{2\pi} \mathrm{d\phi} \int_{0}^{1} \mathrm{d}x \;\left[n_f \hat{P}_{qg}(x)+\frac{1}{2}\hat{P}_{gg}(x)\right]\int \frac{\mathrm{d}q_\perp}{q_{\perp}^{1+2\epsilon}}
\,.\end{align}
Here $\mathrm{d}\Phi_2$ denotes the two-body collinear phase-space integration and $\sigma^{c}_{2,q}$ is the collinear matrix element squared. The two partons have transverse momentum $q_\perp$ with respect to the initial parton and carry a fraction $z$ and $1-z$ of its longitudinal momentum, whose distribution is described by the splitting functions\footnote{We introduce a hat to avoid potential confusion with the splitting functions given in \eq{split}.}
\begin{align}
\hat{P}_{qq}(x)&= C_F \Big[\frac{1+x^{2}}{1-x}-\epsilon (1-x)\Big], \nn \\
\hat{P}_{qg}(x)&= T_F \Big[1-2x(1-x)+2\epsilon\, x (1-x)\Big], \nn \\
\hat{P}_{gg}(x)&= 2C_A \Big[\frac{x}{1-x}+\frac{1-x}{x}+x(1-x)\Big]
.\end{align}
We will make frequent use of the angles of the partons with respect to the initial parton, 
\begin{align} \label{eq:beta}
  \bt_1 = \frac{q_\perp}{x p_T}
  \,, \qquad
  \bt_2 = \frac{q_\perp}{(1-x) p_T}
\end{align}
where $p_T$ is the transverse momentum of the jet.

\subsection{Geometry of the measurement}
\label{sec:geo}

\begin{figure}[t]
    \centering
     \hfill \includegraphics[width=0.3\textwidth]{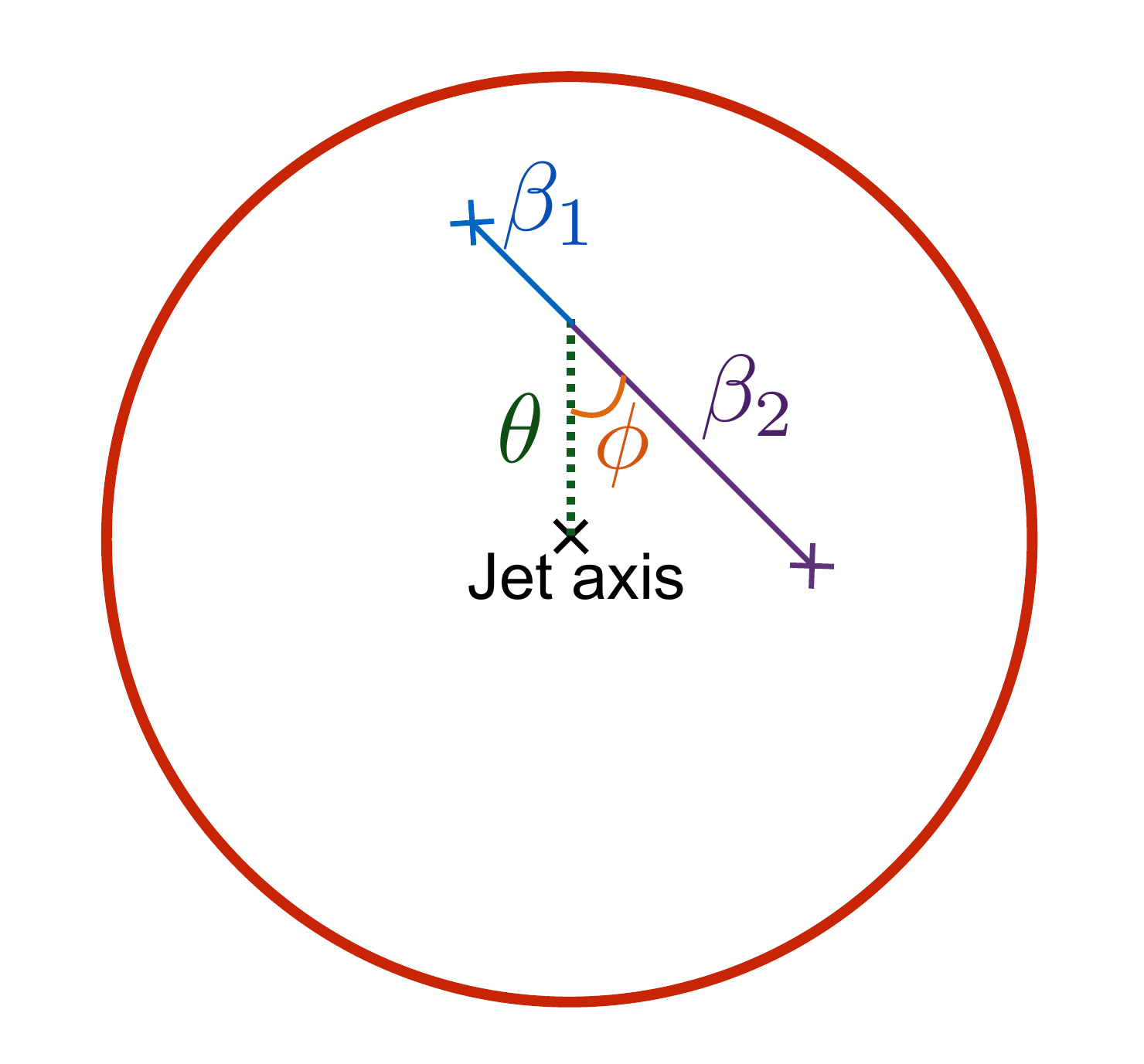} \hfill 
     \includegraphics[width=0.3\textwidth]{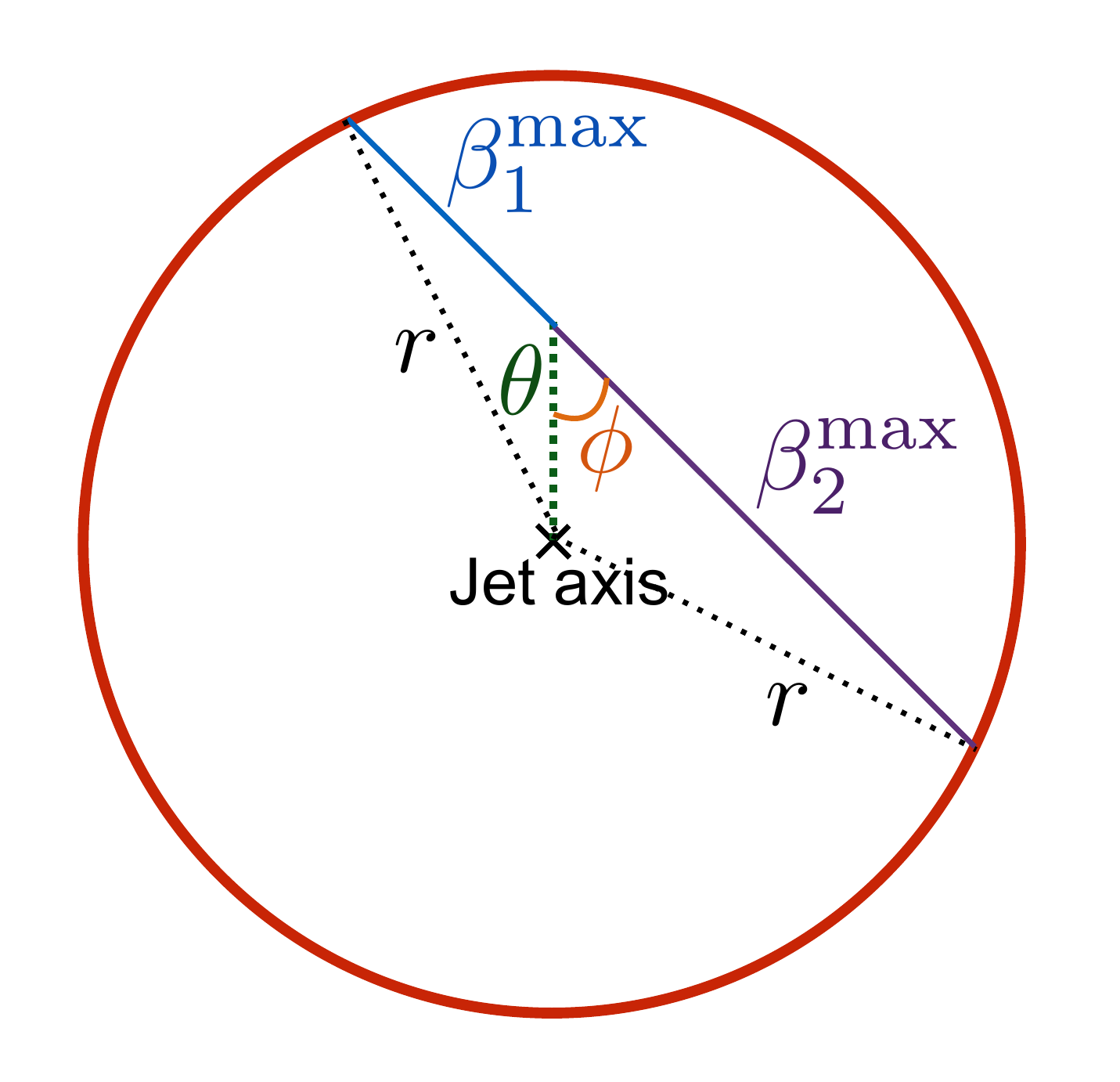} \hfill \phantom{.} \\
    \caption{Projection of the central subjet of radius $r$ onto a plane perpendicular to the jet axis. $\theta$ is the displacement due to recoil and is here assumed to be less than $r$. In the left panel a splitting is shown. In the right panel $\beta_i^{\rm max}$ is shown, which is the upper bound on $\beta_i$ such that the corresponding parton lies inside the central subjet. }  
    \label{fig:geometry}
\end{figure}

We now describe the geometry of the setup, starting with the case where the initial parton is inside the central subjet. Projecting the jet onto a plane perpendicular to the jet axis, as shown in the left panel of \fig{geometry}, we can treat the polar angles as distances since $r \ll 1$. The central subjet corresponds to a circle of radius $r$, and the recoil due to the total transverse momentum $k_\perp$ of soft radiation in the jet corresponds to a displacement of the initial collinear parton by a distance $\theta= k_\perp/p_T$. The angles $\beta_i$ also correspond to distances, though the azimuthal angle $\phi$ remains a true angle.

The condition for each of the partons produced by the splitting to be inside the central subjet is $\beta_i \leq \beta_1^{\rm max}$, displayed in the right panel of \fig{geometry}. Some trigonometry yields
\begin{align} \label{eq:beta_max}
\beta_{1,2}^{\rm max} &=  r \sqrt{1+\frac{\theta^{2}}{r^{2}}-2\frac{\theta}{r}\biggl( \pm \cos\phi \, \sqrt{1-\frac{\theta^{2}}{r^{2}}\sin^{2}\phi }+  \frac{\theta}{r} \sin^{2}\phi\biggr) } 
,\end{align}

\begin{figure}[t]
        \centering
        \hfill
        \includegraphics[scale=0.3]{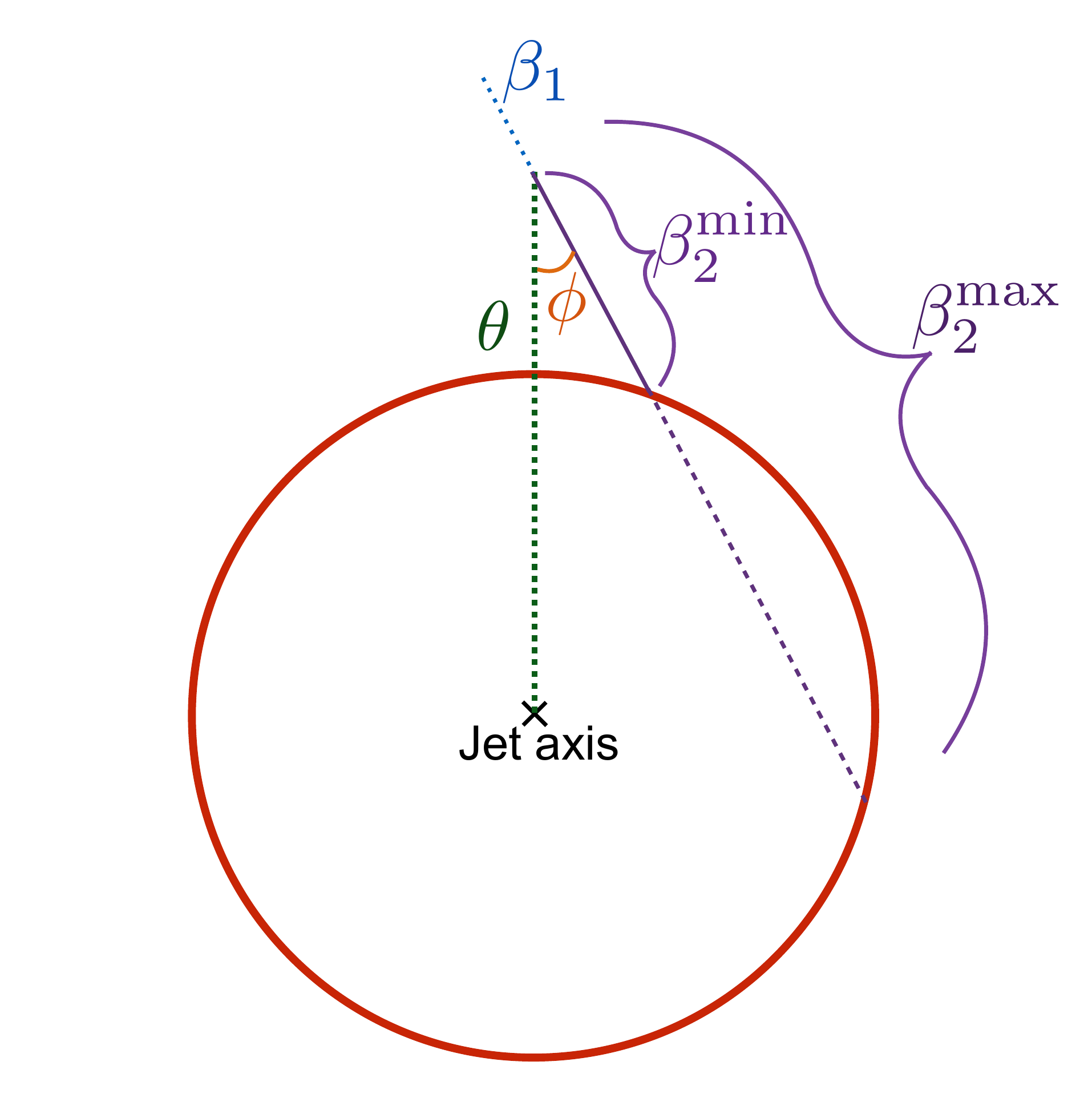} \hfill
        \includegraphics[scale=0.3]{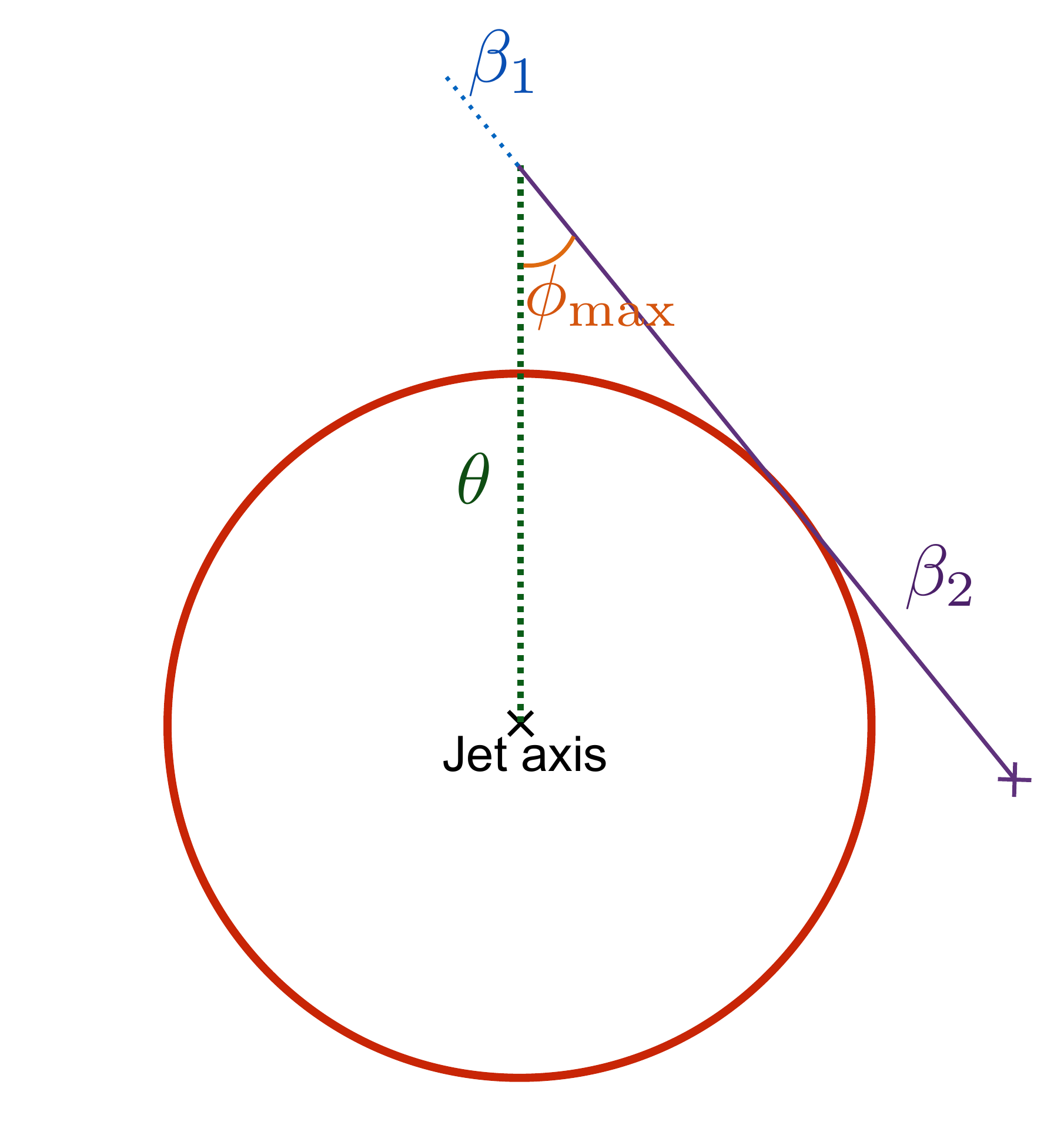} \hfill \phantom{.} \\
    \caption{Geometry of the central subjet when the recoil $\theta$ is larger than the radius $r$. In the left panel the minimum and maximum values of $\beta_2$ are shown such that this parton is inside the central subjet. The right panel depicts the maximum value of $\phi$.}
    \label{fig:geometry2}
\end{figure}

The case $\theta>r$, where the initial parton is outside of the jet, requires the setup depicted in \fig{geometry2}. Here only one parton produced in the splitting can be inside the central subjet, and there is a minimum and maximum value for $\beta_i$. For parton 2,
\begin{align}
\beta_2^{{\rm min}, {\rm max}}= \theta \biggl[\cos \phi \mp \sqrt{\frac{r^2}{\theta^2} - \sin^{2}\phi}\,\biggr]\,,
\label{eq:beta2minmax}
\end{align}
and the expressions for $\beta_1^{{\rm min}, {\rm max}}$ can be obtained by substituting $\phi\rightarrow \phi+\pi $. These expressions are not valid for all $\phi$, breaking down when the square root becomes imaginary. The maximum $\phi$ is shown in the right panel of \fig{geometry2}, and leads to the following  boundaries  
\begin{align}
&\text{parton 1:}&  \quad \pi-\phi_{\rm max}&<\phi<\pi+\phi_{\rm max}\,, & 
\phi_{\rm max}& =\arcsin\Bigl( \frac{r}{\theta}\Bigr) \nn\\
&\text{parton 2:}&  \quad -\phi_{\rm max}&<\phi<\phi_{\rm max}
\,.\end{align}

\subsection{Integrals for the quark collinear function}
\label{sec:coll_q}

We now present the calculation of the collinear function for an initiating quark, separating the two cases:
\begin{align}
C^{(1)}_q(z_r,p_T r, k_\perp, \mu, \nu)= \Theta(k_\perp<p_T r)\, C_q^{(\theta<r)}+\Theta(k_\perp> p_T r)\,C_q^{(\theta>r)}.
\end{align}
In the first case
\begin{align} \label{eq:C_case1}
C_q^{(\theta<r)}&= (\rm A)_{\theta<r}+(\rm B)_{\theta<r}+(\rm C)_{\theta<r}\,, \nn \\
(\rm A)_{\theta<r}&=\delta(1-z_r)\int\! \mathrm{d}\Phi_2\, \sigma^{c}_{2,q}\, \Theta (\beta_1<\beta^{\rm max}_1)\Theta (\beta_2<\beta^{\rm max}_2)\,, \nn\\
(\rm B)_{\theta<r}&=\int\! \mathrm{d}\Phi_2\, \sigma^{c}_{2,q}\, \delta(x-z_r)\, \Theta (\beta_1<\beta^{\rm max}_1)\Theta (\beta_2>\beta^{\rm max}_2)\,, \nn\\
(\rm C)_{\theta<r}&=\int\! \mathrm{d}\Phi_2\, \sigma^{c}_{2,q}\, \delta(1-x-z_r)\, \Theta (\beta_1>\beta^{max}_1)\Theta (\beta_2<\beta^{max}_2),
\end{align}
where (A) refers to the case when both partons end up inside the subjet, (B) when only parton 1 (quark) ends up inside and (C) when only parton 2 (gluon) does (using the same labels as in ref.~\cite{Kang:2017mda}). In the second case only one parton can be inside the subjet, and 
\begin{align} \label{eq:C_case2}
C_q^{(\theta>r)}&= (\rm B)_{\theta>r}+(\rm C)_{\theta>r}\,, \nn \\
({\rm B})_{\theta>r}&= \int\! \mathrm{d}\Phi_2\, \sigma^{c}_{2,q}\, \delta(x-z_r)\,\Theta(\pi-\phi_{\rm max}<\phi<\pi+\phi_{\rm max})\, \Theta( \beta_1^{\rm min}<\beta_1 <  \beta_1^{\rm max})\,, \nn \\
({\rm C})_{\theta>r}&=\int\! \mathrm{d}\Phi_2\, \sigma^{c}_{2,q}\, \delta(1-x-z_r)\, \Theta(-\phi_{\rm max}<\phi<\phi_{\rm max})\, \Theta( \beta_2^{\rm min}<\beta_2 <  \beta_2^{\rm max}) 
\,.\end{align}
The contribution where neither parton is inside is irrelevant, since then $z_r = 0$.

We now set out to calculate these integrals. 
For (A)$_{\theta<r}$ we use \eq{beta} to rewrite the $x$ integral and $\Theta$ functions
\begin{align}
\int_0^1\! \df x\, \Theta (\beta_1<\beta^{\rm max}_1)\Theta (\beta_2<\beta^{\rm max}_2) &=
\int_{0}^{1}\!\mathrm{d}x\, \Theta ( q_\perp < x p_T \beta_1^{\rm max} )\, \Theta ( q_\perp < (1-x) p_T \beta_2^{\rm max} ) \nn \\
&=\int_{0}^{\wtb}\!\mathrm{d}x\, \Theta ( q_\perp < x p_T \beta_1^{\rm max} )\nn \\
&\quad +\int_{\wtb}^{1}\!\mathrm{d}x\, \Theta ( q_\perp < (1-x) p_T \beta_2^{\rm max} ),
\label{integralinx}
\end{align}
where $\wtb$ with $0<\wtb<1$ is given by
\begin{align}
\wtb =  \frac{\beta_2^{\rm max}}{\beta_1^{\rm max}+\beta_2^{\rm max}}
\,.\end{align}
Inserting this in \eq{C_case1}, and performing the $x$ and $q_\perp$ integrals, we obtain 
\begin{align}
(A)_{\theta<r}
&= \frac{\alpha_s C_F}{2\pi^{2}}\, \delta(1-z_r) \int_0^{2\pi} \mathrm{d\phi} \bigg\{ \frac{1}{2 \epsilon ^2}+\frac{1}{\epsilon}\Big(\frac{L_2}{2}+\frac{3}{4}\Big) +\frac{L_2^2}{4}+\frac{3 L_2}{4} 
-\ln^{2}(1-\wtb) 
\nn \\ & \quad
 + 2\ln \wtb \ln(1-\wtb) -\frac{3}{2} \ln (1-\wtb)
 +2 \text{Li}_2\bigl(1-\wtb\bigr) -\frac{\wtb}{2} +2 -\frac{3\pi ^2}{8} \bigg\}
,\end{align}
in terms of $L_i$ defined as
\begin{align}
L_i= \ln \bigg( \frac{\mu^{2}}{p_T^2 (\beta_i^{\rm max})^{2}}\bigg)\,.
\end{align}
We cannot perform the integral over $\phi$ analytically, as it involves the rather complicated expression in \eq{beta_max}. However, this is not a problem as we have already isolated the divergences.

The calculation of  (B)$_{\theta<r}$ involves rapidity divergences, which we regulate using the $\eta$ regulator, resulting in 
\begin{align}
({\rm B})_{\theta<r}&=\frac{\alpha_s C_F}{2\pi^{2}}\, \frac{e^{\epsilon\gamma_E}}{\Gamma(1-\epsilon)}\, \mu^{2\epsilon}\, \int_{0}^{2\pi}\! \mathrm{d}\phi \int_{\wtb}^{1}\mathrm{d}x\, \delta(x-z_r) \Big[\frac{1+x^{2}}{1-x}-\epsilon (1-x) \Big] \Big[ \frac{\nu}{2(1-x)p_T}\Big]^{\eta} \nn \\
&\quad \times \int^{xp_T\beta_1^{\rm max}}_{(1-x)p_T \beta_2^{\rm max} }  \frac{\mathrm{d}q_\perp}{q_{\perp}^{1+2\epsilon}}
\nn \\
&= \frac{\alpha_s C_F}{2\pi^{2}} \int_0^{2\pi}\! \mathrm{d}\phi\, \bigg\{ \delta(1-z_r)\bigg[ 
\frac{1}{\eta}\Big(\frac{1}{\eps} + L_1 \Big)
-\frac{1}{2\epsilon^{2}}+ \frac{1}{\eps}\, \Big( L_\nu -\frac{L_2}{2}\Big) + L_\nu L_1 - \frac{L_2^{2}}{4} +\frac{\pi^{2}}{24} \bigg]  \nn \\
& \quad + \Theta \big( z_r > \wtb\big) \bigg[- (1+z_r^{2}) \bigg(\frac{\ln (1\!-\!z_r)}{1-z_r}\bigg)_{\!\!+} - \ln \bigg(\frac{\wtb}{1-\wtb}\bigg) \frac{1+z_r^{2}}{(1-z_r)}_+ + \frac{(1+z_r^{2})\ln z_r}{1-z_r} \bigg] \bigg\}
.\end{align}
where $L_\nu$ is defined as 
\begin{align}
  L_\nu = \ln \frac{\nu}{2p_T}
\,.\end{align}
Performing a similar calculation for (C)$_{\theta<r}$ we obtain 
\begin{align}
({\rm C})_{\theta<r}= \frac{\alpha_s}{2\pi^{2}}\int_{0}^{2\pi}\! \mathrm{d}\phi \, \Theta\big( z_r > 1-\wtb\big) \biggl\{ P_{gq}(z_r) \biggl[\ln \biggl( \frac{z_r}{1-z_r}\biggr)+\ln \biggl(\frac{\wtb}{1-\wtb}\biggr) \biggr] \biggr\}
\,.\end{align}
Note that the factor $C_F$ is  inside the $P_{gq}$ splitting function, which was defined in \eq{split}.

Moving on to the case $\theta>r$, we also have to regulate rapidity divegences for  $(B)_{\theta>r}$,
\begin{align}
({\rm B})_{\theta>r}=&\frac{\alpha_s C_F}{2\pi^{2}}\, \frac{e^{\epsilon\gamma_E}}{\Gamma(1-\epsilon)}\, \mu^{2\epsilon}\int_{\pi-\phi_{\rm max}}^{\pi+\phi_{\rm max}}\! \mathrm{d}\phi  \int_{0}^{1}\!\mathrm{d}x\, \delta(x-z_r) \Bigl[\frac{1+x^{2}}{1-x}-\epsilon (1-x) \Bigr]\Bigl( \frac{\nu}{2(1-x)p_T}\Bigr)^{\eta}  \nn \\ 
&\times \int^{xp_T\beta_1^{\rm max}}_{x p_T \beta_1^{\rm min}} \frac{\mathrm{d}q_\perp}{q_{\perp}^{1+2\epsilon}} \nn \\
&=\frac{\alpha_sC_F}{2\pi^{2}}\, \biggl[ \delta(1-z_r)\Big(\frac{2}{\eta}+2L_\nu\Big)-\frac{1+z_r^{2}}{(1-z_r)}_+ \biggr] \int_{-\phi_{\rm max}}^{\phi_{\rm max}}\mathrm{d}\phi \ln \biggl(\frac{\beta_2^{\rm min}}{\beta_2^{\rm max}} \biggr) 
\end{align}
Note that we changed the $\phi$ integration region, which is why the final expression involves $\beta_2^{\rm min,\, max}$ rather than $\beta_1^{\rm min,\, max}$.
An analogous computation for (C)$_{\theta>r}$ yields
\begin{align}
({\rm C})_{\theta>r}=-\frac{\alpha_s}{2\pi^{2}}\, P_{gq}(z_r) \int_{-\phi_{\rm max}}^{\phi_{\rm max}}\!\mathrm{d}\phi \ln \biggl( \frac{\beta_2^{\rm min}}{\beta_2^{\rm max}} \biggr).
\end{align}

Adding up all the contributions, 
\begin{align}  \label{eq:QuarkCollinear}
C_q^{(\theta<r)}&= \frac{\alpha_s C_F}{2\pi^{2}} \int_0^{2\pi}\! \mathrm{d}\phi \, \biggl\{  \delta(1-z_r)\bigg[\frac{1}{\eta} \Big(\frac{1}{\epsilon} +L_1 \Big) + \frac{1}{\eps}\Big( L_\nu + \frac34\Big) +L_\nu L_1 +\frac{3 L_1}{4}
 \\ & \quad
- \ln^{2}(1-\wtb) + 2 \ln \wtb \ln(1-\wtb)
 -\frac{3}{2} \ln \wtb
+2{\rm Li}_{2}(1-\wtb)
-\frac{\wtb}{2}-\frac{\pi ^2}{3}+2\bigg] 
\nn \\ & \quad
+ \Theta ( z_r > \wtb) \biggl[-(1+z_r^{2})\bigg(\frac{\ln (1-z_r)}{1-z_r}\bigg)_{\!\!+} 
+ \ln \biggl(\frac{z_r(1-\wtb)}{\wtb} \biggr)\frac{1+z_r^{2}}{(1-z_r)}_+ \biggr] 
\nn \\ &\quad
+\Theta( z_r > 1-\wtb) \biggl[ \frac{1+(1-z_r)^{2}}{z_r}\, \ln \biggl( \frac{z_r \wtb}{(1-z_r)(1-\wtb)}\biggr) \biggr] \biggr\}, \nn \\
C_q^{(\theta>r)}&=\frac{\alpha_s C_F}{2\pi^{2}}  \biggl[ \delta(1-z_r)\Bigl( \frac{2}{\eta}+2L_\nu\Bigr)-\frac{1+z_r^{2}}{(1-z_r)}_+ - \frac{1+(1-z_r)^{2}}{z_r} \biggr] \int_{-\phi_{\rm max}}^{\phi_{\rm max}}\!\mathrm{d}\phi \ln \left(\frac{\beta_2^{\rm min}}{\beta_2^{\rm max}} \right),
\nn \end{align}

\begin{figure}[t]
    \centering
     \hfill \includegraphics[width=0.48\textwidth]{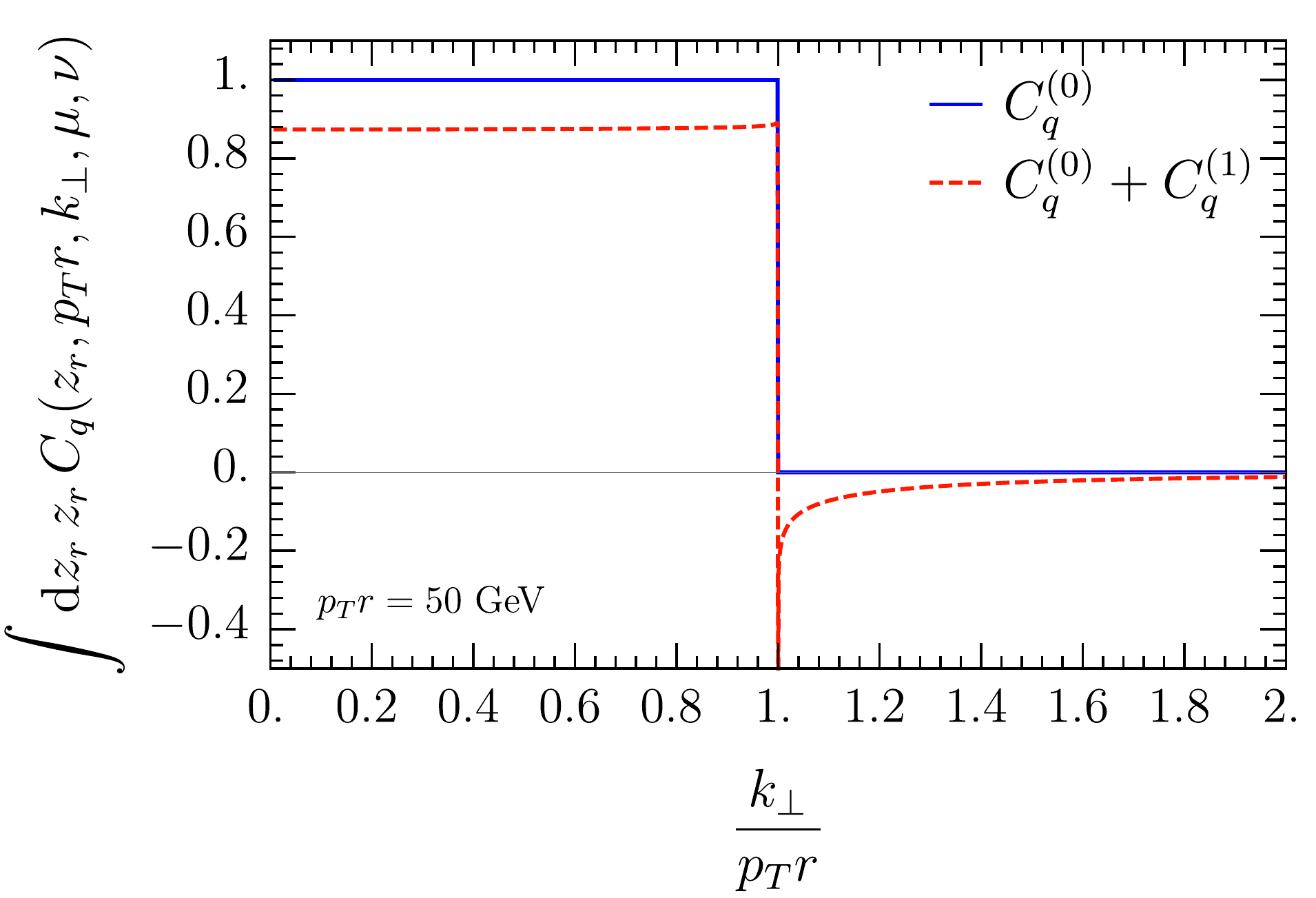} \hfill 
     \includegraphics[width=0.48\textwidth]{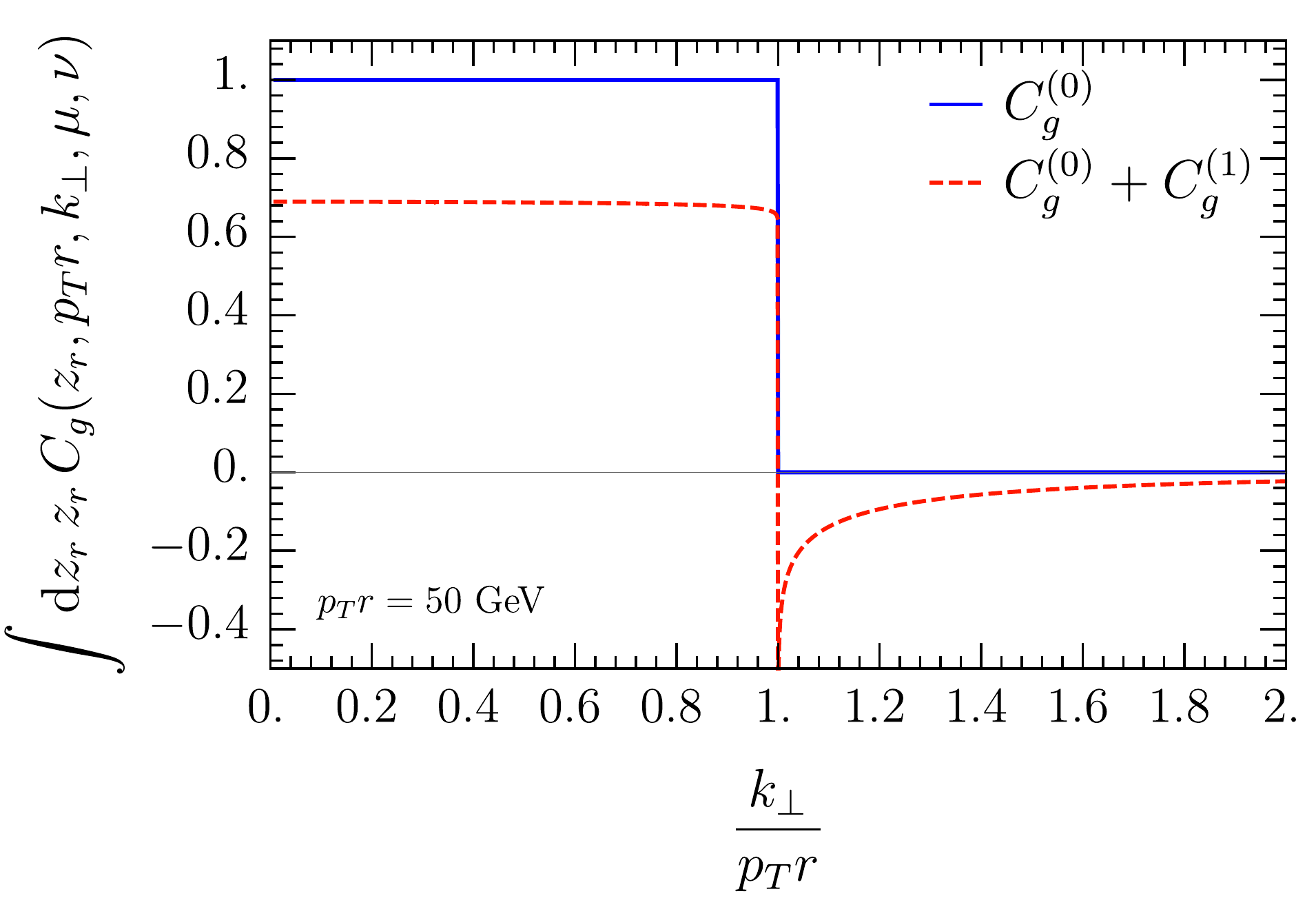} \hfill \phantom{.} \\
    \caption{The linear $z_r$ moment of the collinear function for quark jets (left) and gluon jets (right) at tree level (blue) and including one-loop corrections (red dashed), evaluated at the scales $\mu = p_T r$ and $\nu = p_T$.}
    \label{fig:Ckperp}
\end{figure}

For the jet shape we only need the linear $z_r$ moment  ($\int \df z_r\,z_r$) of the collinear function, which we plot for quark jets in the left panel of \fig{Ckperp}. The result depends on $k_\perp/(p_T r)$, shown on the horizontal axis, and on $p_T r$, through the scale in $\alpha_s$. First we note that the one-loop corrections are of order 10\%, as is typical. At tree level there is a single collinear particle that is simply inside the jet if the recoil due to soft radiation is not too large, $k_\perp < p_T r$. One-loop corrections allow for nonzero values of the collinear function for larger values of $k_\perp$. That the collinear function attains a negative value is not a concern, since it does not correspond to a physical quantity (indeed, this depends on our choice of $\mu$ and $\nu$-scales). We also note that the divergence at $k_\perp = p_T r$ from one-loop corrections is integrable.

\subsection{Gluon collinear function}
\label{sec:coll_g}

The calculation of the collinear function for gluons follows very similar steps. Splitting the calculation into the $\theta<r$ and $\theta>r$ contribution, we obtain
\begin{align}
C_g^{(\theta<r)} &= \frac{\alpha_s}{2\pi^{2}} \int_0^{2\pi} \mathrm{d\phi} \bigg( \delta(1-z_r)\bigg\{C_A\bigg[\frac{1}{\eta} \Big(\frac{1}{\eps} + \frac{L_1}{4} + \frac{L_2}{4} \Big) + \frac{1}{\eps} \Big(L_\nu + \frac{11}{12}\Big) 
\nn \\ & \quad
+ \Big(\frac{L_\nu}{2}  +\frac{11}{24} \Big) (L_1+L_2) - \frac12 \ln^2 \frac{1-\wtb}{\wtb}  -\frac{11}{12} \ln \big(\wtb(1-\wtb)\big) + \frac16 \wtb (1-\wtb)
\nn \\ & \quad
  -\frac{\pi ^2}{6}+\frac{67}{36} \bigg]
  +n_f T_F \bigg[-\frac{1}{3 \epsilon }-\frac{1}{6} (L_1+L_2)+\frac{1}{3} \ln \big(\wtb(1-\wtb)\big) -\frac{1}{3}\wtb(1-\wtb)-\frac{5}{9}\bigg] \bigg\} 
  \nn \\ & \quad
  +\Theta (z_r>\wtb) \bigg\{ 
   - C_A \frac{(1\!-\!z_r\!+\!z_r^2)^2}{z_r} \bigg( \frac{\ln (1\!-\!z_r)}{1-z_r}\bigg)_+ \!\!+ C_A \frac{(1\!-\!z_r\!+\!z_r^2)^2}{z_r(1-z_r)_+} \ln\bigg(\frac{z_r (1-\wtb) }{\wtb}\bigg)
\nn \\ & \quad
   + n_f T_F(z_r^2+(1-z_r)^2)\ln\bigg(\frac{z_r (1\!-\!\wtb) }{(1-z_r)\wtb}\bigg)  \bigg\} \bigg) 
\,,  \nn \\
C_g^{(\theta>r)} &= \frac{\alpha_s}{2\pi^{2}} \bigg\{ 2C_A \bigg[\delta(1-z_r)\Big(\frac{1}{\eta}+L_\nu \Big) - \frac{(1-z_r+z_r^2)^2}{z_r(1-z_r)_+} \bigg]
\nn \\ & \quad 
- 2 n_f T_F (z_r^2+(1-z_r)^2) \bigg\}
\int_{-\phi_{\rm max}}^{\phi_{\rm max}} \mathrm{d\phi}  \ln \bigg(\frac{\beta_2^{\rm min}}{\beta_2^{\rm max}}\bigg) 
\,.\end{align}
The linear moment of the gluon collinear function is shown in the right panel of \fig{Ckperp}. Its features are similar to that for quark jets (left panel), except that the one-loop corrections are larger due to the larger color factor ($C_A$ vs.~$C_F$).

\section{Implementation and first results}
\label{sec:implement}

This section describes how we implement our formulae to obtain numerical predictions. In \sec{scales} our central scale choice is discussed. We then describe how we match our formulae for $r \ll R$ and $r \lesssim R$ in \sec{matching}. The scale variations used to assess the perturbative uncertainty are given in \sec{uncertainties}.  We conclude in \sec{qg_results} with some first, purely perturbative, results for the jet shape for quark and gluons jets.

\subsection{Central scale choice}
\label{sec:scales}

We start by discussing the scales for the regime $r \lesssim R$, described by the factorization in \eq{fact}. We take
\begin{align} \label{eq:canFO}
  \mu_{\cal H} = p_T\,, \qquad
  \mu_{\cG} = p_T R
\,,\end{align}
as our central scale choice. In the regime $r \ll R$, $\cG$ is refactorized in terms of hard, collinear and soft functions, see \eq{refact}. We will take as their central scales
\begin{align} \label{eq:can1}
    \mu_{\cal H} &= p_T\,, &
    \mu_{H} & = p_T R\,, &
    \mu_C &= p_T r\,, &
    \mu_S &= p_T r\,, 
    \nn \\  & & & &
    \nu_C &= p_T\,, &
    \nu_S &= \frac{1}{b_\perp R}
\,.\end{align}
We deviate from \eq{can} by expressing $\nu_S$ in terms of $b_\perp$, which is the Fourier conjugate variable of $k_\perp$ (not $p_T$). Since collinear and soft modes contribute to $k_\perp$, one expects that parametrically $p_T r \sim k_\perp \sim 1/b_\perp$. However, expressing $\nu_S$ in terms of $b_\perp$ avoids a well-known problem with choosing scales in momentum space for transverse momentum resummation~\cite{Frixione:1998dw} (see also refs.~\cite{Monni:2016ktx,Ebert:2016gcn}). Often one also chooses $\mu_C$ and $\mu_S$ in term of $b_\perp$, but this is not required here. In particular, our choice in \eq{can1} ensures that the $\mu$-evolution from $\mu_H$ down to $\mu_C = \mu_S$ is essentially the same as in the jet shape calculation in ref.~\cite{Chien:2014nsa}.

\subsection{Matching predictions for $r \ll R$ and $r \lesssim R$}
\label{sec:matching}

A common approach to matching different regimes in SCET is to use profile scales~\cite{Ligeti:2008ac,Abbate:2010xh}, that would smoothly interpolate between \eqs{canFO}{can1}. Unfortunately, it is challenging to simultaneously obtain good predictions for the integrated and differential jet shape in this way. By choosing scales for $\psi(r)$, i.e.~for the cross section integrated up to $r$ in terms of the upper bound of this integration, one automatically ensures that $\psi(R) =1$, since the scales in \eq{canFO} will be used when $r=R$. However, the corresponding differential jet shape $\rho(r)$  tends to have artifacts, since the scales depend on $r$ and their derivatives enter through the chain rule. Conversely, choosing scales directly for $\rho(r)$, i.e.~for the cross section differential in $r$, avoids these artifacts. However, the resulting jet shape is generically no longer normalized, i.e.~integrating $\rho(r)$ no longer gives $\psi(R)=1$, because the scales in the integrand are not equal to \eq{canFO} but depend on the integration variable $r$. A solution to this problem has been proposed in ref.~\cite{Bertolini:2017eui}, but is not easy to implement.

Rather than using profile scales to interpolate between $r \ll R$ and $r \lesssim R$, we will directly interpolate between them:
\begin{align}
 \psi(r) = \Big[1-g\Big(\frac{r}{R}\Big)\Big]\psi_{r \ll R}(r) + g\Big(\frac{r}{R}\Big)\, \psi_{r \lesssim R}(r)
\,,\end{align}
which was inspired by ref.~\cite{Echevarria:2018qyi}. Here $\psi_{r \ll R}$ and $\psi_{r \lesssim R}$ correspond to \eqs{psi_fact}{psi_fo}, respectively.
The function $g(r)$ is zero for $r \ll R$ and one for $r$ close to $R$, smoothly interpolating between the two cases. We implement this using the following double-quadratic function
\begin{align} \label{eq:g_match}
g(x) &= \left\{
\begin{tabular}{ll}
$0$ & $\phantom{x_1 \leq}\, x \leq x_1 $
\\
$\frac{(x-x_1)^2}{(x_2-x_1)(x_3-x_1)}$ & $x_1 \leq x \leq x_2$
\\
$1 - \frac{(x-x_3)^2}{(x_3-x_1)(x_3-x_2)}$ & $x_2 \leq x \leq x_3$
\\
$1$ & $x_3 \leq x$ \,.
\end{tabular}\right.
\end{align}
An advantage of this method over the use of profile scales is that we can determine $\psi_{r \ll R}$ and $\psi_{r \lesssim R}$ once and for all before creating the interpolation. In addition, it makes it possible to match LL resummation of $r/R$ to the NLO result in the region where $r \sim R$. (Normally this breaks down because the single logarithmic term $\alpha_s \ln(r/R)$ in the NLO result is not resummed at LL accuracy.)

\begin{figure}[t]
    \centering
     \hfill \includegraphics[width=0.48\textwidth]{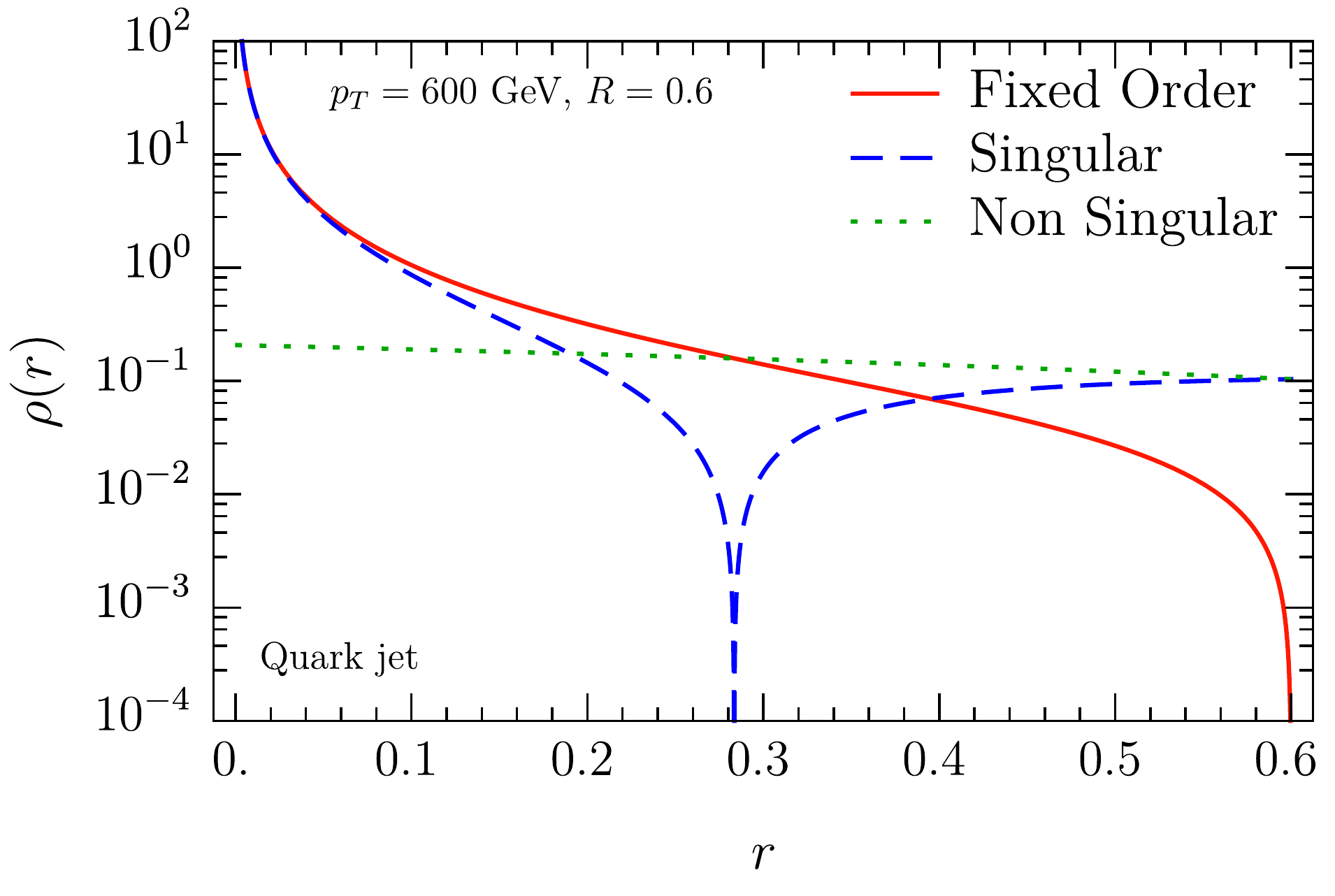} \hfill 
     \includegraphics[width=0.48\textwidth]{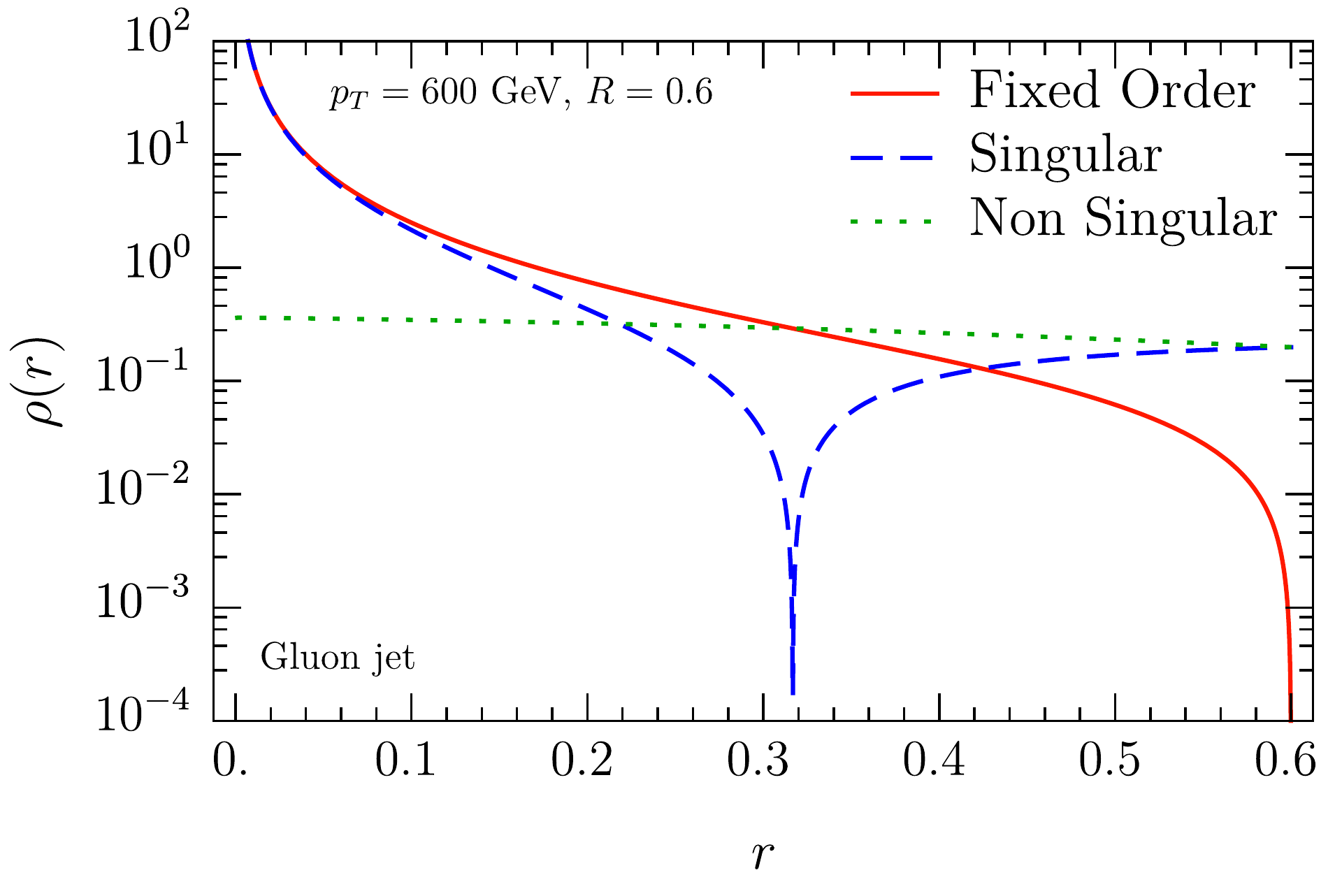} \hfill \phantom{.} \\
    \caption{The differential jet shape at order $\al_s$ for quark jets (left) and gluon jets (right) at $p_T = 600$ GeV and $R=0.6$. Shown are $\rho_{r \lesssim R}$ (red solid), decomposed in its singular $\rho_{r \ll R}$ (blue dashed) and nonsingular (green dotted) contribution that correspond to the power corrections in \eq{psi_fact}.}
    \label{fig:factorizationHCS}
\end{figure}

To determine the transition points $x_i$ in \eq{g_match}, we assess the numerical size of the corrections to the factorization of $\psi_{r\ll R}$ at $\ord{\al_s}$, shown in \fig{factorizationHCS}. The solid red curve shows the differential jet shape, the blue dashed curve shows its singular contribution obtained from the factorization in \eq{psi_fact}, and the green dotted curve is the difference (often called the nonsingular), all at $\ord{\al_s}$. It is essential that the resummation of logarithms of $r/R$ is turned off before we enter the region where the singular is no longer a good approximation to the full jet shape, leading us to choose 
\begin{align}
  x_1 = 0.15 
  \,, \qquad
  x_2 = (x_1 + x_3)/2
  \,, \qquad
  x_3 = 0.38 
.\end{align}
Specifically, $x_1$ corresponds roughly to the point where the nonsingular is 10\% of the cross section, and $x_3$ is chosen close to the point where the singular and nonsingular are equal in size. These points are somewhat arbitrarily chosen, and will be varied as part of the uncertainty estimate.

\subsection{Scale variations and perturbative uncertainties}
\label{sec:uncertainties}

For our uncertainty estimate, we take the envelope of the following variations
\begin{enumerate*}
\item Vary all scales in \eq{can1} simultaneously up and down by a factor of 2. 
\item Vary $\mu_{\cal H}$ and $\mu_H$ simultaneously up and down by a factor of 2.
\item Vary $\mu_C$ and $\mu_S$ simultaneously up and down by a factor of 2.
\item Vary $\nu_C$ up and down by a factor of 2.
\item Vary $\nu_S$ up and down by a factor of 2.
\item Vary the transition point $x_3$ between 0.33 to 0.48.
\end{enumerate*}
We can interpret the first as a fixed-order uncertainty, since it only changes the overall scales and not the ratios between them, which are the logarithms that are being resummed. The second and third variation probe the $\mu$ resummation, since they vary the hierarchy between $\mu_H$ and $\mu_C=\mu_S$, and similarly the fourth and fifth probe the rapidity resummation. While the evolution kernels are very similar for the variations in 2 and 3, the scales of different fixed-order ingredients are probed: The change in the evolution kernel is (partially) cancelled by a corresponding change in $H$ for variation 2 and by $C \otimes S$ for variation 3, so they are not redundant. The same is true for variations 4 and 5. Variation 6 probes the uncertainty from our interpolation between the resummation and fixed-order region. We also explored the dependence on the transition point $x_1$, but it has a small effect and thus would not impact the total uncertainty.

\subsection{First results for quark and gluon jets}
\label{sec:qg_results}

\begin{figure}[t]
    \centering
     \hfill \includegraphics[width=0.48\textwidth]{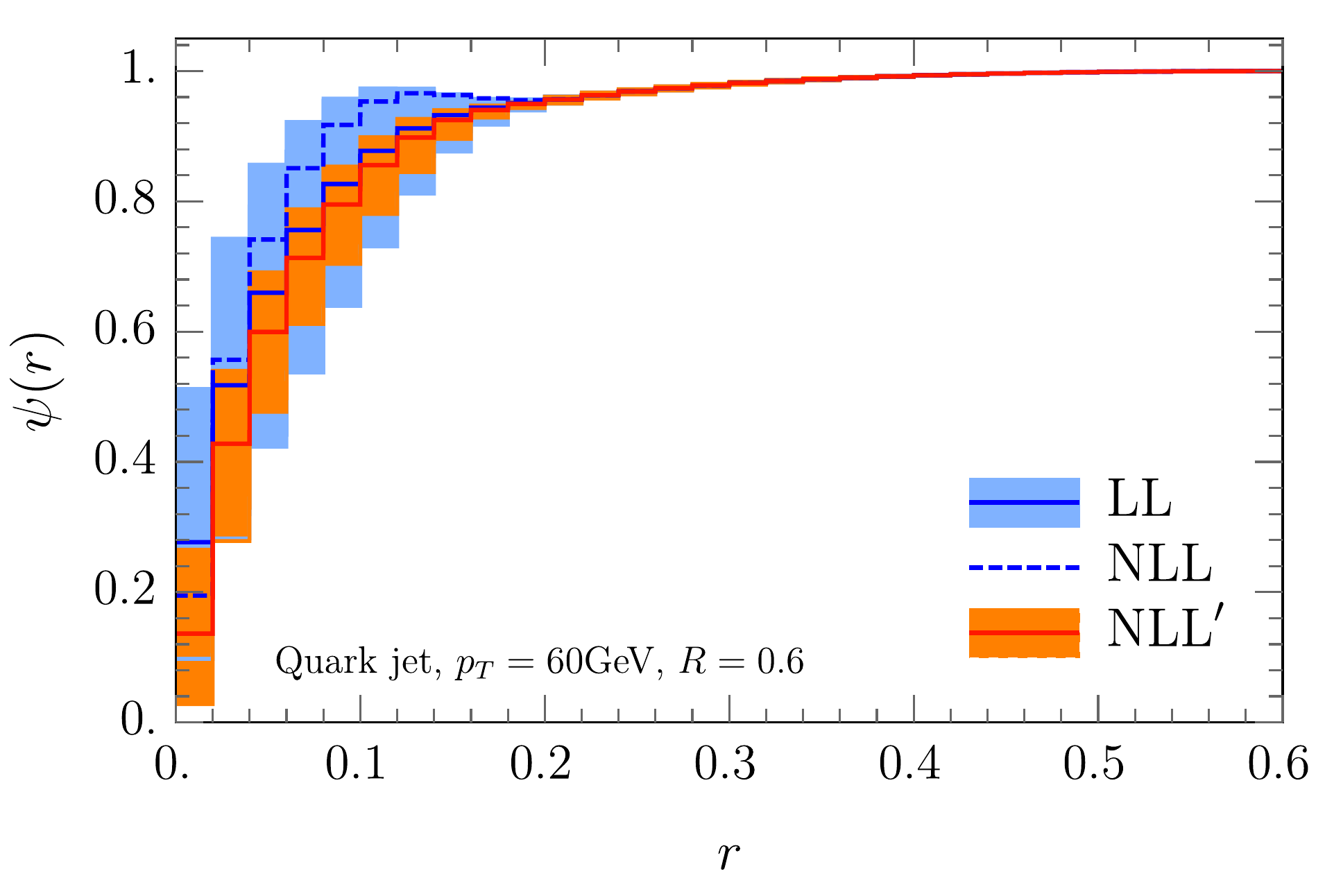} \hfill 
     \includegraphics[width=0.48\textwidth]{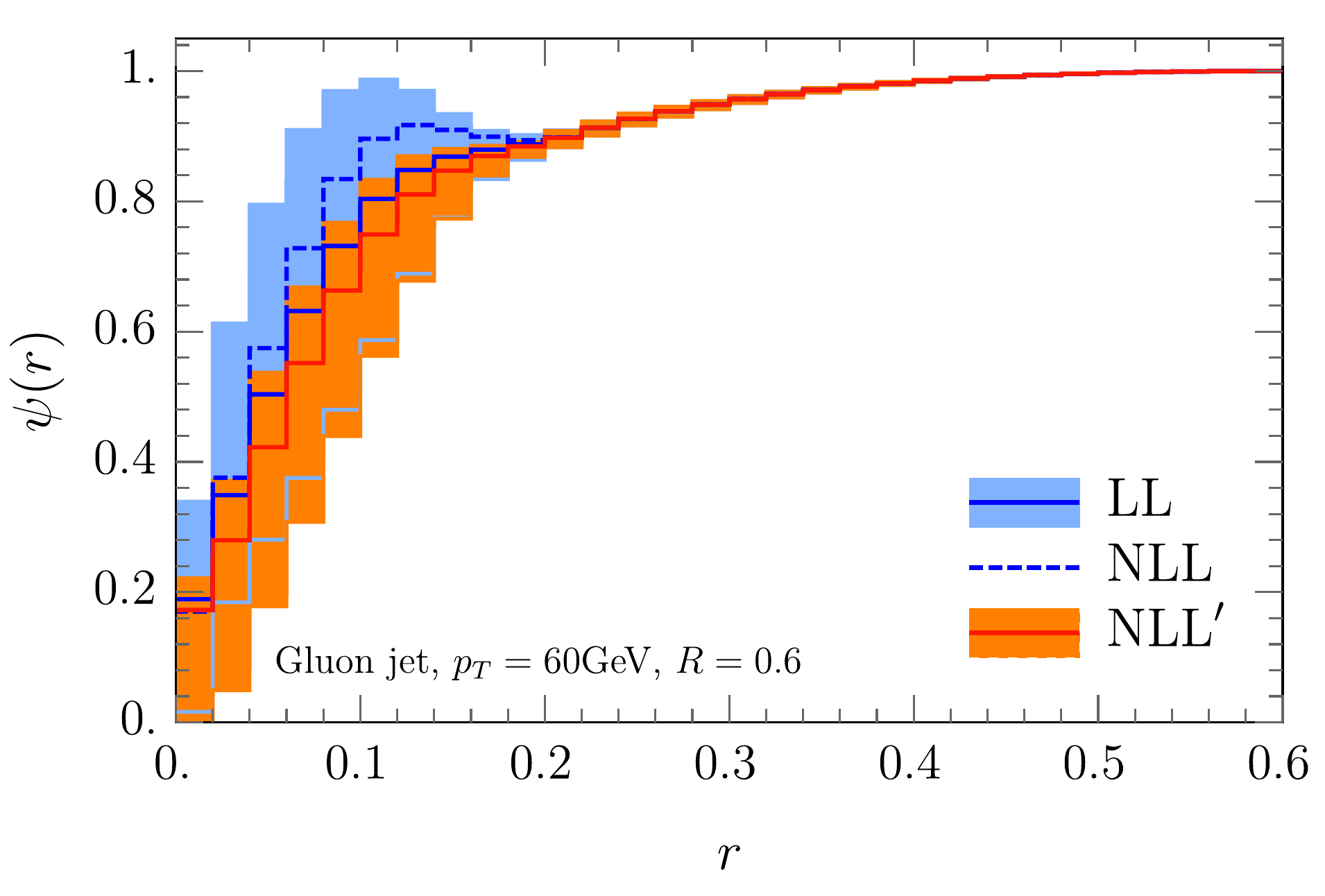} \hfill \phantom{.} \\
     \hfill \includegraphics[width=0.48\textwidth]{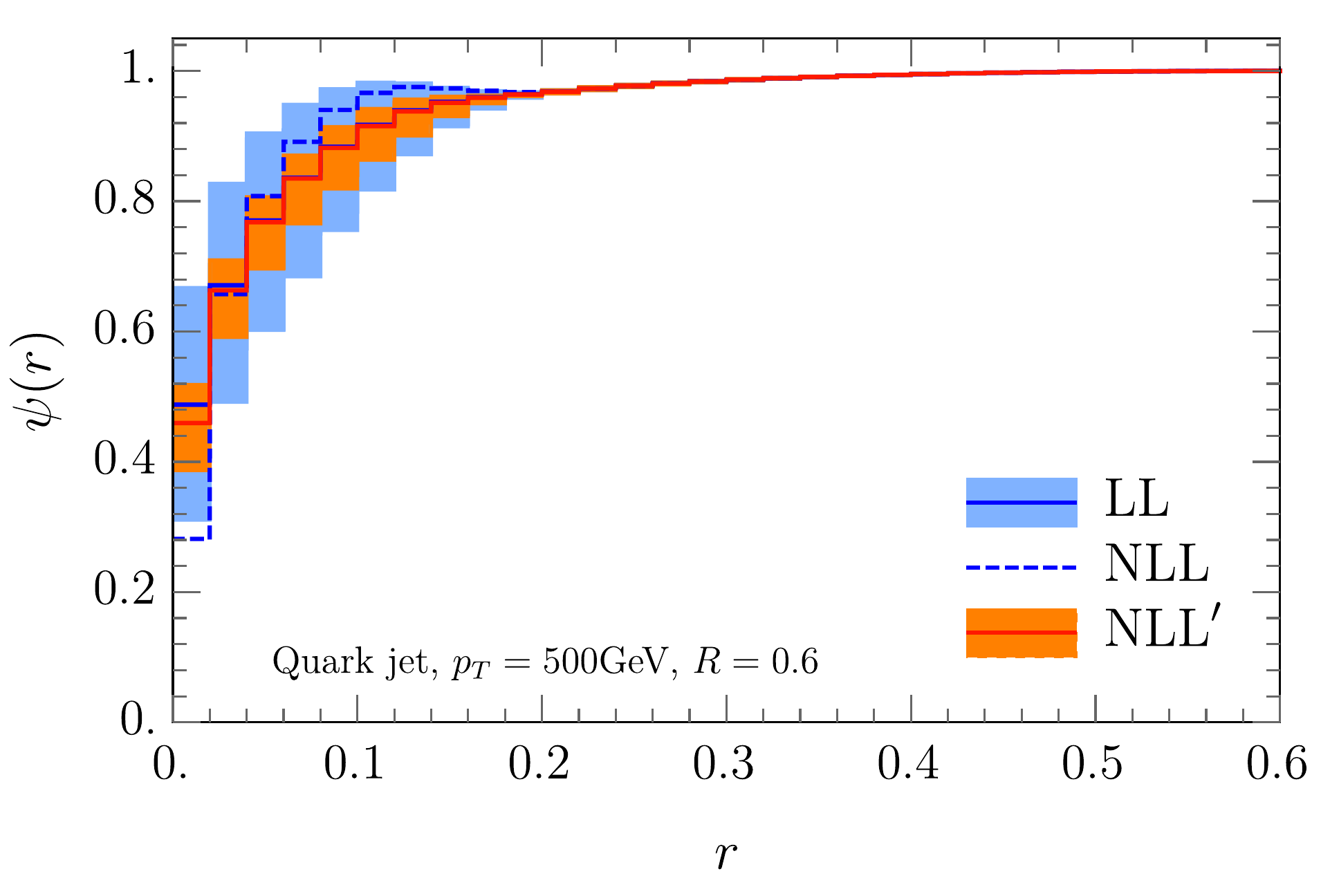} \hfill 
     \includegraphics[width=0.48\textwidth]{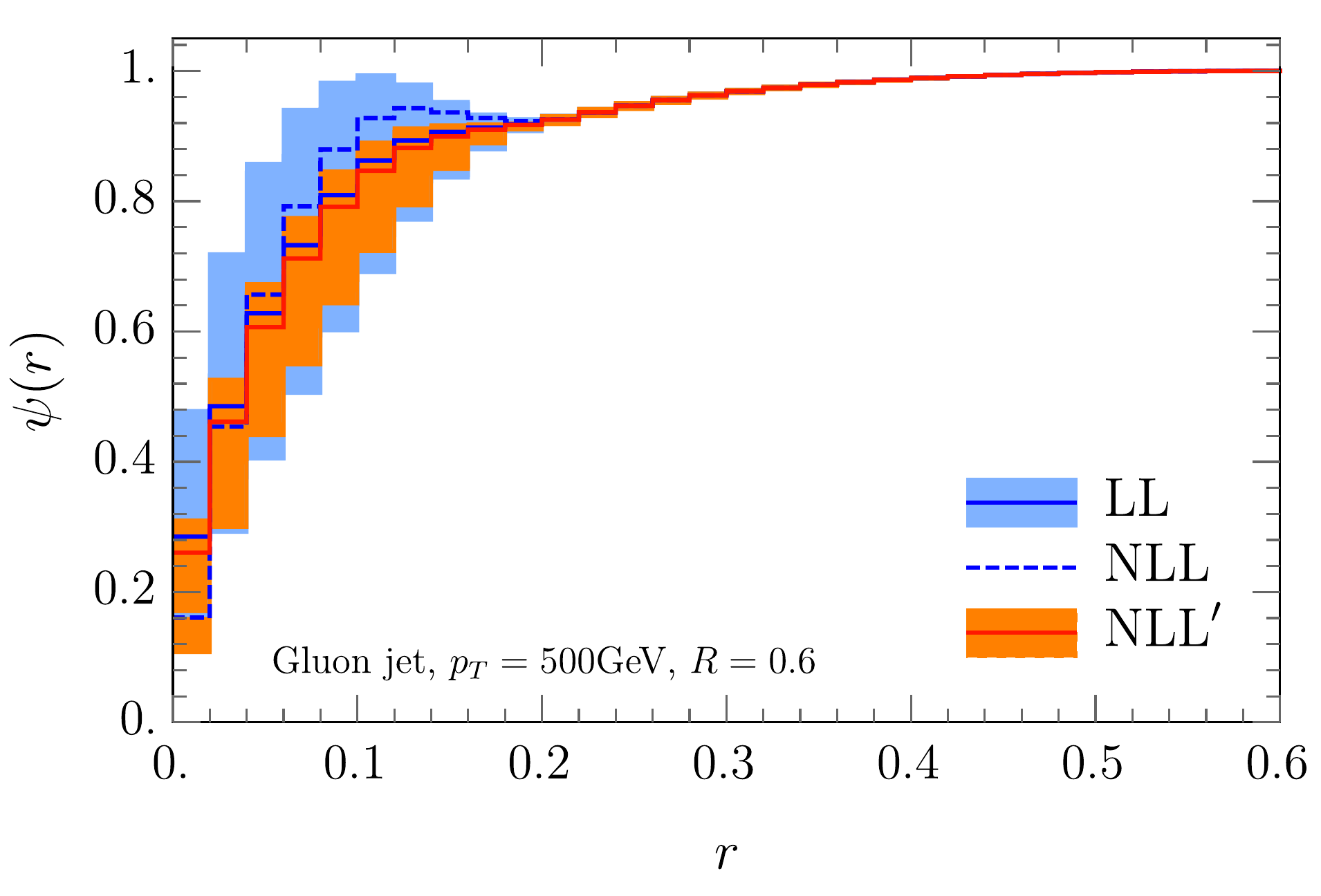} \hfill \phantom{.} \\
    \caption{The integrated jet shape for quark jets (left) and gluon jets (right) for $p_T =$ 60 GeV (top) and 500 GeV (bottom) and $R=0.6$. Shown are LL (blue solid curve and band), NLL (blue dashed curve) and NLL$'$ (orange solid curve and band), all matched to NLO. The bands indicate the perturbative uncertainties estimated using the procedure in \sec{uncertainties}. }
  \label{fig:convergence}
\end{figure}

\begin{figure}[t]
    \centering
     \hfill \includegraphics[width=0.48\textwidth]{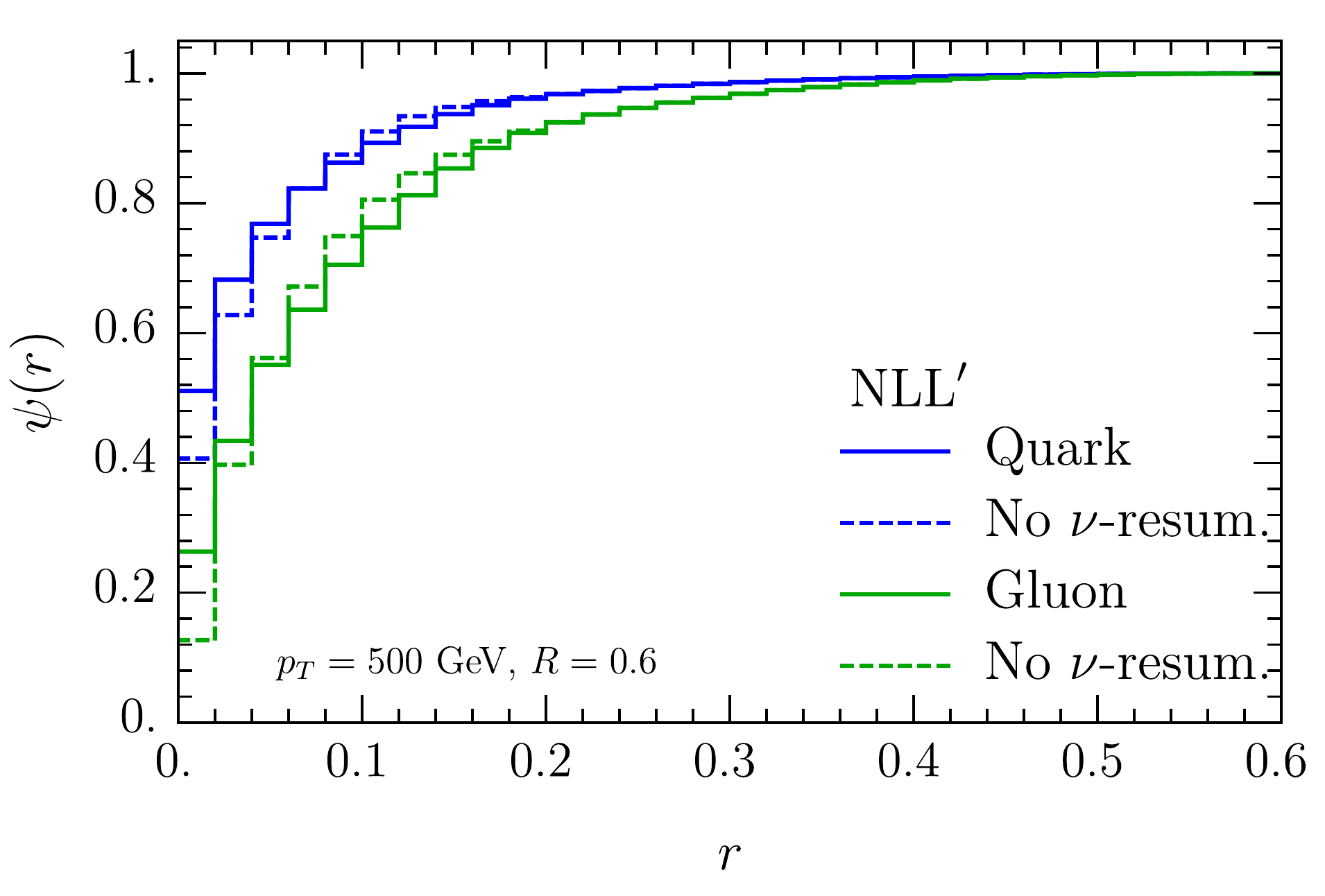} \hfill 
     \includegraphics[width=0.48\textwidth]{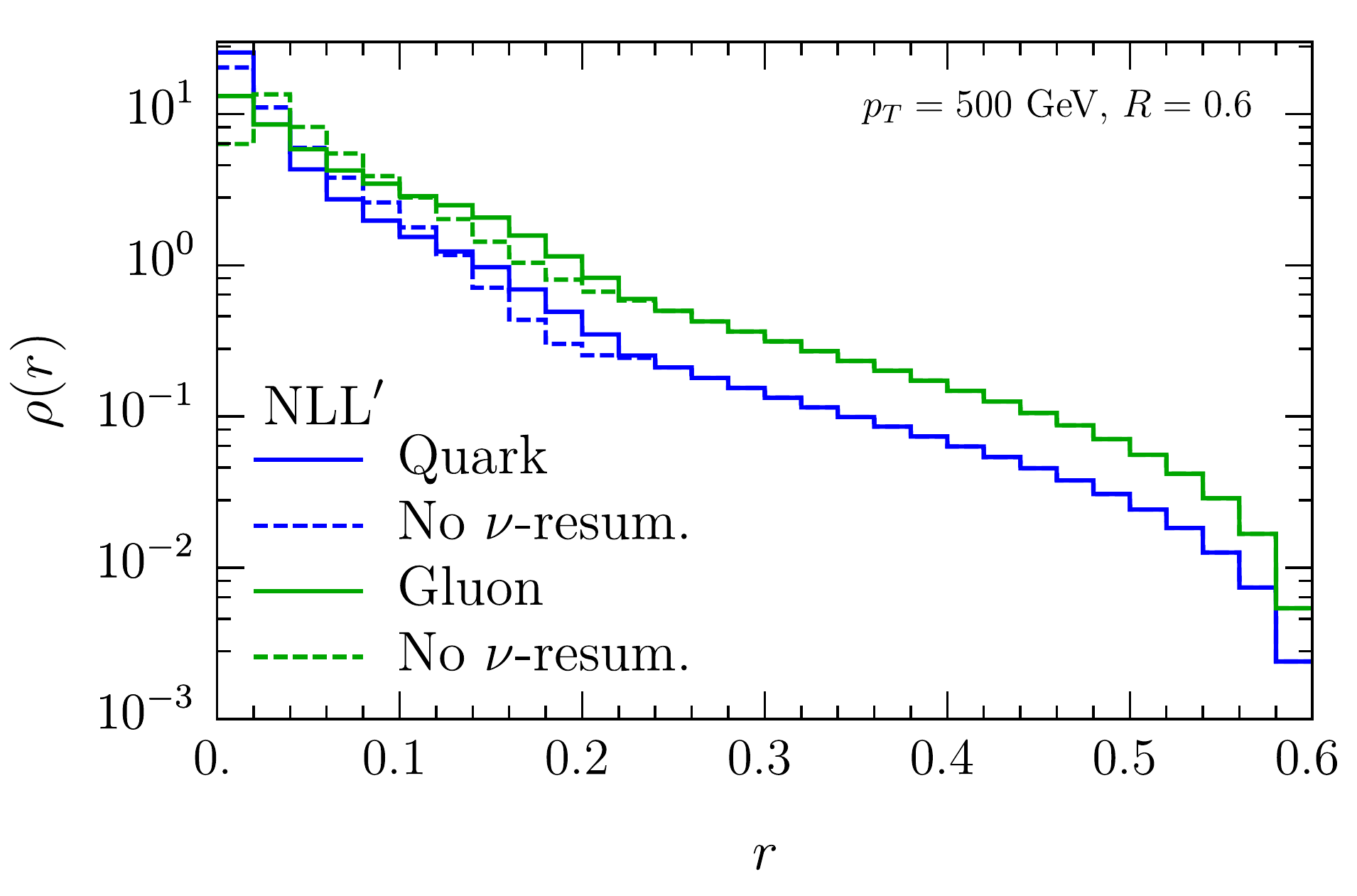} \hfill \phantom{.} \\
     \hfill \includegraphics[width=0.48\textwidth]{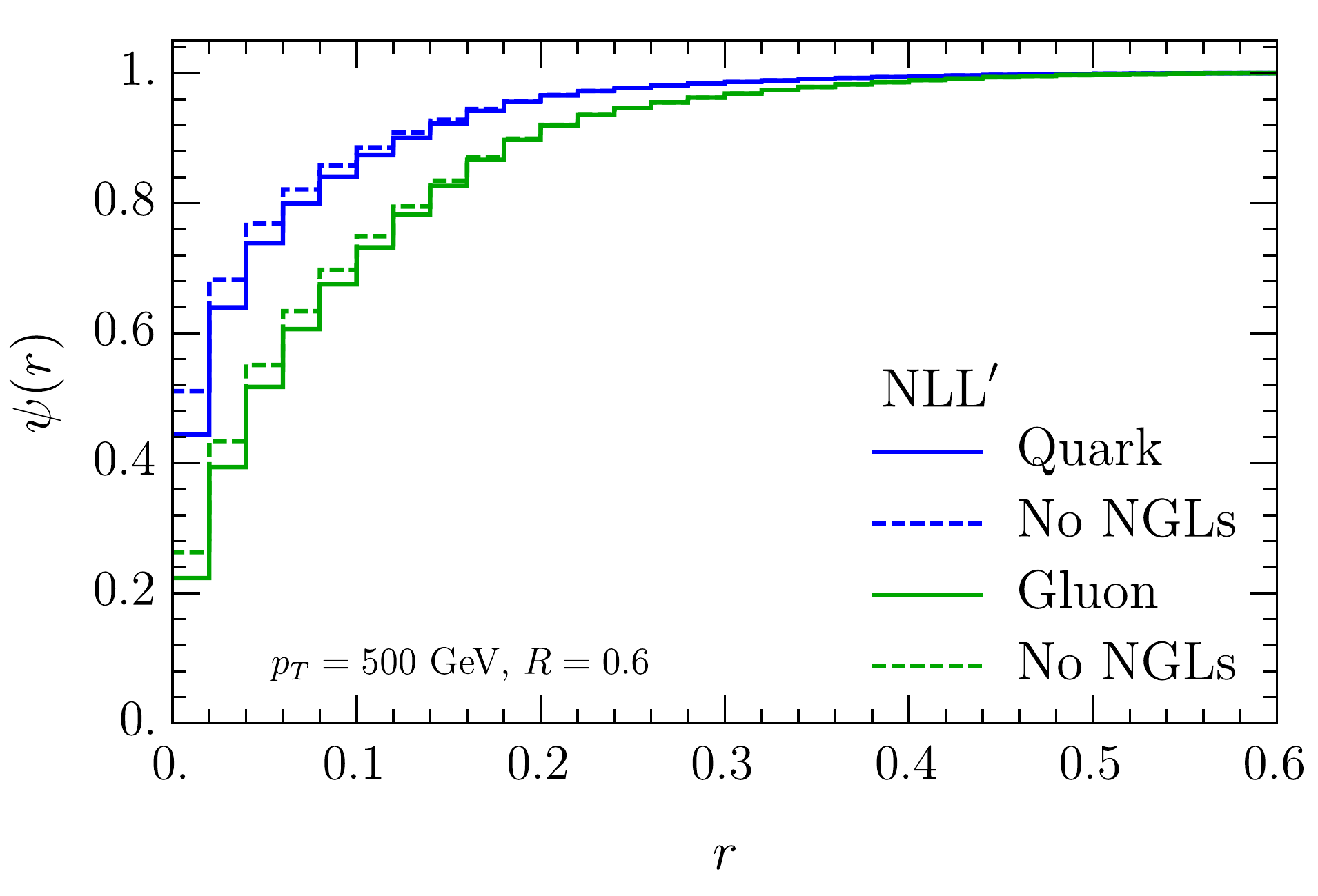} \hfill 
     \includegraphics[width=0.48\textwidth]{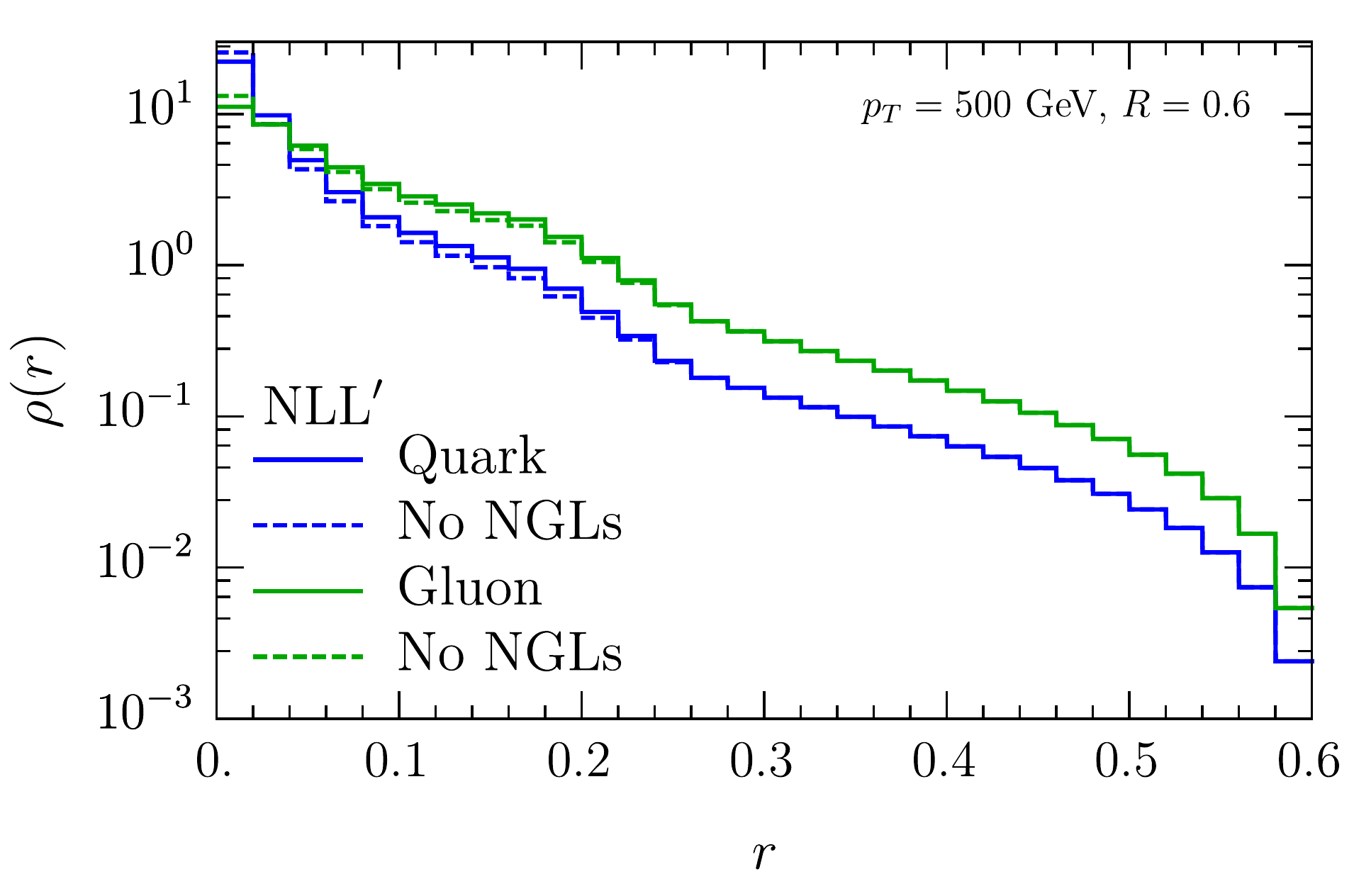} \hfill \phantom{.} \\
    \caption{The integrated (left) and differential (right) jet shape for quark jets (blue) and gluon jets (green) at NLL$'$ with $p_T =500$ GeV and $R=0.6$. The top row shows the result with (solid) and without (dashed) rapidity resummation. The bottom row shows the result with (solid) and without (dashed) non-global logarithms.}
  \label{fig:RR_NGL}
\end{figure}

In \fig{convergence} we show the integrated jet shape for quark and gluon jets with transverse momentum 60 and 500 GeV, comparing LL, NLL and NLL$'$, all matched to NLO. We start by noting that our resummed calculations converge, i.e.~the bands overlap and are smaller at NLL$'$ than at LL. As these plots show, the uncertainty bands are larger at small $p_T$ and for gluon jets, which is not surprising since the perturbative corrections are simply larger in these cases due to the size of $\alpha_s$ and the color factor ($C_A$ vs.~$C_F$), respectively. At LL the bands are quite large, and they remain sizable at NLL$'$, even though this exceeds the accuracy of previous jet shape calculations. We believe that this is due to our more conservative uncertainty estimate, described in \sec{uncertainties}. Indeed, for gluon jets with $p_T = 60$ GeV, the NLL$'$ central curve lies only just inside the LL uncertainty, warranting these large bands. For large $r$ all predictions overlap, since in each case we match to NLO. In this region there are no large logarithms of $r/R$, so it is not surprising that the uncertainties are small. It is hard to gauge whether their size is reasonable, since we only have one perturbative order in this region (at LO $\psi(r)$ is simply 1). Lastly we note that the NLL curve is not monotonic. That this unphysical behavior arises at NLL but not LL is due to the non-cusp anomalous dimension, whose contribution becomes equal and opposite in size to that of the cusp anomalous dimension. At NLL$'$ this is remedied by the inclusion of one-loop corrections to the ingredients in the factorization formula.

Fig.~\ref{fig:RR_NGL} illustrates the effect of the rapidity resummation and non-global logarithms on the jet shape for quark and gluon jets, by showing the integrated and differential jet shape with and without them. Both effects enter for the first time at NLL and were not taken into account in previous calculations. They are clearly important for $r \ll R$, where their size is about half that of the NLL$'$ uncertainty band in \fig{convergence}, but are of course irrelevant in the region where the resummation of logarithms of $r/R$ is turned off. The effect of the rapidity resummation is still significant in the transition region and is only turned off by the matching to fixed order, whereas the effect of NGLs is already small before the transition. The NGLs push the energy distribution in the jet out to larger values of $r$, in agreement with the trend observed in e.g.~ref.~\cite{Dasgupta:2012hg} for the jet mass distribution. This is clear from the integrated jet shape, and in the differential jet shape this arises through a decrease in the first bin (not as visible due to the logarithmic scale) and an increase in subsequent bins. 

\section{Nonperturbative effects}
\label{sec:nonp}

At the LHC, nonperturbative corrections to the jet shape can be large, particularly for smaller values of the jet $p_T$. These arise from the underlying event and hadronization effects. In this section we will explore simple models to account for these effects in our predictions. We will use \Pythia 8.2~\cite{Sjostrand:2014zea} to assess how reasonable these models are, by applying them to parton level predictions and comparing the result to hadron level predictions, including multi-parton interactions and initial-state radiation\footnote{This is not a nonperturbative contribution. However, we include it, because in our approach initial-state radiation is formally $\ord{R^2}$ suppressed.}. When we later apply this model to our predictions, we will use LHC data to fit the model parameter. 

In the first model, we treat these effects by simply adding a uniform energy density to the jets. Explicitly
\begin{align} \label{eq:mod1}
  \text{Model 1:} \quad  \rho (r) \to \frac{2f}{1+f} \frac{r}{R^2} + \frac{1}{1+f} \rho (r)
\,,\end{align}
where $\rho(r)$ is the differential jet shape and the fraction $f$ describes the size of the nonperturbative radiation compared to the perturbative contribution from our calculation. The dependence on $r/R$ is linear because it scales with the circumference of the circle of radius $r$.

\begin{figure}[t]
    \centering
     \hfill \includegraphics[width=0.48\textwidth]{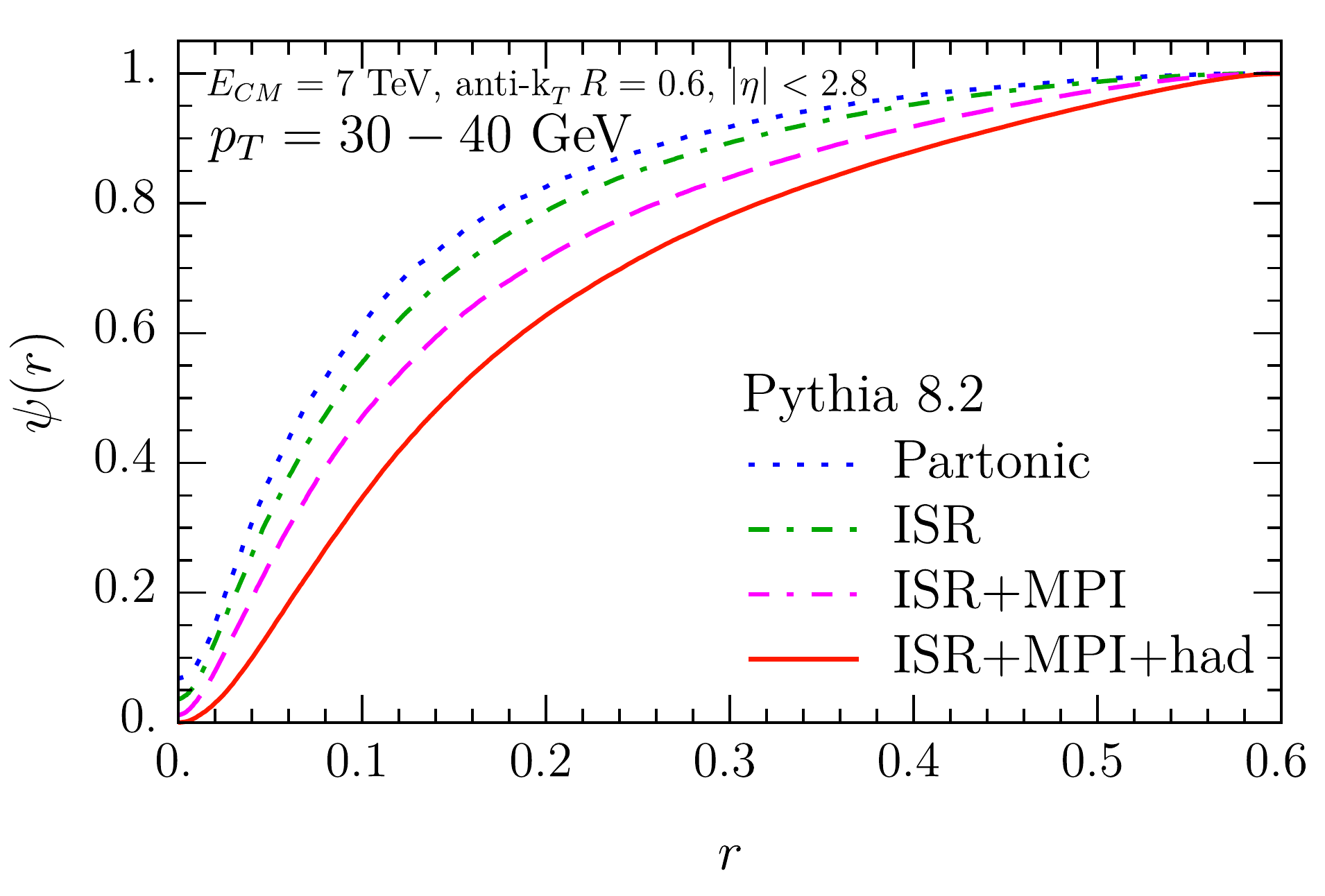} \hfill 
     \includegraphics[width=0.48\textwidth]{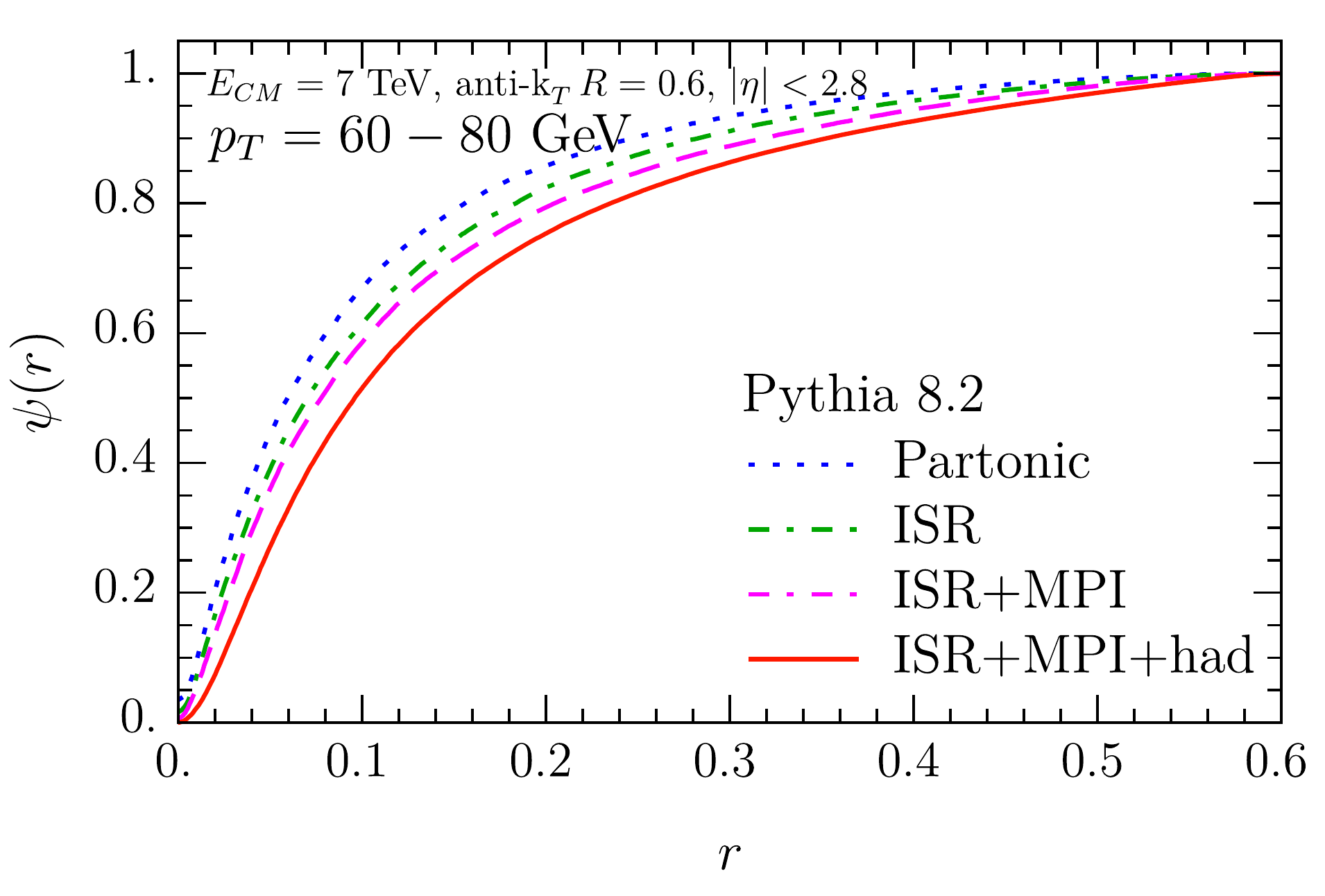} \hfill \phantom{.} \\
     \hfill \includegraphics[width=0.48\textwidth]{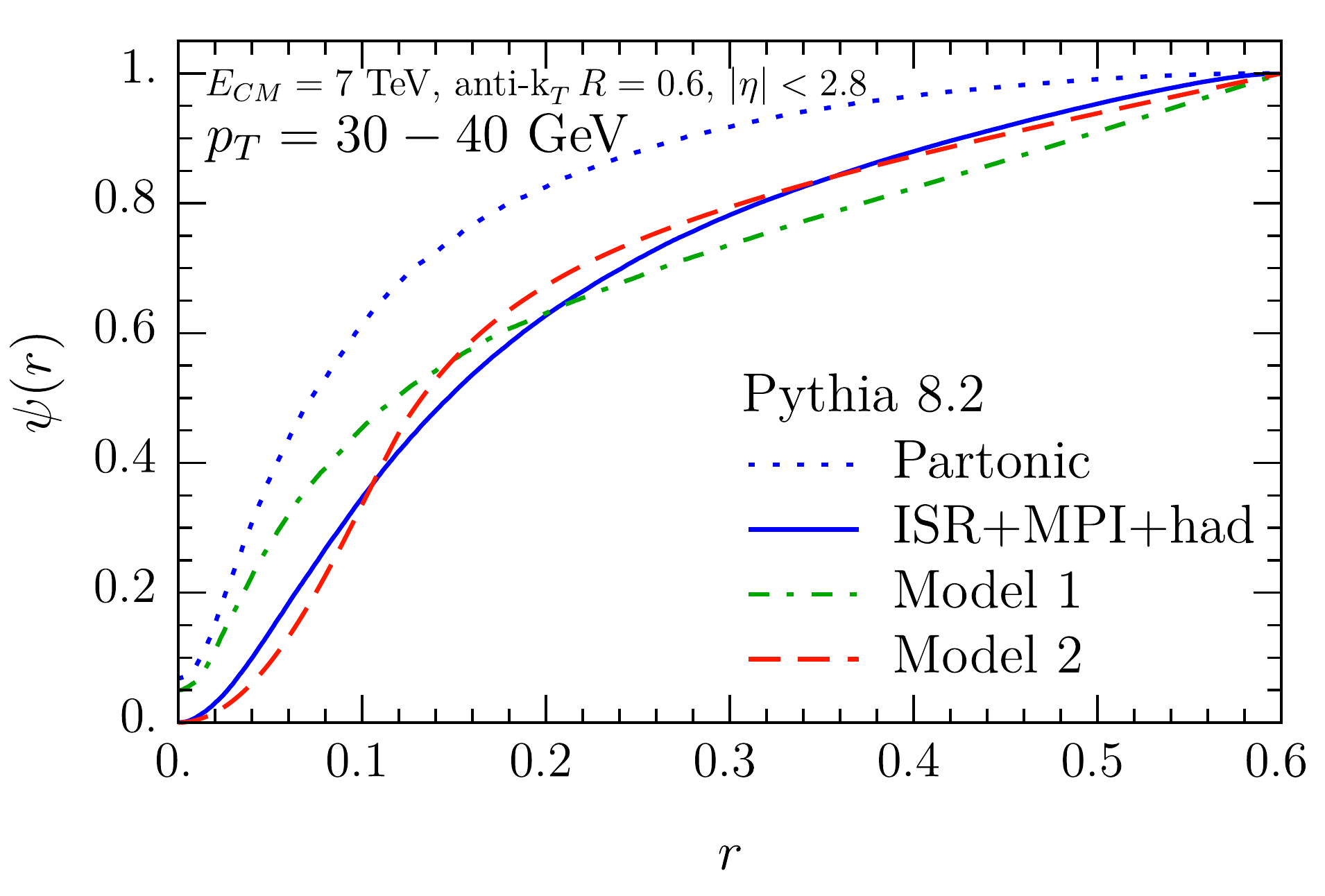} \hfill 
     \includegraphics[width=0.48\textwidth]{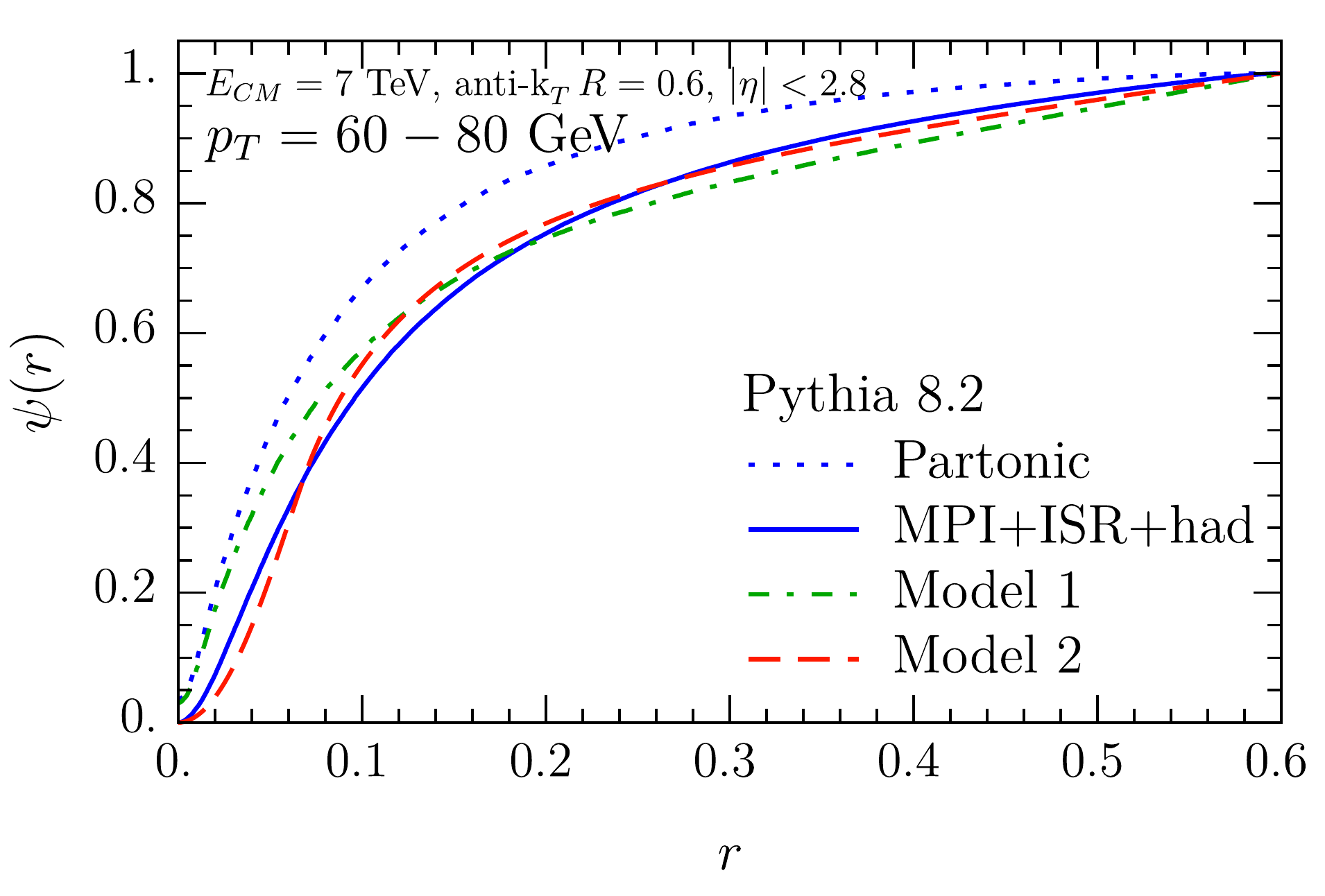} \hfill \phantom{.} \\
    \caption{Jet shape in \Pythia at $E_{\rm cm} = 7$ TeV for anti-$k_T$ jets with radius $R=0.6$, rapidity $|\eta|<2.8$ and transverse momentum $p_T = 30 - 40$ GeV (left) and $p_T = 60-80$ GeV (right). The top row shows results at parton level (blue dotted), including initial-state radiation (green dot-dashed), also multi-parton interactions (magenta dashed), and also hadronization (blue solid). The bottom row repeats the blue dotted and solid curves, where the effect of ISR, MPI and hadronization are either modeled using \eq{mod1} (green dot-dashed) or \eq{mod2} (orange dashed), and the parameter $f$ is fitted.}
    \label{fig:nonp}
\end{figure}

In the second model, the nonperturbative effects are again treated by adding energy. However, rather than considering a uniform distribution, we will assume this additional energy is completely localized, thus displacing the jet axis. The location $(r', \phi')$ of this extra energy is then (uniformly) integrated over the jet. In terms of the differential jet shape $\rho(r)$ this amounts to
\begin{align} \label{eq:mod2}
  \text{Model 2:} \quad  \rho(r) \to &\frac{1}{\pi (1+f)^2 R^2} \int_0^{(1+f)R}\! \df r'\, r'\! \int_0^{2\pi}\! \df \phi'
  \bigg\{ \frac{f}{1+f}\, \de\Big( r - \frac{r'}{1+f} \Big)
 \\ & +
  \frac{1}{1\!+\!f}\,\int_0^R\! \df r''\,  \rho'(r'')\, \de\bigg[r \!-\! \bigg((r'')^2 \!+\! \frac{2 r' r'' f}{1\!+\!f} \cos \phi \!+\! \frac{(r')^2 f^2}{(1\!+\!f)^2} \bigg)^{1/2} \bigg] \bigg\}
  \nn\\ &
  = \frac{2f}{1+f}\, \frac{r}{R^2} +   \frac{4r}{\pi f^2(1+f) R^2} \int_{\max\{0, r- fR\}}^{\min\{R,r+f R\}}\! \df r''\, \rho(r'')
  \nn \\ & \quad
    \int_{|r_*-r''_*|}^{\min\{1, r_*+r''_* \}} \, \df r'_*\, \frac{r'_*}{\sqrt{2(r_*^2+(r'_*)^2) (r''_*)^2 - (r_*^2 - (r'_*)^2)^2 - (r''_*)^4}}
\,,\nn \end{align}
The first term in the curly brackets (on the first line) corresponds to the contribution to the jet shape from the extra energy. Upon integration this results in the first term on the third line, in agreement with \eq{mod1}. However, the jet shape also gets smeared by the displacement of the jet axis. Specifically, the $r''$ integral in the second term integrates over the jet shape, taking into account the displacement from the extra energy at $(r',\phi')$ through the delta function. In the final expression we eliminated the $\phi'$ integral using the delta function, and introduced 
\begin{align}
  r_* = \frac{r}{f R}
  \,, \qquad
 r'_* = \frac{r'}{(1+f) R}
 \,, \qquad
 r''_* = \frac{r''}{f R}
\end{align}
to remove any explicit dependence on $f$ and $R$ in the $r'$ integral. Thus we can determine this integral numerically once and for all, to obtain the kernel against which we integrate the jet shape in the remaining $r''$ integral.

In \fig{nonp} we show the partonic and hadronic jet shape obtained from \Pythia, and investigate to what extent multiparton interactions, initial-state radiation and hadronization effects can be described by one of the models above. The top row shows the contribution of each of these effects, the bottom row shows a fit (of the coefficient $f$) to these effects using one of the models proposed here. We find better agreement using model 2, suggesting that these nonperturbative effects are at least fairly localized in the jet. We therefore use model 2 when we compare our results to LHC data in the next section.

\section{Results for the LHC}
\label{sec:results}

\begin{figure}[t!]
    \centering
     \hfill \includegraphics[width=0.48\textwidth]{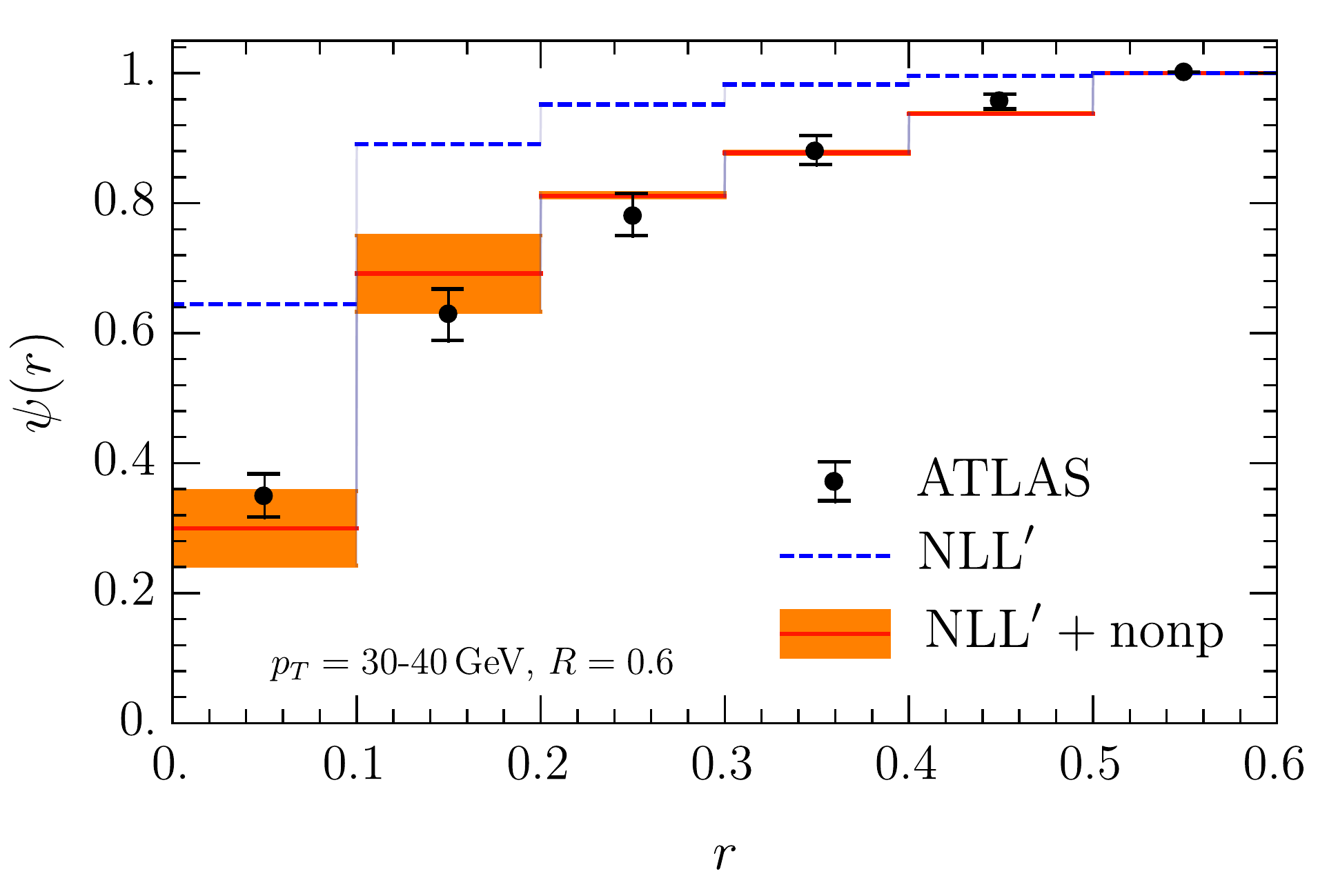} \hfill 
     \includegraphics[width=0.48\textwidth]{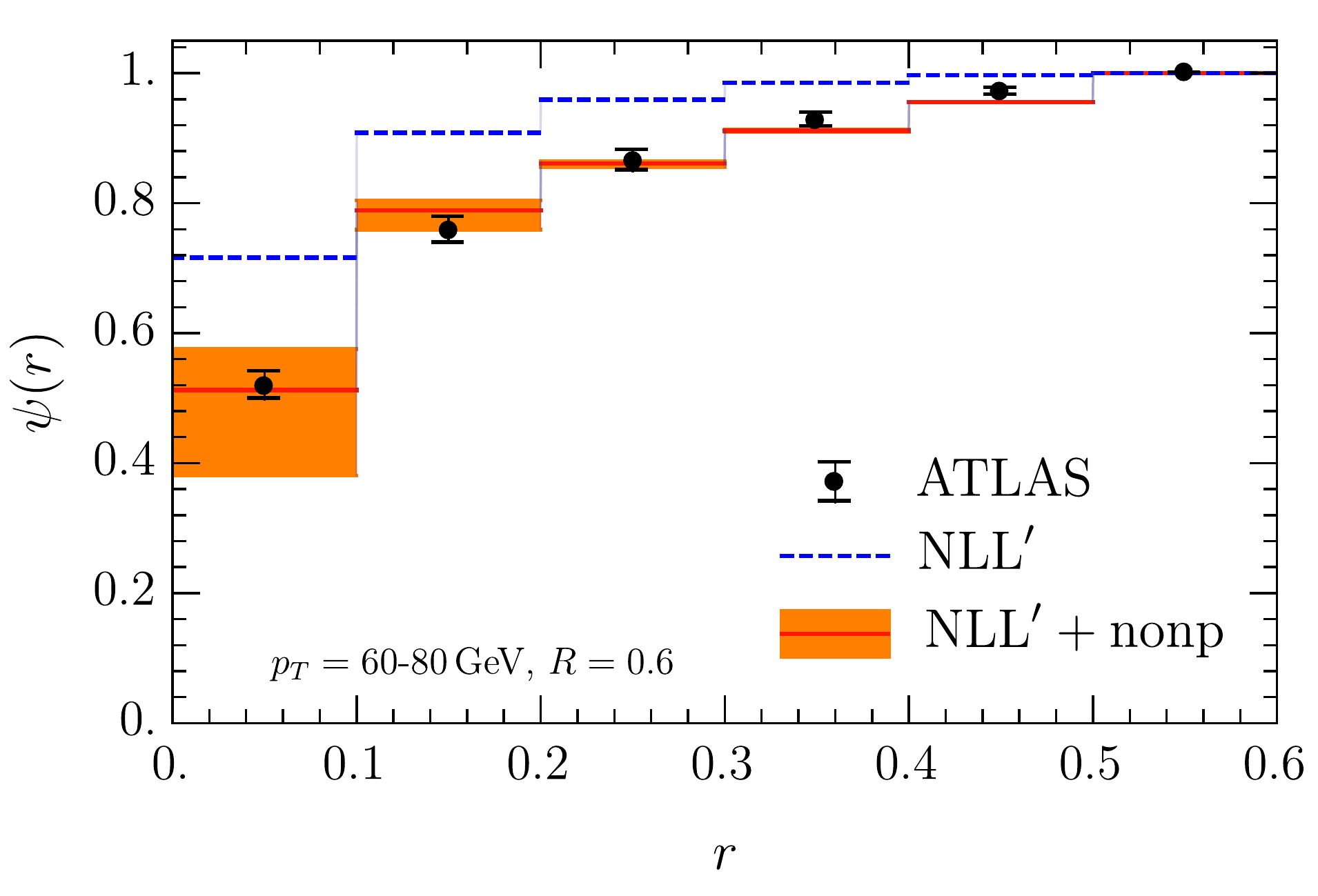} \hfill \phantom{.} \\
     \hfill \includegraphics[width=0.48\textwidth]{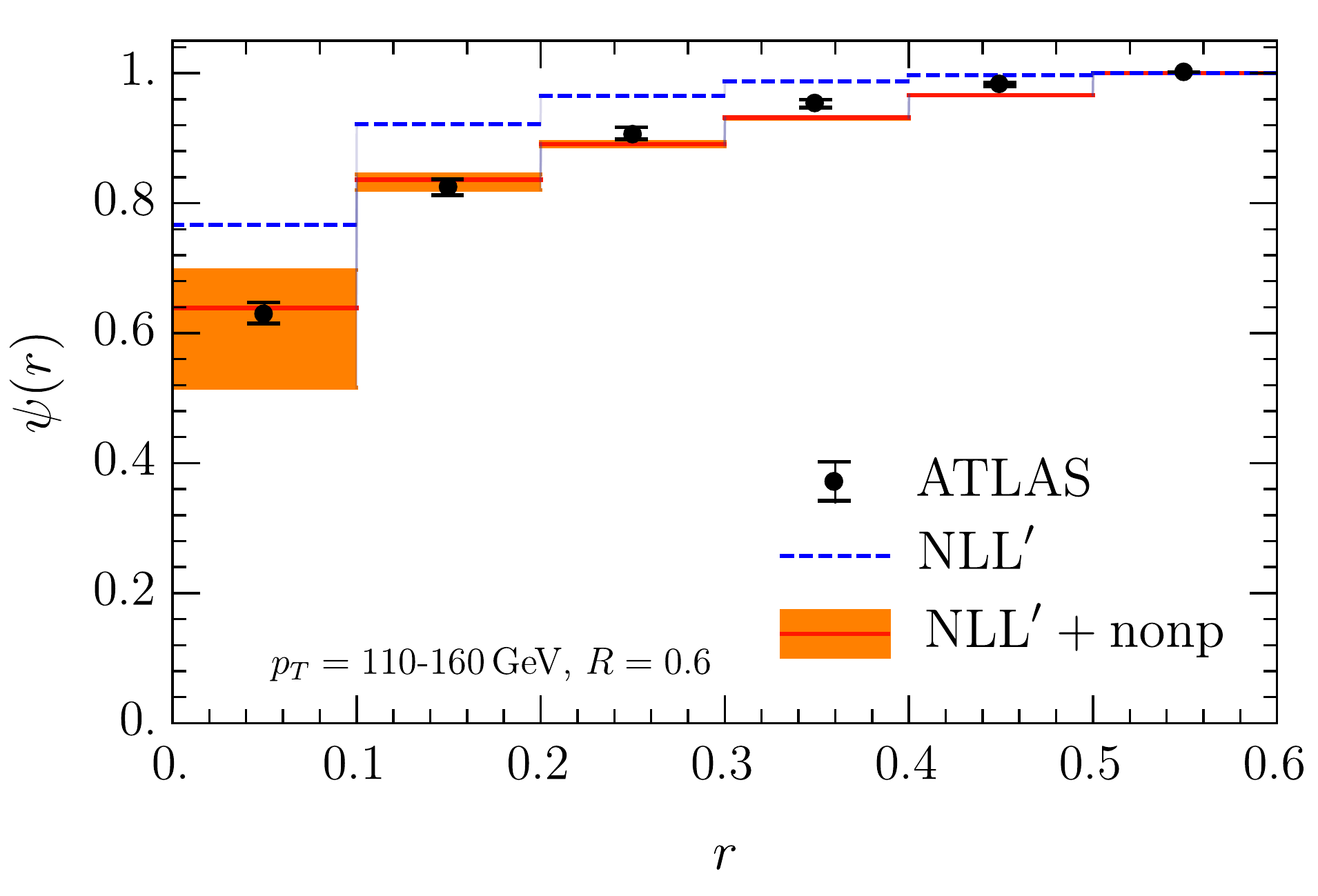} \hfill 
     \includegraphics[width=0.48\textwidth]{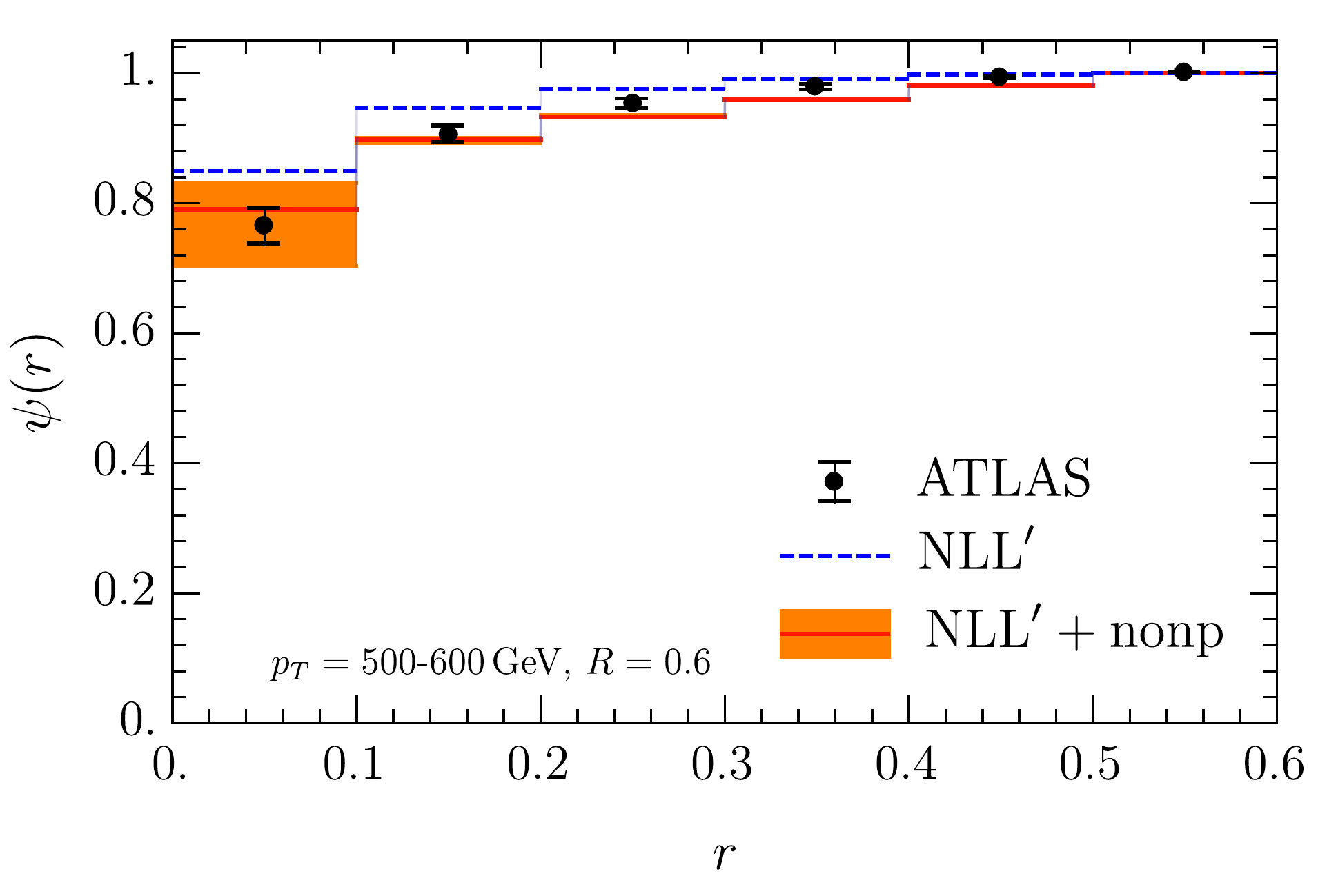} \hfill \phantom{.} \\
    \caption{Comparison of our theoretical calculation at NLL$'$ accuracy for the inclusive jet shape $\psi(r)$ to the data from ATLAS~\cite{Aad:2011kq}. The inclusive anti-k$_T$ jet sample is reconstructed for a jet radius of $R=0.6$ and $|\eta|<2.8$ measured at $E_{\rm cm}=7$~TeV. Four representative jet $p_T$ intervals in the range  $30-600$~GeV are shown, as indicated in the panels. Our purely perturbative calculation is shown (dashed blue), along with the results that include the nonperturbative effects through a model (orange) and perturbative uncertainty band. \label{fig:ATLAS1}}
\end{figure}

In this section, we present comparisons of our theoretical results to data from ATLAS and CMS. We start with the ATLAS data for the integrated and differential jet shape of ref.~\cite{Aad:2011kq}. Although the central values of the differential and integrated jet shape are directly related, their uncertainties are not because of correlations, which is why we show results for both. The inclusive jet sample $pp\to{\rm jet}+X$ was obtained using the anti-k$_T$ algorithm with $R=0.6$. The jet rapidity was restricted to $|\eta|<2.8$, and different intervals for its transverse momentum $p_T$ in the range of $30-600$~GeV were considered. In fig.~\ref{fig:ATLAS1}, we show a comparison of our numerical results to the ATLAS data for the integrated jet shape $\psi(r)$ for four representative $p_T$ intervals, as indicated in the different panels. For all phenomenological results presented in this section, we use the CT14 NLO PDF set of ref.~\cite{Dulat:2015mca}. We show both the purely perturbative calculation (dashed blue curve) as well as the results after including the nonperturbative ``model 2'' (orange curve and band), described in section~\ref{sec:nonp}. The QCD scale uncertainty is shown only for the final result including the nonperturbative contribution, following the procedure discussed in section~\ref{sec:uncertainties}. The parameter $f$ of the nonperturbative model is determined from fitting our central curve to the central value of the data. Its value is tabulated in table~\ref{tab:NPparamf} for each $p_T$ interval. At smaller values of the jet transverse momentum, the purely perturbative result and the data disagree significantly, where the perturbative calculation predicts that there is more radiation close to the center of the jet and less near the jet boundary. Nevertheless, we find that the data is well described after including a nonperturbative correction. As is clear from the plots and the table, at higher values of $p_T$ the nonperturbative correction becomes smaller. Indeed, for the highest transverse momentum interval $500-600$~GeV we already find a good agreement between the data and our purely perturbative calculation obtained from the QCD factorization theorem. Because the nonperturbative model parametrizes the effect of nonperturbative physics relative to the perturbative prediction for the jet, rather than through some absolute energy scale, it is not surprising that approximately $f \sim 1/p_T$.

\begin{figure}[t]
    \centering
     \hfill \includegraphics[width=0.48\textwidth]{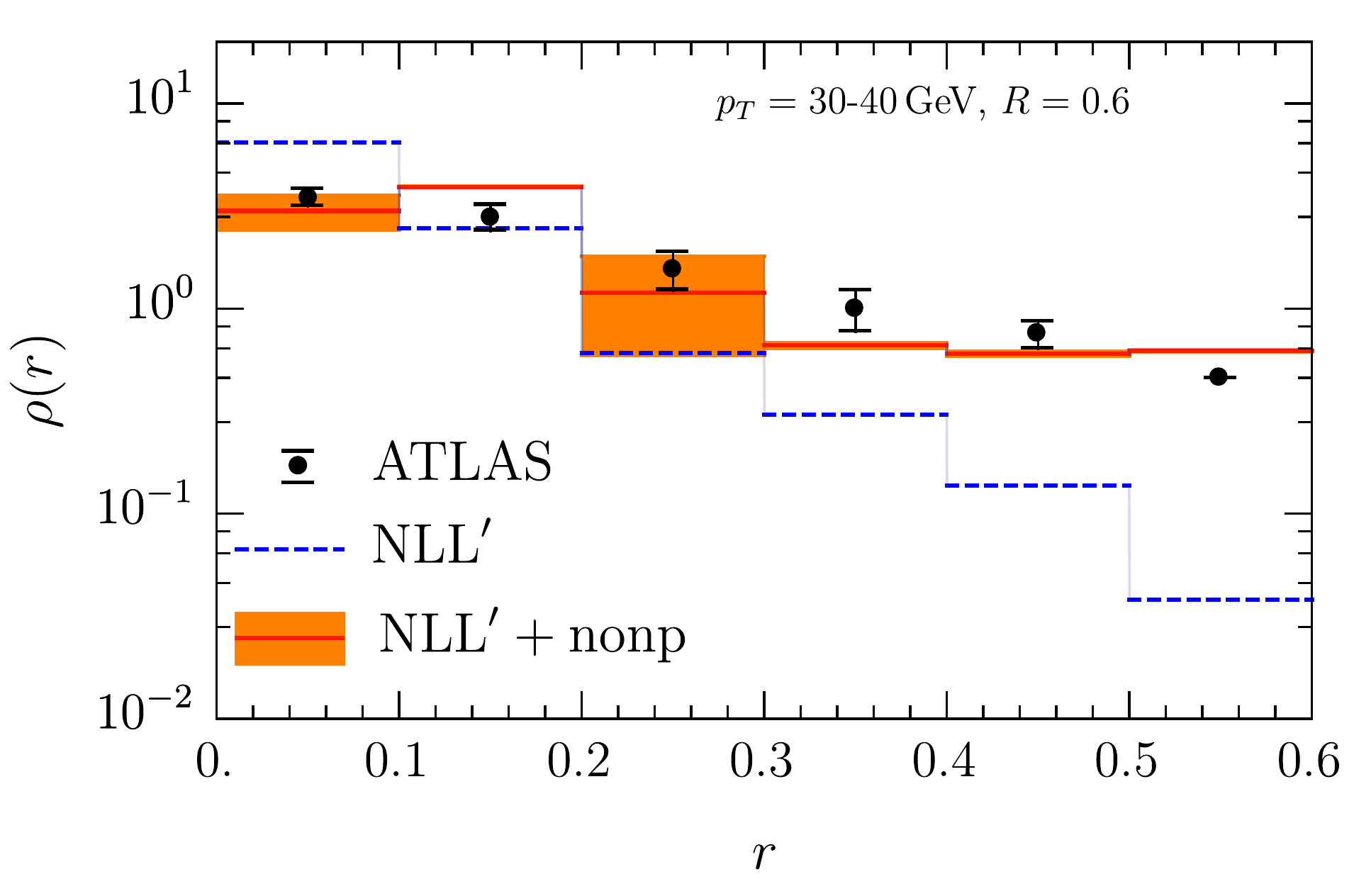} \hfill 
     \includegraphics[width=0.48\textwidth]{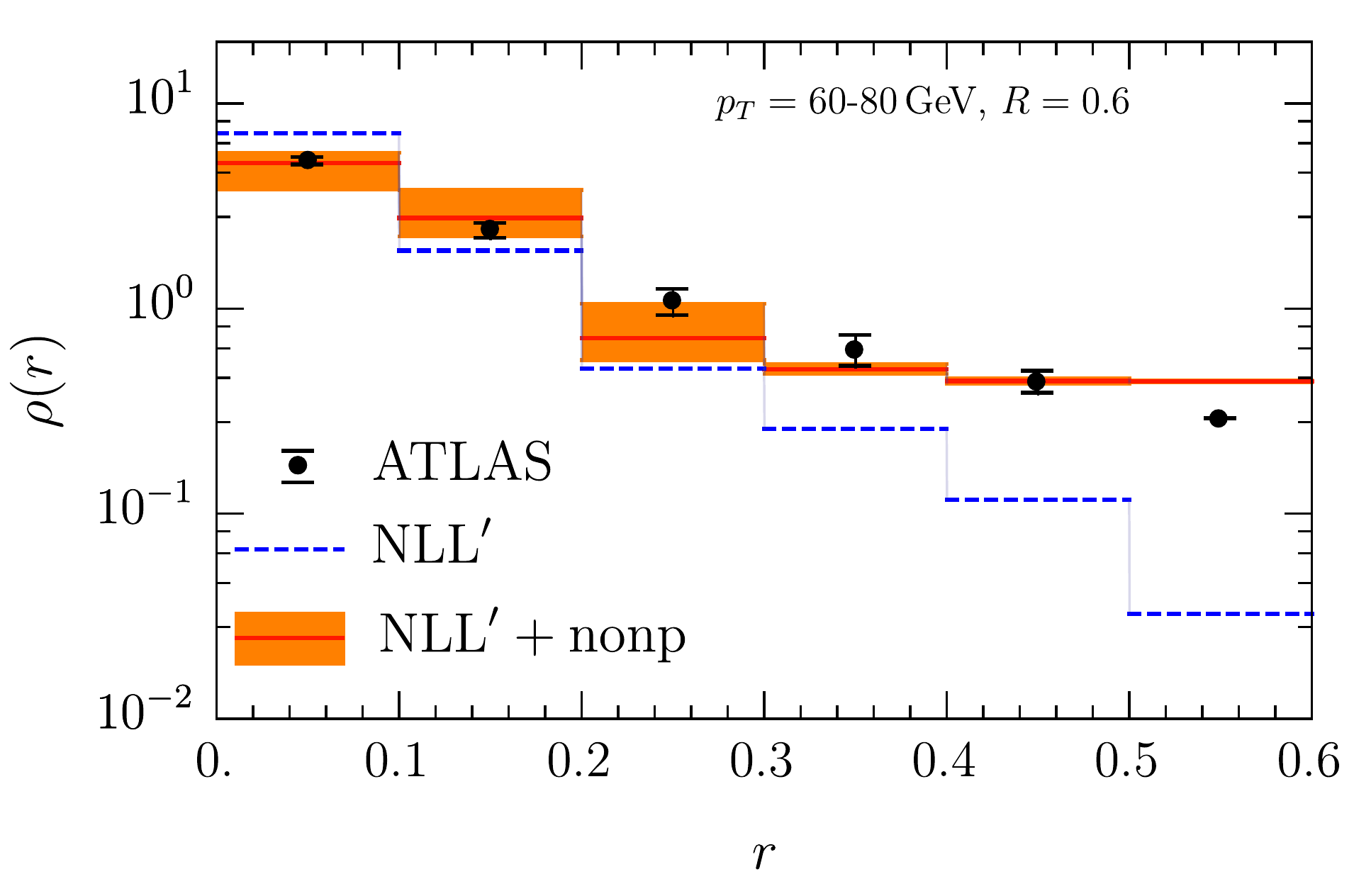} \hfill \phantom{.} \\
     \hfill \includegraphics[width=0.48\textwidth]{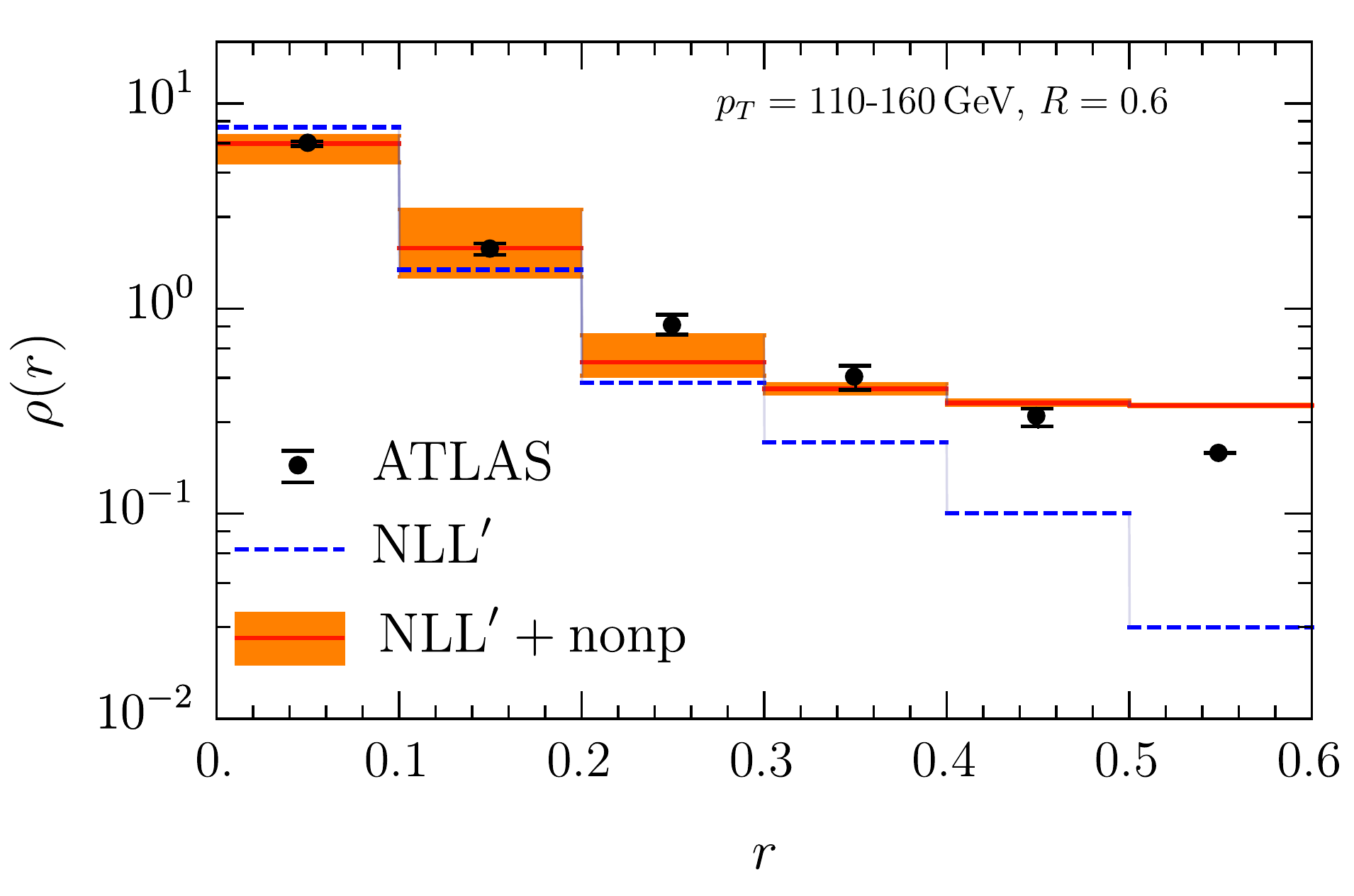} \hfill 
     \includegraphics[width=0.48\textwidth]{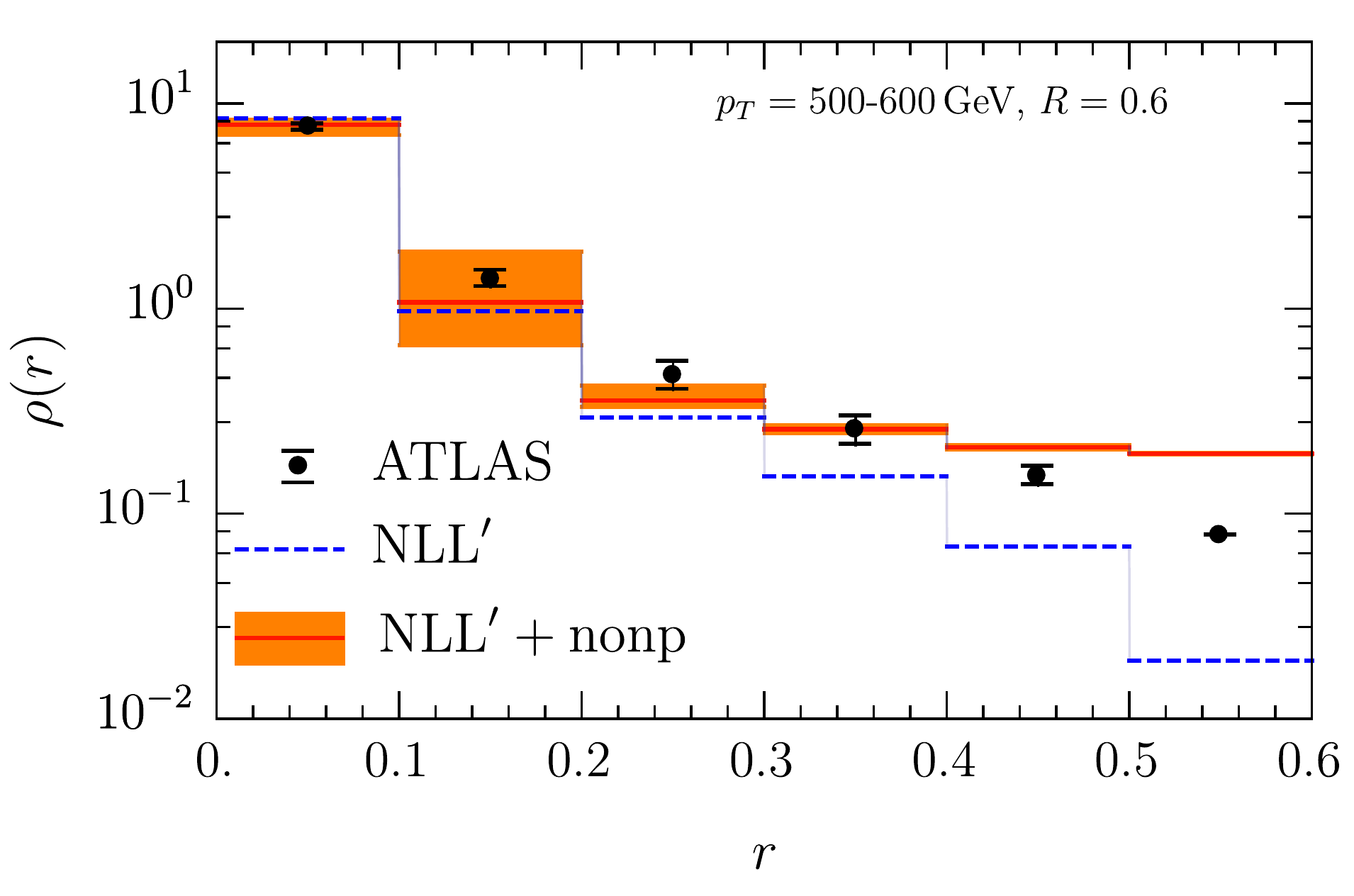} \hfill \phantom{.} \\
    \caption{Comparison of our numerical results for the differential jet shape $\rho(r)$ and the ATLAS data of~\cite{Aad:2011kq}, for the same kinematics as in fig.~\ref{fig:ATLAS1}. \label{fig:ATLAS2}}
\end{figure}

\begin{table}[b]
  \centering
  \begin{tabular}{l l | c c c c c }
  \hline \hline
  ATLAS~\cite{Aad:2011kq} & $p_T$~[GeV] & 30-40 & 60-80 & 110-160 & 500-600 \\ \hline
 $E_{\rm cm}=7$~TeV & $f$ & 0.23 & 0.16 & 0.11 & 0.062  \\
    \hline \hline
  CMS~\cite{Chatrchyan:2012mec} & $p_T$~[GeV] & 30-40 & 500-600 & 600-1000  \\ \hline
 $E_{\rm cm}=7$~TeV & $f$ & 0.28 & 0.068 & 0.054   \\
    \hline \hline
 CMS~\cite{Chatrchyan:2013kwa} & $p_T$~[GeV] & $>100$   \\ \hline
 $E_{\rm cm}=2.76$~TeV  & $f$ & 0.050   \\  
  \hline\hline
  \end{tabular}
  \caption{Fitted values for the parameter $f$ of the nonperturbative ``model 2'', for the jet transverse momentum intervals of the ATLAS~\cite{Aad:2011kq} and CMS~\cite{Chatrchyan:2012mec,Chatrchyan:2013kwa} data that we show here. Note that the jet radius $R$ and the rapidity intervals $\eta$ differ between the different data sets. \label{tab:NPparamf}}
\end{table}

In fig.~\ref{fig:ATLAS2}, we show our numerical results for the differential jet shape $\rho(r)$ in comparison to the corresponding ATLAS data for the same jet kinematics. We use the same $f$ values for the nonperturbative model as for the integrated jet shape. Again we observe that a large nonperturbative correction is needed to describe the differential jet shape data for jets with low transverse momentum. This is particularly pronounced close to the jet boundary, where no resummation of $r/R$ is required, and the fixed-order expressions in \eq{psi_fo} clearly undershoot the data. (This was not so visible for the integrated jet shape, since everything is close to one.) For the higher transverse momentum bins, the agreement between the data and the purely perturbative results is improved and only a smaller nonperturbative correction is needed. However, close to the jet boundary, the effect of the nonperturbative correction remains substantial, even for the highest $p_T$ interval. We note that for $p_T = 30 - 40$ GeV, the nonperturbative model causes the second bin to be higher than the first. This feature is already visible in \fig{nonp}, both for our model 2 and the curve for Pythia at hadron level, since the integrated jet shape is not the steepest at $r=0$.

\begin{figure}[t]
    \centering
     \hfill \includegraphics[width=0.48\textwidth]{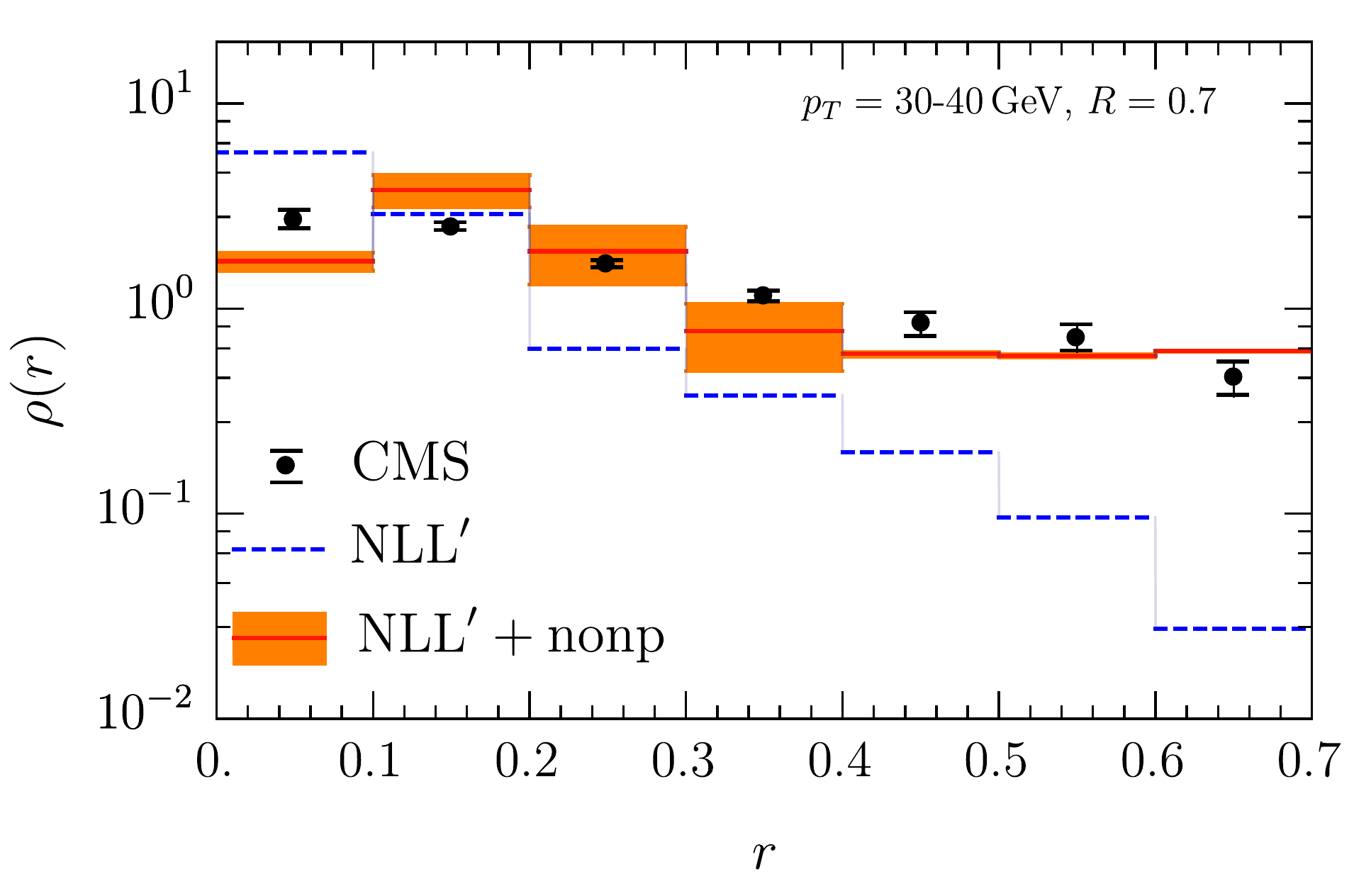} \hfill 
     \includegraphics[width=0.48\textwidth]{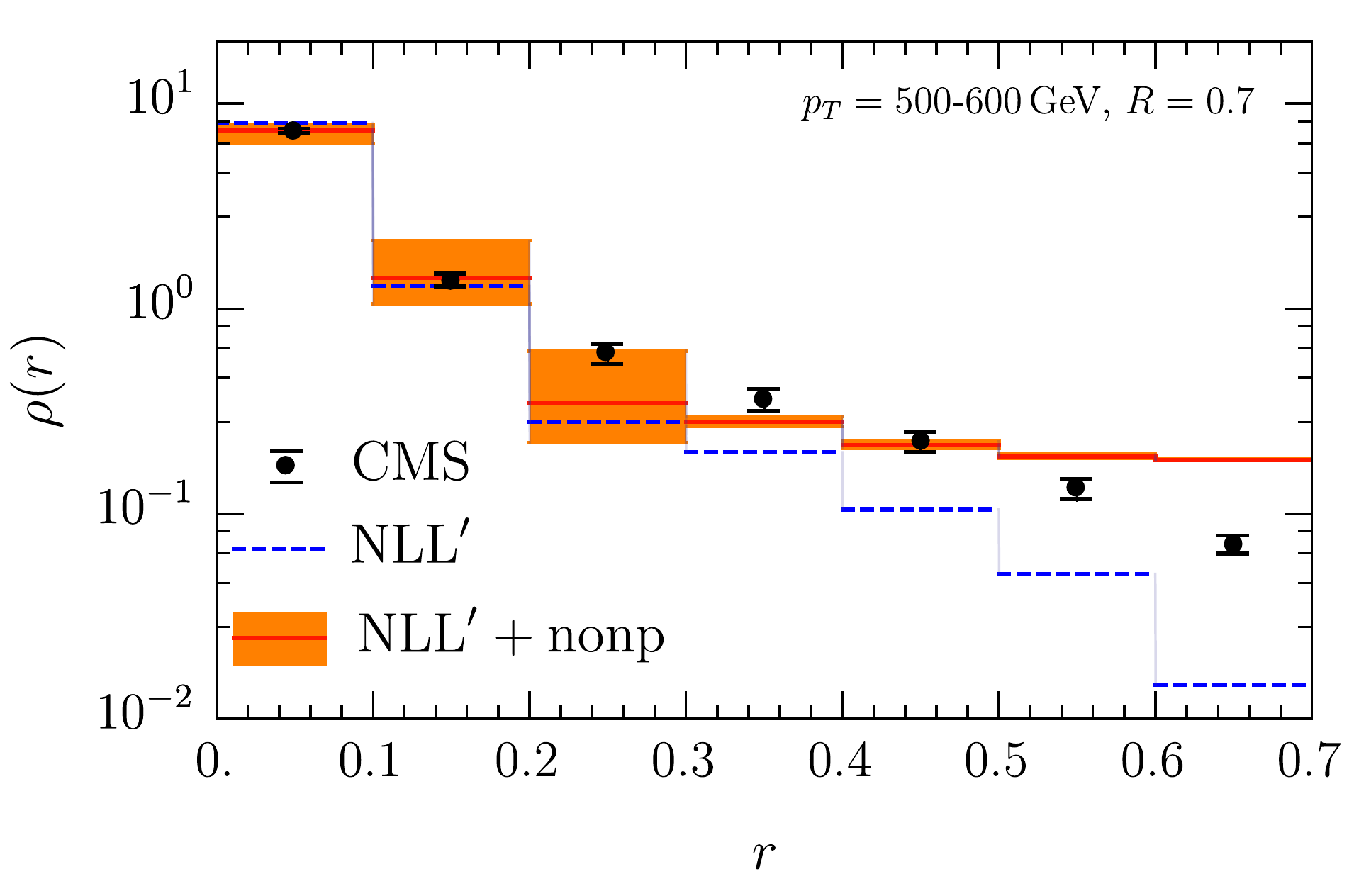} \hfill \phantom{.} \\
     \hfill
     \includegraphics[width=0.48\textwidth]{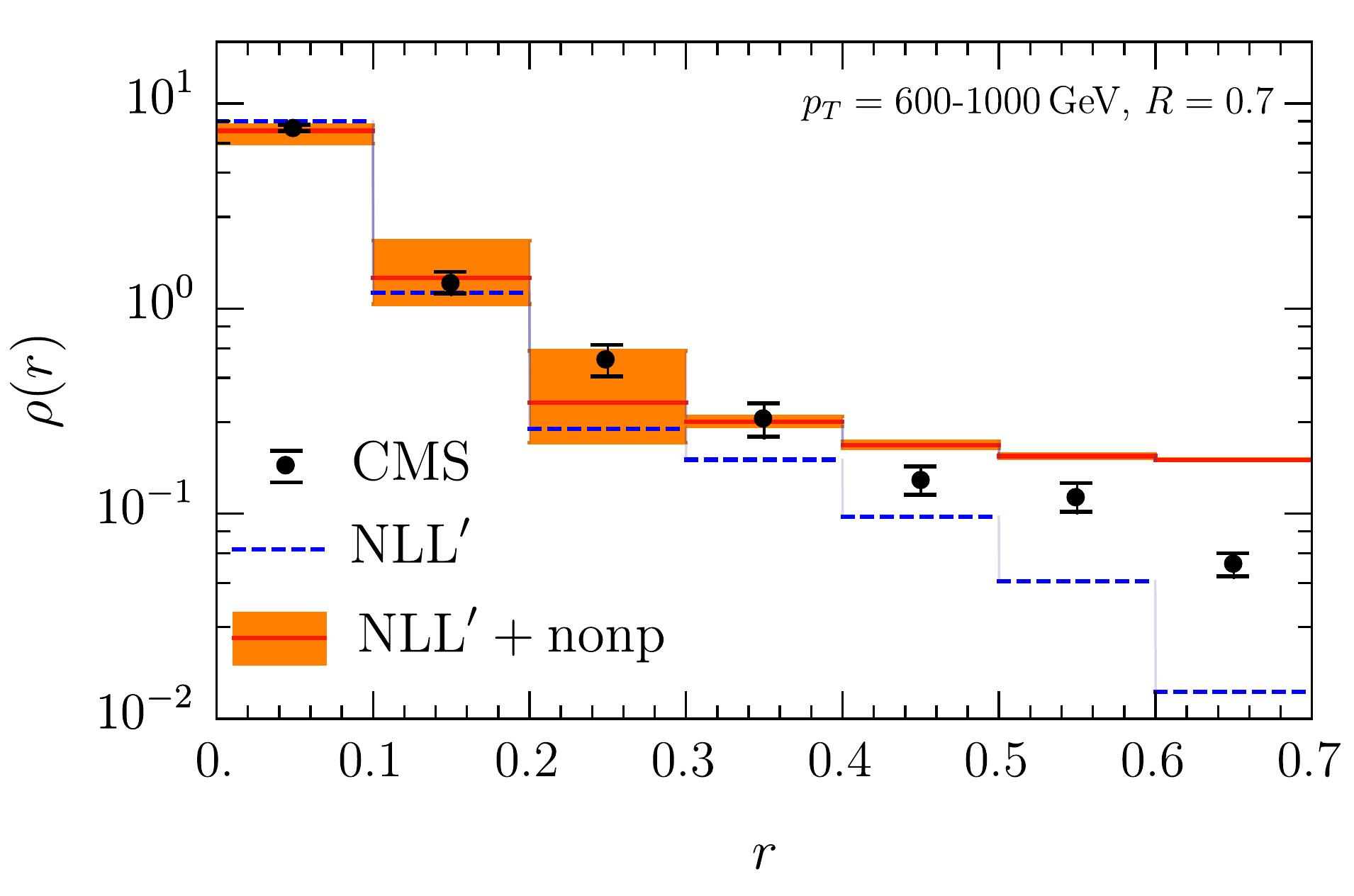} \hfill \phantom{.} \\
    \caption{Comparison of our numerical results for the differential jet shape $\rho(r)$ to the CMS data of ref.~\cite{Chatrchyan:2012mec}. The jets are obtained using anti-k$_T$ with $R=0.7$ and, $|\eta|<0.5$ at $E_{\rm cm}=7$~TeV. We show the result for three intervals of the jet transverse momentum in the range of $p_T=30-1000$~GeV, as labeled in the different panels. \label{fig:CMS}}
\end{figure}

Next, we consider the CMS data set of ref.~\cite{Chatrchyan:2012mec} for the differential jet shape $\rho(r)$, which was also measured at a center-of-mass energy of $E_{\rm cm}=7$~TeV. The kinematics of the reconstructed inclusive jet sample is similar to the setup from ATLAS, but now using anti-k$_T$ with $R=0.7$, $|\eta|<0.5$ and dividing the jet transverse momentum $p_T$ into several intervals in the range of $30-1000$~GeV. In fig.~\ref{fig:CMS} we show the comparison of our numerical results with and without the nonperturbative contribution. We again fit the nonperturbative parameter $f$ of ``model 2'', listing their values in table~\ref{tab:NPparamf}. These values are consistent with those obtained for ATLAS, in the cases for which we have a corresponding $p_T$ interval. (They do not need to be exactly the same, given the difference in jet radius and rapidity interval.) We note again that even for the highest jet transverse momentum interval $p_T=600-1000$~GeV a significant nonperturbative correction is needed in order to describe the energy distribution close to the edge of the jet, $r\lesssim R$, as we also observed for the ATLAS data. 

\begin{figure}[t]
    \centering
    \includegraphics[width=0.52\textwidth]{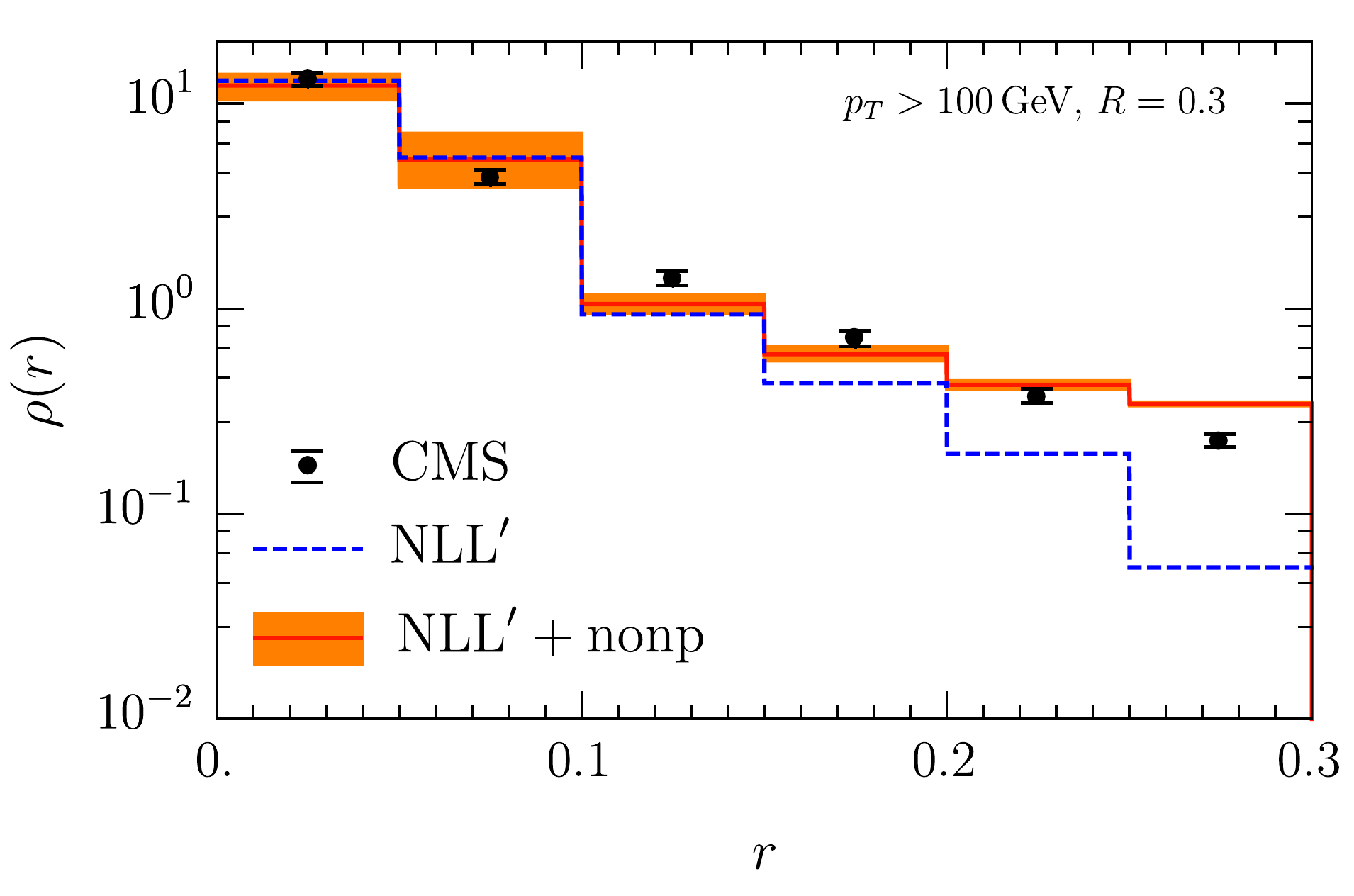}
    \caption{Comparison of the differential jet shape $\rho(r)$ and the CMS data of ref.~\cite{Chatrchyan:2013kwa}. The jets are obtained using anti-k$_T$ with $R=0.3$, $p_T>100$~GeV, $0.3<|\eta|<2$ at $E_{\rm cm}=2.76$~TeV. \label{fig:CMS2}}
\end{figure}

Finally, we compare to the differential jet shape measurement of CMS in ref.~\cite{Chatrchyan:2013kwa}. This data set was taken at $\sqrt{s}=2.76$~TeV as a baseline measurement for a heavy-ion analysis. The jets were reconstructed with $p_T>100$~GeV and $0.3<|\eta|<2$ using the anti-k$_T$ algorithm with $R=0.3$. Analogous to the two data sets discussed above, the jet shape data is separated into six bins with a distance in $r$ of 0.05. An additional cut of $p_{Ti}>1$~GeV was imposed on the transverse momentum of each particle in the jet, and only charged particles are used to determine the jet shape even though full jets are reconstructed. Furthermore, the jet $p_T$ was smeared to account for the difference of the jet energy resolution between proton-proton and heavy-ion collisions. This should be kept in mind when comparing to our calculation, since we do not include these effects. Nevertheless, we find good agreement as shown in fig.~\ref{fig:CMS2}. The edge of the jet is still dominated by nonperturbative physics for $R=0.3$. However, the data is closer to the purely perturbative result compared to the jet shape for a similar $p_T$ interval in fig.~\ref{fig:ATLAS2} for $R=0.6$ jets. Indeed, the fitted value of the nonperturbative parameter $f$, see table~\ref{tab:NPparamf}, is about half of that for the $p_T$ interval 110-160 GeV in the ATLAS data. This is not surprising, given the lower center-of-mass energy and the smaller jet radius. In general, we conclude that our numerical results agree well with the data, once the nonperturbative correction is taken into account. 

\section{Conclusions}
\label{sec:conclusions}

In this paper we calculate the jet shape at next-to-leading logarithmic accuracy (NLL$'$). Specifically, we account for the single logarithms of the jet radius $R$ and the double logarithms of $r/R$ at next-to-leading logarithmic order, and match to next-to-leading order. To achieve this accuracy, the recoil of soft radiation on the jet axis must be included when $r \ll R$. This involves the one-loop calculation of a recoil-sensitive collinear function and rapidity logarithms, that we resum using the rapidity renormalization group, as well as non-global logarithms from in vs.~out-of-jet soft radiation. Our calculation constitutes the first extension of this classic jet substructure observable beyond leading logarithmic accuracy.

The inclusion of higher-order corrections significantly reduces the uncertainty of the perturbative predictions. However, we also find that the effect of nonperturbative contributions are substantial at the LHC, in particular for low jet transverse momenta and close to the edge of the jet. To enable a comparison to the available data from ATLAS and CMS, we explored two one-parameter models, choosing the model that best captures these effects in \Pythia. Including this, we find good agreement between our predictions and LHC data. One interesting future direction is to obtain predictions that can be compared to LEP and HERA data, where these nonperturbative effects should be substantially smaller. On the other hand, for the LHC it is natural to consider grooming, to reduce the contamination from these effects, which will be the topic of a forthcoming publication~\cite{CKLRW}.

\acknowledgments
We thank Y.-J.~Lee, B.~Nachman, D.~Neill, N.~Sato and F.~Yuan for discussions and K.~Lee for feedback on the manuscript. This work is supported by the U.S.~Department of Energy under Contract No.~DE-AC02-05CH11231, the LDRD Program of Lawrence Berkeley National Laboratory, by the ERC grant ERC-STG-2015-677323 and the D-ITP consortium, a program of the Netherlands Organization for Scientific Research (NWO) that is funded by the Dutch Ministry of Education, Culture and Science (OCW).

\appendix

\section{Jet function for the jet shape when $\mathbf{r \lesssim R}$}
\label{app:large_r}

The jet function (or central subjet function) in \eq{G_def} for the anti-$k_T$ algorithm is up to one-loop order given by~\cite{Kang:2017mda}
\bea 
&\cG_q^{\mathrm{jet}}(z, z_r, p_T R , r/R, \mu)
\nnu
& = \delta(1\!-\!z)\delta(1\!-\!z_r)+\f{\as}{2\pi}\bigg\{\delta(1\!-\!z_r)L_R\left[P_{qq}(z)+P_{gq}(z)\right]
-\delta(1\!-\!z_r)\bigg[2C_F(1\!+\!z^2)\Big(\f{\ln(1\!-\!z)}{1-z}\Big)_+ 
\nnu & \quad
+2P_{gq}(z)\ln(1-z) +C_F\bigg] 
+ \delta(1-z)\theta\Big(z_r>\f12\Big)[P_{qq}(z_r)+P_{gq}(z_r)](L_{r/R}+2\ln z_r)
\nnu & \quad
+\theta(r<R/2) \bigg[\delta(1-z)\delta(1-z_r)C_F\bigg(\f72+3\ln 2-\f{\pi^2}{3} \bigg)
\nnu & \qquad
-\delta(1-z)\theta\Big(\f12<z_r<1-\f{r}{R} \Big)[P_{qq}(z_r)+P_{gq}(z_r)]\big(L_{r/R}+2\ln(1-z_r)\big)\bigg]
\nnu & \quad
+ \theta(r>R/2) \bigg[
 \delta(1\!-\!z)\delta(1\!-\!z_r)\, C_F\bigg(-\f{1}{2}L_{r/R}^2+\f{3}{2}L_{r/R}-2L_{r/R}\ln\Big(1\!-\!\f{r}{R}\Big)
  +4\text{Li}_2\Big(1\!-\!\f{r}{R}\Big)
\nnu & \qquad 
+\f12-\f{2\pi^2}{3}+6\f{r}{R} \bigg)
-\delta(1-z)\theta\Big(\f12<z_r<\f{r}{R} \Big)[P_{qq}(z_r)+P_{gq}(z_r)]\big(L_{r/R}+2\ln z_r\big)\bigg]
\bigg\} \,,
\nnu
&\cG_g^{\mathrm{jet}}(z, z_r, p_T R , r/R, \mu)
\nnu
& = \delta(1-z)\delta(1-z_r)+\f{\as}{2\pi}\bigg\{\delta(1-z_r)L_R\big[P_{gg}(z)+2n_fP_{qg}(z)\big]
\nnu & \quad
+ \delta(1-z)\theta\Big(z_r>\f12\Big)[P_{gg}(z_r)+2n_fP_{qg}(z_r)](L_{r/R}+2\ln z_r)
\nnu & \quad
-\delta(1-z_r)\bigg[4C_A\f{(1-z+z^2)^2}{z}\Big(\f{\ln(1-z)}{1-z}\Big)_+ +4n_f\big(P_{qg}(z)\ln(1-z) +T_Fz(1-z)\big)\bigg]
\nnu & \quad
 +
 \theta(r<R/2) \bigg[\delta(1-z)\delta(1-z_r)\bigg(C_A\bigg(\f{137}{36}+\f{11}{3}\ln 2-\f{\pi^2}{3}\bigg)-T_F n_f\bigg(\f{23}{18}+\f43\ln 2\bigg) \bigg)
\nnu & \qquad
-\delta(1-z)\theta\Big(\f12<z_r<1-\f{r}{R} \Big)[P_{gg}(z)+2n_fP_{qg}(z)]\big(L_{r/R}+2\ln(1-z_r)\big)\bigg] 
\nnu & \quad
 +\theta(r>R/2) \bigg[\delta(1-z)\delta(1-z_r)\bigg[-\f{C_A}{2}L_{r/R}^2+\f{\beta_0}{2}L_{r/R}-2C_A L_{r/R}\ln\Big(1-\f{r}{R}\Big)
 \nnu & \qquad
 +4 C_A \text{Li}_2\Big(1\!-\!\f{r}{R}\Big)-C_A\f{2\pi^2}{3}+C_A\Big(\f{8r}{R}-\f{r^2}{R^2}+\f{4r^3}{9R^3}\Big)
 + T_F n_f\Big(\f{1}{3}-\f{4r}{R}+\f{2r^2}{R^2}-\f{8r^3}{9R^3}\Big)\!\bigg)
\nnu & \qquad
-\delta(1-z)\theta\left(\f12<z_r<\f{r}{R} \right)[P_{gg}(z)+2n_fP_{qg}(z)]\left(L_{r/R}+2\ln z_r\right)\bigg]\bigg\}\,.
\eea
The splitting functions $P_{ij}$ are given in \eq{split}, and we use the following short-hand notation 
\begin{align} \label{eq:L_def_app}
 L_R = \ln\Big(\f{\mu^2}{p_{T}^2 R^2} \Big) 
  \,, \qquad
  L_r = \ln\Big(\f{\mu^2}{p_{T}^2 r^2}\Big)  
 \,, \qquad 
   L_{r/R} \equiv L_r - L_R = \ln \Big(\frac{R^2}{r^2}\Big)
\,.\end{align}

\section{Anomalous dimensions}
\label{app:anom}

The one-loop anomalous dimensions directly follow from the expressions in secs.~\ref{sec:hard}, \ref{sec:soft} and \ref{sec:coll}, 
\begin{align}
 \ga^H_{qq}(z,p_{T} R,\mu)
  &= \f{\as}{\pi} \bigg[C_F \Big(-  L_R - \frac32\Big)\, \delta(1-z)
+ P_{qq}(z)  \bigg]
, \nn \\
 \ga^H_{qg}(z,p_{T} R,\mu) 
 &= \f{\as}{\pi}\, P_{gq}(z)
, \nn \\
\ga^H_{gg}(z,p_{T} R,\mu)
  &= \f{\as}{\pi} \bigg[\Big(- C_A  L_R - \frac12 \beta_0 \Big)\, \delta(1-z)
+ P_{gg}(z)  \bigg]
, \nn \\
 \ga^H_{gq}(z,p_{T} R,\mu) 
 &= \f{\as}{\pi}\, P_{qg}(z)
, \nn \\
\ga^C_q(\mu,\nu/p_T) 
&= \frac{\al_s C_F}{\pi}\, 
  \Big(2  \ln \frac{\nu}{2p_{T}} + \frac32\Big)
\,, \nn \\
\ga^C_g(\mu,\nu/p_T) 
&= \frac{\al_s}{\pi}\, 
  \Big(2C_A  \ln \frac{\nu}{2p_{T}} + \frac12 \beta_0\Big)
\,, \nn \\
\ga^S_q(\mu,\nu R)
&= \frac{\al_s C_F}{\pi}\, \ln \frac{4\mu^2}{\nu^2 R^2}
\,, \nn \\
\ga^S_g(\mu,\nu R)
&= \frac{\al_s C_A}{\pi}\, \ln \frac{4\mu^2}{\nu^2 R^2}
\,, \nn \\
\ga^\nu_q(k_\perp,\mu)
&= 4\al_s C_F\, \frac{1}{\mu^2}\,\frac{1}{(k_\perp^2/\mu^2)}_+
\,, \nn \\
\ga^\nu_g(k_\perp,\mu)
&= 4\al_s C_A\, \frac{1}{\mu^2}\,\frac{1}{(k_\perp^2/\mu^2)}_+
\,.\end{align}
Here $L_R$ is defined in \eq{L_def_app} and the splitting functions are given in \eq{split}.
To achieve NLL$'$ accuracy, we include the two-loop cusp anomalous dimensions for the $\ln \mu$ and $\ln \nu$ terms in the anomalous dimensions. This amounts to multiplying these terms by
\begin{align}
   1 + \frac{\al_s}{4\pi} \Bigl[\Bigl( \frac{67}{9} -\frac{\pi^2}{3} \Bigr)\,C_A  -
   \frac{20}{9}\,T_F\, n_f \Bigr]
\,.\end{align}

\section{Rapidity anomalous dimension of the one-loop collinear function}
\label{app:check}

We will now check that our expression for the collinear function satisfies the rapidity evolution equation in \eq{RGE}, which expanded up to one-loop order reads
\begin{align} \label{eq:satisfies}
\f{\df}{\df \ln \nu}\,C_i^{(1)}(z_r,p_T r,k_\perp,\mu,\nu) &= -\int\! \frac{\df^2 k_\perp'}{(2\pi)^2}\,
\gamma^{\nu(1)}_i(k_\perp-k_\perp',\mu)\,C_i^{(0)}(z_r,p_T r,k_\perp',\mu,\nu)
\,.\end{align}
Note that $\gamma_i^{\nu(1)}(k_\perp)$ was already obtained from the soft function, so this provides a cross check. From \eq{QuarkCollinear} we find that the left-hand side of \eq{satisfies} is given by
\begin{align} \label{eq:Lnu_LHS}
\frac{\mathrm{d}C^{(1)}_{i,\theta<r}}{\mathrm{d}\ln \nu} &=\Theta(k_\perp<p_T r)\,\delta(1-z_r)\, \frac{\alpha_s C_i}{2\pi^{2}} \int_{0}^{2\pi}\mathrm{d}\phi\, L_1
\nn \\ & \quad
+ \Theta(k_\perp>p_T r)\,\,\delta(1-z_r)\, \frac{\alpha_s C_i}{\pi^{2}}\int_{-\phi_{\rm max}}^{\phi_{\rm max}}\mathrm{d}\phi\, \ln \Bigl(\frac{\beta_{2}^{\text{min}}}{\beta_{2}^{\text{max}}} \Bigr),
\end{align}
where the color factor is given by $C_q = C_F$ for quarks and $C_g = C_A$ for gluons.

\begin{figure}[t]
\centering
\hfill \includegraphics[scale=0.3]{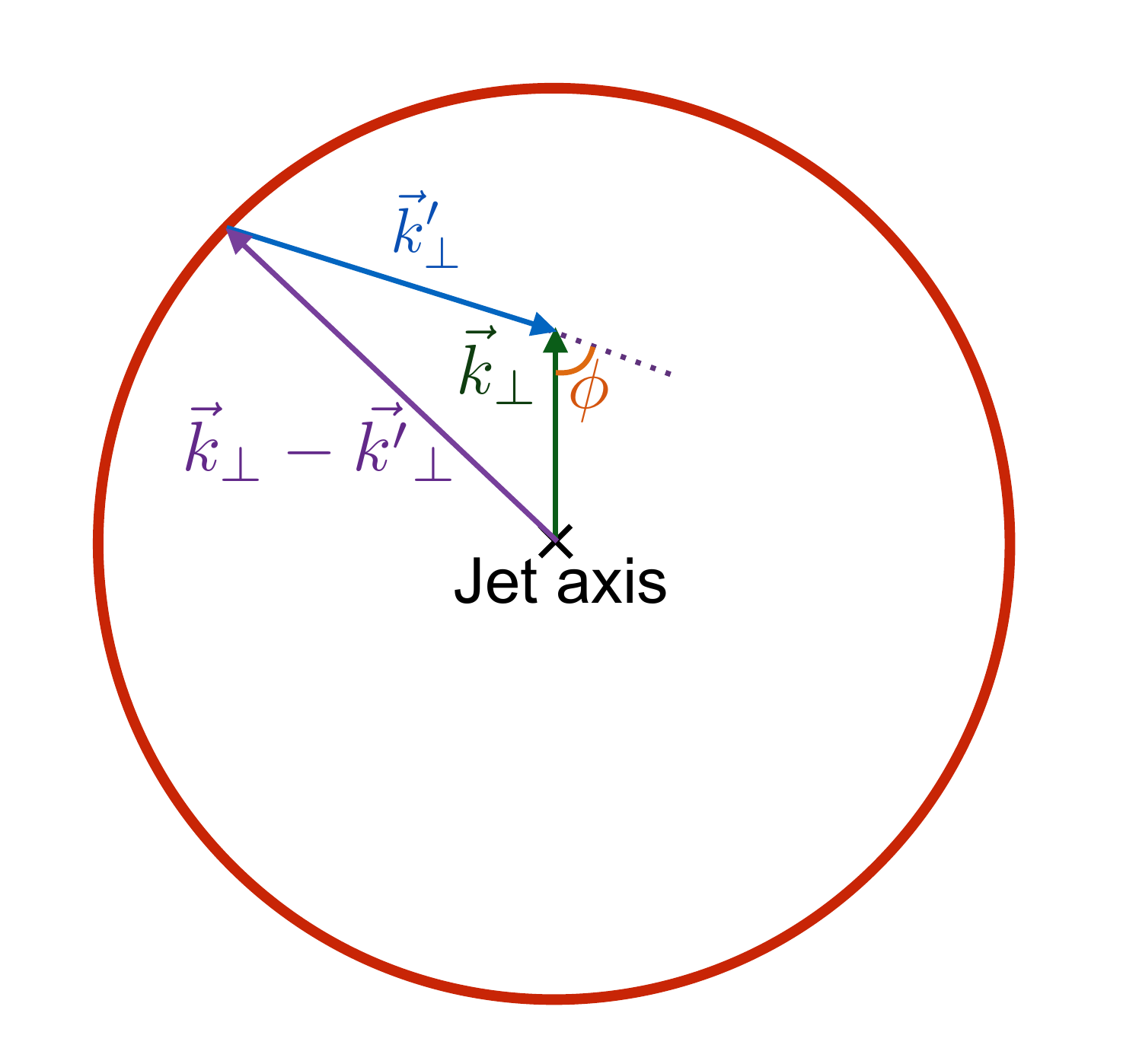} \hfill 
\includegraphics[scale=0.3]{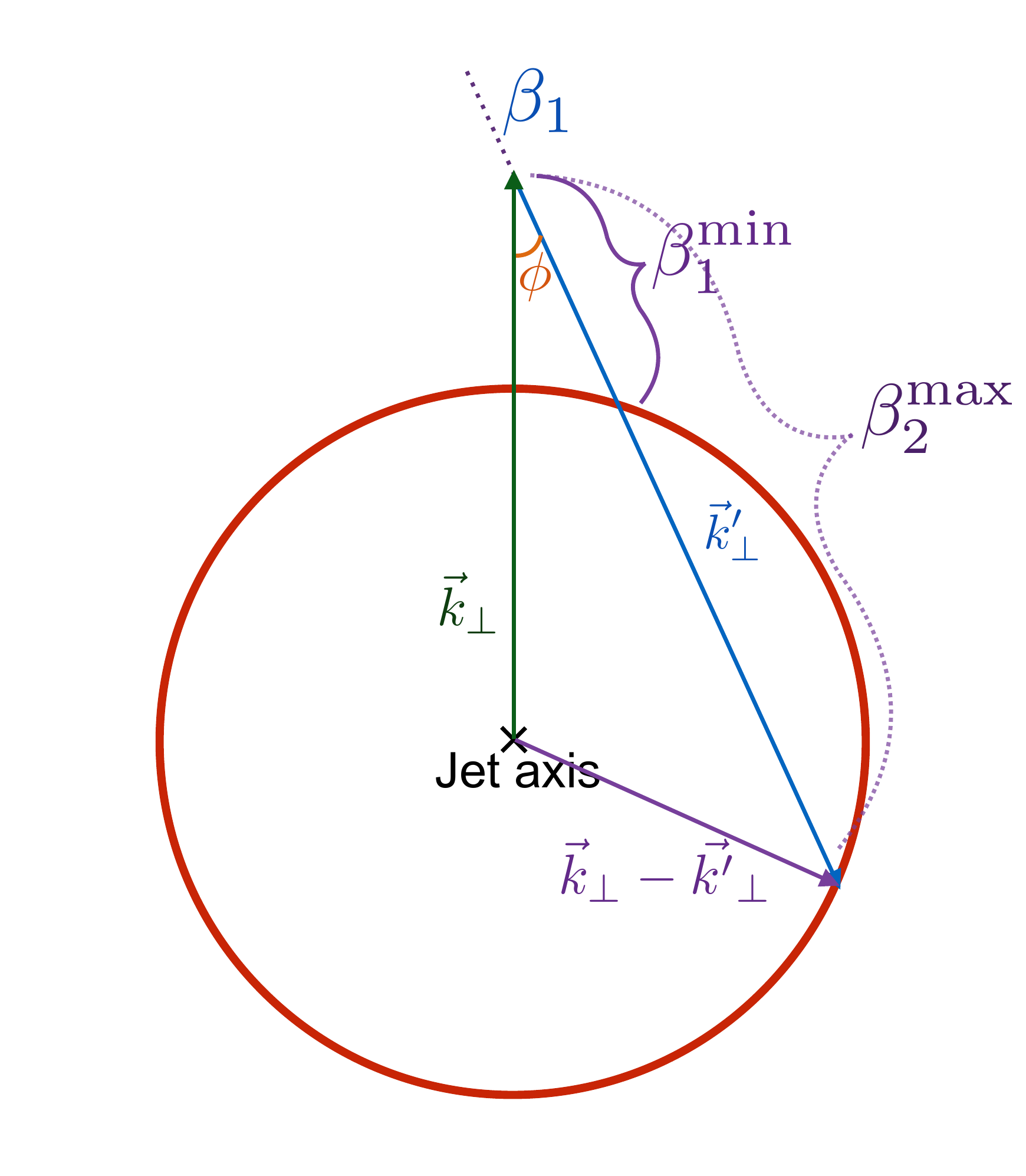} \hfill \phantom{.} \\
\caption{The geometry of $|k_\perp-k'_\perp|< p_T r$ for $|k_\perp| < p_Tr$ (left) and $|k_\perp| > p_Tr$ (right).}
\label{fig:anomdim}
\end{figure}

Since the terms in \eq{Lnu_LHS} correspond to $\theta<r$ and $\theta>r$, respectively, it is natural to split up the calculation of the right-hand side of \eq{satisfies} in the same pieces. For $\theta<r$ this requires computing,
\begin{align}
  &\Theta(k_\perp<p_T r)\int \frac{\mathrm{d}^{2}k'_\perp}{(2\pi)^{2}}\frac{1}{\mu^{2}}\frac{1}{(k_\perp'^{2}/\mu^{2})}_+\Theta\bigl( |k_\perp-k'_\perp|< p_T r\bigr)
   \nn \\ & \quad
  =
\Theta(k_\perp<p_T r) \int_{0}^{2\pi} \mathrm{d}\phi \int_{0}^{\big( \frac{p_T \beta_1^{\text{max}}}{\mu}\big)^{2}} \frac{\mathrm{d}(k'^{2}_\perp/\mu^{2})}{8\pi^{2}}\frac{1}{(k_\perp'^{2}/\mu^{2})}_+ 
  \nn \\ & \quad
= -\Theta(k_\perp<p_T r)\,\frac{1}{8\pi^2} \int_{0}^{2\pi} \mathrm{d}\phi\, L_1(k_\perp)
\,,\end{align}
which agrees with the first term in \eq{Lnu_LHS} when including the overall factor of $-4 \al_s C_F$. Here we used that $|k_\perp-k'_\perp|< p_T r$ is equivalent to $|k'_\perp|< p_T \beta_1^{\rm max}$, as it involves the same geometry used in the computation of the collinear function, see the left panel of \fig{anomdim}.

Similarly, for $\theta<r$ we can rewrite the condition $|k_\perp-k'_\perp|< p_T r$ as  $p_T \beta_2^{\rm min} < |k'_\perp| < p_T \beta_2^{\rm max}$ and $-\phi_{\rm max} < \phi < \phi_{\rm max}$, as is clear from the right panel of \fig{anomdim}. This leads to
\begin{align}
  &\Theta(k_\perp>p_T r)\int \frac{\mathrm{d}^{2}k'_\perp}{(2\pi)^{2}}\frac{1}{\mu^{2}}\frac{1}{(k_\perp'^{2}/\mu^{2})}_+\Theta\bigl( |k_\perp-k'_\perp|< p_T r\bigr)
   \nn \\ & \quad
  =
\Theta(k_\perp>p_T r) \int_{-\phi_{\rm max}}^{\phi_{\rm max}} \mathrm{d}\phi \int_{\big( \frac{p_T \beta_2^{\text{min}}}{\mu}\big)^2}^{\big( \frac{p_T \beta_2^{\text{max}}}{\mu}\big)^{2}} \frac{\mathrm{d}(k'^{2}_\perp/\mu^{2})}{8\pi^{2}}\frac{1}{(k_\perp'^{2}/\mu^{2})}_+ 
  \nn \\ & \quad
= -\Theta(k_\perp<p_T r)\,\frac{1}{4\pi^2} \int_{0}^{2\pi} \mathrm{d}\phi\, \ln \Bigl(\frac{\beta_{2}^{\text{min}}}{\beta_{2}^{\text{max}}} \Bigr)
\,,\end{align}
yielding the second term on the right-hand side of \eq{Lnu_LHS}.

\bibliographystyle{JHEP}
\bibliography{bibliography}

\end{document}